\documentclass[a4paper,fleqn]{cas-dc}

\usepackage{soul} 
\usepackage{color, xcolor} 
\usepackage{lineno}
\usepackage[numbers]{natbib}
\usepackage[boxed,commentsnumbered,ruled,linesnumbered]{algorithm2e}
\usepackage{hyperref}
\hypersetup{
    colorlinks=true,
    linkcolor=blue,
    urlcolor=blue,
    anchorcolor=blue,
    citecolor=blue
}

\newtheorem{definition}{Definition}  

\begin{document}
\shorttitle{Dual Pruning and Sorting-Free Overestimation for Average-Utility Sequential Pattern Mining}

\shortauthors{K. Cao et al.}


\title [mode = title]{Dual Pruning and Sorting-Free Overestimation for Average-Utility Sequential Pattern Mining}   

\author[1,2]{Kai Cao}
\ead{caokai.pds@gmail.com}
\address[1]{School of Cyberspace Security, Hainan University, Haikou 570228, China}
\address[2]{School of Cryptology, Hainan University, Haikou 570228, China}

\author[3]{Yucong Duan}
\cortext[cor1]{Corresponding author}
\ead{duanyucong@hotmail.com}
\address[3]{School of Computer Science and Technology, Hainan University, Haikou 570228, China}
\cormark[1]

\author[4]{Wensheng Gan}
\ead{wsgan001@gmail.com}
\address[4]{College of Cyber Security, Jinan University, Guangzhou 510632, China}

\begin{abstract}
    In a quantitative sequential database, numerous efficient algorithms have been developed for high-utility sequential pattern mining (HUSPM). HUSPM establishes a relationship between frequency and significance in the real world and reflects more crucial information than frequent pattern mining. However, high average-utility sequential pattern mining (HAUSPM) is deemed fairer and more valuable than HUSPM. It provides a reasonable measure for longer patterns by considering their length. In contrast to scenarios in retail business analysis, some pattern mining applications, such as cybersecurity or artificial intelligence (AI), often involve much longer sequences. Consequently, pruning strategies can exert a more pronounced impact on efficiency. This paper proposes a novel algorithm named HAUSP-PG, which adopts two complementary strategies to independently process pattern prefixes and remaining sequences, thereby achieving a dual pruning effect. Additionally, the proposed method calculates average utility upper bounds without requiring item sorting, significantly reducing computational time and memory consumption compared to alternative approaches. Through experiments conducted on both real-life and synthetic datasets, we demonstrate that the proposed algorithm could achieve satisfactory performance.
\end{abstract}


\maketitle

\section{Introduction}  \label{sec: introduction}

A large scale of smart devices facilitates the collection of a wide range of data and information in real-world applications, creating an urgent need for efficient techniques to extract valuable patterns from such massive datasets. Among the various techniques for discovering interesting patterns based on their occurrence frequency in transaction databases, frequent pattern mining (FPM) \cite{Ref1, Ref2} stands out as a representative method. The pioneering FPM method was proposed by Agrawal et al. in 1993 \cite{Ref3}. Over time, FPM has found extensive applications in diverse domains such as biology, environment, and social networks \cite{Ref6}. In our daily lives, sequential pattern mining (SPM) \cite{Ref43} has been utilized to analyze consumer behavior \cite{Ref7}, cybersecurity, and to extract meaningful knowledge from the web clickstream data. SPM aims to derive frequent sequential patterns from the sequential database. 
However, it is important to note that there is an underlying assumption in knowledge discovery problems: high occurrence frequency implies high relevance. In real-life scenarios, various interestingness criteria hold more significance than the support of a pattern. As is well known, in the retail business, the pursuit of profit often outweighs sales volume. Consequently, while some patterns are indeed important, only those with high frequency are likely to be identified. To address this issue, some research explored the concept of relative item importance \cite{Ref9, Ref10}, incorporated the notion of utility, and proposed high utility patterns or itemsets mining (HUPM or HUIM) \cite{Ref11} to identify patterns with higher utility \cite{Ref13, Ref12}.

The relative importance of an item is defined by a specified constant, namely external utility, while its quantity is captured by internal utility. The utility of a pattern could be measured using a user-specified utility function, which is typically defined by both external and internal utilities. The research has shown that the utility measure is more comprehensive than the support measure \cite{Ref14}. In a given traditional transaction, an item or an itemset appears no more than once and has a unique utility value. However, in quantitative sequential databases where sequences are ordered (e.g., by time), an itemset may have multiple occurrences and multiple utilities. To tackle this scenario, the high-utility sequential pattern mining (HUSPM) task was developed \cite{Ref14} and has been applied in various common scenarios, including m-commerce \cite{Ref16}, web services, gene regulation \cite{Ref20}, and healthcare activity-cost events \cite{Ref21}.

Compared to the previously mentioned FPM, SPM, and HUPM, HUSPM incorporates the consideration of chronological item ordering and associated utility values \cite{Ref22}. HUSPM has demonstrated the effectiveness of efficient pruning strategies and novel data structures \cite{Ref24}. However, several challenging problems remain. In HUSPM, a typical challenge arises from the absence of the downward closure property. This property, observed in the frequency measures, does not hold for the utility measures. One solution to solve this problem is to design upper bounds (UBs) on utility, such as sequence-weighted utilization (SWU) \cite{Ref14}, that satisfy the property. When dealing with low minimum utility threshold values, in particular, tighter UBs are particularly effective in the early pruning of low utility generated sequences. Another drawback of HUSPM is that longer patterns have an advantage over shorter ones in terms of higher utility. But, as the number of patterns increases, understanding the association rules between them becomes challenging \cite{Ref29}. To overcome this drawback, Hong et al. \cite{Ref30} proposed a new measurement, called average utility, which provides a better assessment than itemset utility. By considering both the utility of patterns and their length \cite{Ref33, Ref31}, the high average utility itemset or pattern mining (HAUIM or HAUPM) approach \cite{Ref32, Ref31} extracts the specific patterns corresponding to a higher average utility value. The average utility offers a more objective assessment of a pattern or an itemset compared to the utility, and high average utility patterns or itemsets (HAUPs or HAUIs) are more useful in real-world applications than high utility patterns or itemsets (HUPs or HUIs). Recently, several emerging research directions have focused on addressing practical limitations in real-world applications by building upon the foundations of HUPM or HUIM. For example, HUOPM \cite{Ref4} introduced the occupancy metric, which measures the proportion of a pattern's utility within its supporting transactions, aiming to discover patterns that are both profitable and representative of their transaction contexts. To solve the problem that high-utility patterns may require excessive investment, HEPM \cite{Ref8, Ref17} was proposed based on the utility-investment ratio \cite{Ref70}, which is used to balance utility gains against related costs. Considering the temporal properties of utilities in dynamic data scenarios, HUOPS \cite{Ref71} was developed for utility pattern mining in data streams, while ILTFU-Miner \cite{Ref72} focuses on discovering fuzzy utility patterns considering timestamps. Additionally, DUOS \cite{Ref78}, a new anomaly detection framework, was proposed to identify utility-aware outlier sequential rules (UOSRs) corresponding to anomalous movement behaviors in network intrusion detection, thereby detecting intrusions amid a wide range of cyber threats. However, even these HUPM-based extensions face similar challenges. For instance, HUOMI \cite{Ref15} was developed to handle dynamic datasets, and MHEINU \cite{Ref19} was proposed to deal with datasets containing negative utilities. In particular, to tackle the limitation of pattern length bias, HAUOPM \cite{Ref40} and HAEIM \cite{Ref23} were put forward for incremental databases, and IIMHAUP \cite{Ref73} was proposed for the streaming environment. Besides, DMAUP \cite{Ref75} and SHAUPM \cite{Ref74} also incorporate explicit considerations of pattern length into their utility evaluation metrics, which improves the practical relevance of the discovered results.

As previously mentioned, it is the UBs used in HUPM or HUSPM that are anti-monotonic. However, the average utility measure used in HAUPM or high average-utility sequential pattern mining (HAUSPM) is neither anti-monotonic nor monotonic. While designing a tighter UB for HAUSPM is more complicated, there are more possibilities compared to HUSPM. Some preliminary work had been undertaken to capture HAUIM. The research topic remains in its nascent stage of development, and these strategies are incapable of addressing sequential databases. An approach to identify HAUSPs in sequential databases was proposed, yet the approach is not applicable to a general case \cite{Ref35}. It hinges on the presumption that each distinct item appears only once in a single sequence within the dataset. EHAUSM is the first approach to identify HAUSPs in a quantitative sequential database \cite{Ref35}. However, existing research on upper bound (UB) models for average utility problems has primarily followed two design paradigms. The first paradigm relies on the maximum utility of items, with examples including auub \cite{Ref30, Ref47}, lub, rtub \cite{Ref52}, and mfuub \cite{Ref76}. Nevertheless, when the utility of most items in a sequence is far lower than this maximum utility, the gap between the overestimated utility and the actual utility becomes substantial. The second commonly adopted paradigm estimates utility based on the utilities of the top-\textit{k} items after descending sorting. This design avoids the interference of a few items with extremely high utilities on the overall estimation result, yielding utility estimates that are closer to the actual utility and thus demonstrating distinct advantages. Examples of such models include lubau \cite{Ref76}, krtmuub \cite{Ref76}, BiUB, and twaub \cite{Ref52}. However, to obtain estimates closer to the actual utility, repeated sorting of the changing remaining sequence is required alongside pattern growth. Notably, in contrast to retail business analysis scenarios where sequence lengths tend to be shorter, many pattern mining applications (such as cybersecurity, behavioral analysis, and knowledge discovery) often involve much longer sequences. Obviously, the longer the pattern, the more significant the impact of sorting operations on algorithm efficiency. This raises a critical question: is there an alternative approach that can achieve utility estimates closer to the actual value without relying on sorting operations with high computational complexity?

To ameliorate aspects related to memory consumption and execution time, we combine HAUSPM with a pattern-growth method and design a novel algorithm termed HAUSP-PG. We summarize the major contributions of this study as follows:

\begin{itemize}
    \item  We propose a novel algorithm to identify HAUSPs by considering the sequence utility and its length. We adopt two complementary strategies to prune the irrelevant items from the prefix sequence and to remove the unpromising items from the remaining subsequence independently.
	
    \item  The proposed algorithm incorporates the pattern-growth method into HAUSPM. Based on the concept of average utility, we design strict upper bounds and some novel variants of upper bounds to optimize pruning strategies. The proposed method avoids sorting multiple items in the remaining sequence.
	
    \item  Comprehensive experiments conducted on both synthetic and real-world datasets demonstrate that the proposed algorithm HAUSP-PG has a good performance for HAUSPM. Its strengths lie particularly in runtime efficiency, memory usage optimization, filtration of promising generated sequences, and scalability.
\end{itemize}

The subsequent sections of this paper are organized as follows. A concise review of related work is stated in Section \ref{sec: relatedWork}. Section \ref{sec: preliminary} delineates the formulation of HAUSPM problem and gives some requisite definitions. The proposed algorithm HAUSP-PG is expounded upon, elucidating several strategies and data structures, in Section \ref{sec: algorithm}. Then, the performance evaluation of the proposed algorithm is conducted through experimental results in Section \ref{sec: experiment}. Section \ref{sec: conclusion} brings a succinct summary and the future work.

\section{Related Work} \label{sec: relatedWork}

The related work is structured around three main elements: HUSPM, HAUIM, and HAUSPM.

\subsection{High-utility sequential pattern mining}
\label{sec: huspm}

The problem of mining all itemsets whose utilities are not lower than a predefined minimum is defined as HUIM \cite{Ref36}. The utility in HUIM is not anti-monotone \cite{Ref1}, and it is difficult to develop a powerful pruning strategy to limit its search space. In Ref. \cite{Ref37}, a two-phase algorithm was proposed to mine high transaction-weighted utilization itemsets and prove the corresponding downward closure property. The Two-Phase algorithm required extra scanning to filter the overestimated ones, so the number of candidates affected its efficiency. Further cuts in candidate sequences were considered. The tree-based algorithm UP-Growth \cite{Ref38} was developed to reduce generated sequences. HUI-Miner \cite{Ref39} adopted a vertical representation and discovered HUIs without generating candidate sequences. A single-phase algorithm called d$^2$HUP \cite{Ref41} was designed for enumerating patterns by prefix extensions. EFIM \cite{Ref42} and two UBs were designed to effectively reduce its search space.

SPM was defined as a task to analyze customer consumption records by Agrawal and Srikant in 1995 \cite{Ref43}. In Ref. \cite{Ref44}, Ahmed et al. defined HUSPM as incorporating the concept of utility into SPM. Utility Span and Utility Level, both two-phase algorithms, were proposed, using SWU as the utility upper bound. Utility Level is a level-wise approach and is more straightforward than Utility Span, but Utility Span adopted a pattern-growth method to control the growth of sequences. The projection method for the database was used in the one-phase algorithm UM-Span to avoid multiple scans. USpan \cite{Ref14} adopted the lexicographic quantitative sequence tree (LQS-tree) to avoid multiple scannings. LQS-tree is an efficient prefix-tree structure and could be extended by the extension operation. Sequence projected utilization (SPU) \cite{Ref14} was designed as the pruning strategy to stop traversing deeper in the LQS-tree. Extensive studies have convincingly demonstrated that combining a more accurate upper bound with the corresponding pruning strategies is effective in reducing the search space. The projection-based algorithm PHUS \cite{Ref45} was presented with a sequence-utility upper bound (SUUB) and an efficient indexing strategy. The maximum utility was used as the measure method to simplify the evaluation for HUSPM \cite{Ref32}. Alkan et al. \cite{Ref26} designed the tighter upper bound cumulate rest of match (CRoM) in HuspExt to reduce the search space. Two tighter utility UBs, prefix extension utility (PEU) and reduced sequence utility (RSU), were designed in HUS-Span \cite{Ref27}. An upper bound sequence extension utility (SEU) was designed in ProUM \cite{Ref24}. Based on the upper bound PEU, Gan et al.  \cite{Ref22} proposed the pruning strategies of irrelevant item pruning (IIP) and look ahead removing (LAR), to remove unpromising sequences quickly. Besides the tighter upper bound, the compressed index and storage structure are also effective in improving efficiency. The utility-array data structure was designed to avoid multiple scannings of the original database in ProUM \cite{Ref24}. Another compact data structure, UL-list, was designed in HUSP-ULL \cite{Ref22}. A seqPro structure \cite{Ref46} was designed in HUSP-SP to facilitate sequence generation and calculate utility efficiently.

\subsection{High Average Utility Itemset Mining}
\label{sec: hauim}

HUIM and HUSPM do not take into account the length of a pattern, which leads to short patterns being missed. The length of the high utility pattern is generally larger. The HAUI mining algorithm took length into account when calculating the utility \cite{Ref30}. The average utility upper bound (auub) was proposed to maintain the downward closure property in the first two-phase HAUIM algorithm, TPAU \cite{Ref47}. It overestimated the average utility of the pattern and improved the performance of the pruning process. However, the main drawback of TPAU is the cost of multiple scans and generating a large number of sequences. HAUI-Growth designed a HAUI-tree structure to maintain the average utility and to avoid scanning the original database repeatedly. The projection-based algorithm PBAU \cite{Ref31} utilized a tree structure and index tables. HAUL-Growth \cite{Ref49} introduced a HAUL-Tree structure and employed the pattern growth approach for mining. Instead of the tree structure, HAUI-Miner \cite{Ref50}, MHAI \cite{Ref51}, and EHAUPM \cite{Ref52} adopted the list structure.

HUSPM combines upper bound and pruning strategies to limit the number of generated sequences, and HAUPM improves performance likewise. In addition to the list structure, the new UBs, transaction maximum utility, and maximum average utility were designed in HAUI-Miner and MHAI. Two UBs and a MAU-list were designed in EHAUPM \cite{Ref52}. The \textit{lub} is one of them, a looser upper bound; the other is the \textit{rtub}, a revised tighter upper bound. The top-\textit{k} revised transaction maximum utility upper bound (krtuub) and the maximum following utility upper bound (\textit{mfuub}) were presented in TUB-HAUPM \cite{Ref53}.  In VMHAUI \cite{Ref54, Ref55}, three tighter UBs were designed, which introduce vertical information among transactions instead of horizontal information from one transaction. A list-based structure was employed in Ref. \cite{Ref56}, and two tighter UBs—the maximum remaining average utility upper bound and the tight maximum average utility upper bound—were designed. An early abandoning strategy, namely EA, was designed for terminating the KFAU-List construction of unpromising itemsets in EMAUI \cite{Ref57}.

\subsection{High Average Utility Sequential Pattern Mining}
\label{sec: hauspm}

The utility function tends to be biased toward finding long patterns, which is a limitation of HUSPM. A weak upper bound on the au measure and the other two UBs were introduced in EHAUSM \cite{Ref35}. Four pruning strategies are proposed to speed up HAUS mining. A novel framework of SPM was proposed to mine the set of potential HAUSPs from the uncertain dataset \cite{Ref58}. HANP-Miner \cite{Ref59} was proposed to search for sequential patterns under the one-off condition. It utilized a simplified Nettree in search and backtracking strategies for greater efficiency. HAOP-Miner \cite{Ref60} was designed to find non-overlapping sequential patterns. The reverse filling method was designed and utilized to avoid redundant calculations. C-FHAUSPM \cite{Ref61} was designed for discovering frequent high minimum average utility sequences with constraints. In addition, some bi-objective algorithms \cite{Ref66} were proposed to discover patterns from a multi-objective perspective. FLCHUSM problem was formalized, targeting both high average utility and low average cost, and FLCHUSPM \cite{Ref67} with an upper bound and a lower bound was proposed.

HUPM is considered an unfair measurement of utility for mining interesting patterns in some cases. It neglects sequence length and the potential existence of numerous low-utility items within a dataset. HAUPM was proposed as one approach to address these limitations. Nevertheless, in the quantitative sequential dataset, the average utility is neither monotonic nor anti-monotonic. Typically, an upper bound is devised to satisfy the anti-monotone property and to prune the search space. However, overestimating the actual utility to establish an upper bound inevitably introduces gaps between the estimated and actual utilities. Larger gaps result in more sequences being generated. Therefore, it is imperative to develop algorithms that achieve good performance while avoiding excessive overestimation. The task of estimating tighter upper bounds based on the total sequence utility has proven difficult. Approaches utilizing maximum or top-\textit{k} item utilities require the sorting of all items, incurring additional computational costs in terms of memory usage and runtime overhead. Our research aims to devise anti-monotonic measures for quantitative sequential data while minimizing overestimation and the associated computational burdens.

\section{Preliminaries} \label{sec: preliminary}

In this section, the notations and definitions in this paper are detailed and are used to express the research issue and proposed method clearly. The remainder of this section shows some examples.

Let $I$ = \{$i_1$, $i_2$, $\cdots$, $i_M$\} be a set of distinct items. An itemset $X$ is a nonempty subset of $I$, and $|X|$ represents the size of $X$. A sequence $S$ is an ordered list of itemsets, and the items in each itemset are sorted lexicographically. The size of $S$ is the number of itemsets in the sequence. The length of $S$ is the total number of items in this sequence, and $S$ is called a $l$-sequence when its length is $l$. Assume that there exist integers 1 $\leqslant$ $k_1$ < $k_2$ < $\cdots$ < $k_m$ $\leqslant$ $n$ such that ${X_v}'$ $\subseteq$ $X_{k_v}$, (1 $\leqslant$ $v$ $\leqslant$ $m$). The sequence $S$: $\langle$$X_1$, $X_2$, $\cdots$, $X_n$$\rangle$ contains the subsequence $s'$: $\langle$${X'_v}$, ${X'_{v+1}}$, $\cdots$, ${X'_m}$$\rangle$, denoted as  $s'$ $\subseteq$ $S$. For example, in the sequence $s$ = $\langle$\{$a$\}, \{$a, b$\}, \{$c, d, e$\}, \{$f, g$\}$\rangle$, there are 4 itemsets and 7 distinct items. The size of $s$ is 4, and its length is 8. The sequence $s'$ = $\langle$\{$b$\}, \{$c, d$\}, \{$f$\}$\rangle$ is contained in $s$, or $s$ contains the subsequence $s'$.

\begin{table}[ht]
    \caption{An example of a quantitative sequential database} 
    \renewcommand{\arraystretch}{1.35}
	\label{table: database}
	\centering
    \resizebox{0.49\linewidth}{!}{ 
	\begin{tabular}{|c|ccccc|}
	\hline
	\textbf{SID}         & \multicolumn{5}{c|}{\textbf{Q-sequence}}                                                                                       \\ \hline
	\multirow{2}{*}{${QS}_1$} & \multicolumn{1}{c|}{itemset}  & \multicolumn{1}{c|}{$a, c$} & 		\multicolumn{1}{c|}{$a, b, e$} & \multicolumn{1}{c|}{$c, d$}    & $e, f$ \\ \cline{2-6} 
	                     & \multicolumn{1}{c|}{quantity} & \multicolumn{1}{c|}{2, 8} & 			\multicolumn{1}{c|}{1, 4, 5} & \multicolumn{1}{c|}{1, 1}    & 1, 1 \\ \hline
	\multirow{2}{*}{${QS}_2$} & \multicolumn{1}{c|}{itemset}  & \multicolumn{1}{c|}{$a, d$} & 		\multicolumn{1}{c|}{$b, f$}    & \multicolumn{1}{c|}{$a, d, f$} & $b, d$ \\ \cline{2-6} 
	                     & \multicolumn{1}{c|}{quantity} & \multicolumn{1}{c|}{1, 8} & 			\multicolumn{1}{c|}{2, 3}    & \multicolumn{1}{c|}{1, 3, 1} & 1, 1 \\ \hline
	\multirow{2}{*}{${QS}_3$} & \multicolumn{1}{c|}{itemset}  & \multicolumn{1}{c|}{$a, c$} & 		\multicolumn{1}{c|}{$c, b, g$} & \multicolumn{1}{c|}{$b, d$}    & $e, f$ \\ \cline{2-6} 
	                     & \multicolumn{1}{c|}{quantity} & \multicolumn{1}{c|}{1, 3} & 			\multicolumn{1}{c|}{4, 9, 5} & \multicolumn{1}{c|}{1, 7}    & 6, 2 \\ \hline
	\end{tabular}}
\end{table}

\begin{definition}[\textit{q}-item, \textit{q}-itemset, \textit{q}-sequence, and QSDB]\label{def: q-Item}
  \rm A quantitative sequential database (QSDB) consists of a quantitative sequence (\textit{q}-sequence) and the corresponding unique identifier (SID). Each quantitative sequence (\textit{q}-sequence) is an ordered list of the quantitative itemsets (\textit{q}-itemsets). In a certain QSDB, each distinct item $i$ corresponds with its external utility $eu(i)$ \cite{Ref44}, called a unity utility. The quantitative item (\textit{q}-item) in the \textit{q}-itemset is a pair $(item, quantity)$, and the internal utility of each \textit{q}-item is its quantity, denoted as $q(i, j, QS)$, where $i$ is the label of the item, and $j$ is the numerical order of the quantitative itemset which contains this item in the quantitative sequence $QS$. 
\end{definition} 

For example, there are three quantitative sequences in $\mathcal{D}$ in Table \ref{table: database}. The distinct item and its external utility are \{$a$:2, $b$:1, $c$:4, $d$:3, $e$:6, $f$:5, $g$:8\} in this QSDB. $a$ is a quantitative item in quantitative sequence ${QS}_1$ which appears in the $1^{st}$ quantitative itemset, and its internal utility is $q(a, 1, {QS}_1)$ = 2.

\begin{definition}[utility of \textit{q}-item, \textit{q}-itemset, and \textit{q}-sequence]\label{def: q-ItemUtility}
  \rm Consider a given quantitative sequential database $\mathcal{D}$, let $QS$: $\langle$$Y_1$, $Y_2$, $\cdots$, $Y_n$$\rangle$ denote a \textit{q}-sequence, and $Y_j$ is the $j^{th}$ \textit{q}-itemset in $QS$. The $(i, q)$ denotes one of the \textit{q}-items within $Y_j$. The internal utility of the \textit{q}-item $i$ is $q(i, j, s)$ and its external utility is $eu(i)$. The utility of \textit{q}-item $(i, q)$ is defined as $u(i, j, QS)$ = $q(i, j, QS) \times eu(i)$. The utility of \textit{q}-itemset is defined as the sum of the utilities of all \textit{q}-items contained in it and is denoted as $u(Y_j, j, QS)$ = $\sum_{\forall{(i, q)\in {Y_j}}}{q(i, j, QS) \times eu(i)}$. The utility of the quantitative sequence $QS$ is defined as $u(QS)$ = $\sum_{\forall {Y_j}\in {QS}}{u(Y_j, j, QS)}$.
\end{definition} 

For example, ${QS}_1$: $\langle$\{($a$, 2), ($c$, 8)\}, \{($a$, 1), ($b$, 4), ($e$, 5)\}, \{($c$, 1), ($d$, 1)\}, \{($e$, 1), ($f$, 1)\}$\rangle$ is one of the \textit{q}-sequences of the given quantitative sequential database $\mathcal{D}$ in Table \ref{table: database}. The $2^{nd}$ \textit{q}-itemset is \{($a$, 1), ($b$, 4), ($e$, 5)\}, the \textit{q}-item of $a$ in this \textit{q}-itemset is ($a$, 1), and its internal utility is 1. We set its external utility as 2, as shown in Definition \ref{def: q-Item}. The utility of ($a$, 1) within the $2^{nd}$ \textit{q}-itemset in ${QS}_1$ is $u(a, 2, {QS}_1)$ = 1 $\times$ 2 = 2. The utility of \{($a$, 1), ($b$, 4), ($e$, 5)\} in ${QS}_1$ is $u(\{a, b, e\},2,{QS}_1)$ = 1 $\times$ 2+4 $\times$ 1+5 $\times$ 6 = 36. The utility of ${QS}_1$ is $u({QS}_1)$ = 36+36+7+11 = 90.

\begin{definition}[average utility of \textit{q}-item, \textit{q}-itemset, and \textit{q}-sequence]\label{def: q-averageUtility}
  \rm Let \textit{QS}: $\langle$$Y_1$, $Y_2$, $\cdots$, $Y_n$$\rangle$ denotes a \textit{q}-sequence within the given quantitative sequential database, which we denote by $\mathcal{D}$. Let $(i, q)$ denote one of the \textit{q}-items in the $j^{th}$ \textit{q}-itemset $Y_j$ in $QS$. The size of $Y_j$ is the total count of \textit{q}-items in $Y_j$, denoted as $|Y_j|$. The size of $QS$ is $|QS|$ = $n$. The length of $QS$ is $|QS|$ = $\sum_{\forall{Y_j\in {QS}}}{|Y_j|}$. The average utility of \textit{q}-itemset $Y_j$ is defined as $au(Y_j, j, QS)$ = $\frac{\sum_{\forall{(i, q)\in Y_j}}{q(i, j, QS) \times eu(i)}}{|Y_j|}$. The average utility of \textit{q}-item $(i, q)$ is defined as $au(i, j, QS)$ = $\frac{q(i, j, QS) \times eu(i)}{1}$ = $u(i, j, QS)$. The average utility of \textit{q}-sequence $QS$ is defined as $au(QS)$ = $\frac{\sum_{\forall{Y_j\in {QS}}}{u(Y_j, j, QS)}}{|QS|}$.
\end{definition} 

For example, the size of itemset $\{a, b, e\}$ is 3, and the average utility of \textit{q}-itemset \{$(a, 1)$, $(b, 4)$, $(e, 5)$\} in the $2^{nd}$ \textit{q}-itemset in ${QS}_1$ is $au(\{a, b, e\}$, 2, ${QS}_1)$ = $\frac{(1 \times 2+4 \times 1+5 \times 6)}{3}$ = 12. The average utility of ${QS}_1$ is $au({QS}_1 )$ = $\frac{(36+36+7+11)}{9}$ = 10.

\begin{definition}[match and contain] \label{def: match&contain}
  \rm We say that the itemset $X$: \{$i_1$, $i_2$, $\cdots$, $i_m$\} $matches$ the \textit{q}-itemset $Y$: \{(${i'}_1$, $q_1$), (${i'}_2$, $q_2$), $\cdots$, (${i'}_n$, $q_n$)\} if and only if $m$ = $n$ such that $i_k$ = ${i'}_k$, (1$\leqslant$ $k$$\leqslant$ $n$). It could be notated as $X$ $\sim$ $Y$. Let $X'$ denote a subset of $X$. We could say that $Y$ contains $X'$, it is notated as $X'$ $\sqsubseteq$ $Y$.
\end{definition}

\begin{definition}[instance] \label{def: instance}
  \rm Consider the \textit{q}-sequence $QS$: $\langle$$Y_1$, $Y_2$, $\cdots$, $Y_n$$\rangle$ and the sequence $S$: $\langle$$X_1$, $X_2$, $\cdots$, $X_m$$\rangle$, where $m$ $\leqslant$ $n$. Assume that there exist integer $j_v$, if and only if 1 $\leqslant$ $j_1$ < $j_2$ < $\cdots$ < $j_m$ $\leqslant$ $n$ and $X_v$ $\sqsubseteq$ $Y_{j_v}$, (1 $\leqslant$ $v$ $\leqslant$ $m$). We say that in $QS$, there is an \textit{instance} of $S$ at position $p$: $\langle$$j_1$, $j_2$, $\cdots$, $j_m$$\rangle$.
\end{definition}

Consider any \textit{instance} of the sequence $S$ in $QS$, the sum of all \textit{q}-items utilities is the instance utility. It is defined as $u(S, p, QS)$ = $\sum_{\forall{X_v \in S}}{u(X_v, j_v, QS)}$, where ${u(X_v, j_v, QS)}$ = $\sum_{\forall{\{{Y'}_{j_v}|{ {Y'}_{j_v} \sim X_v} \wedge {X_v \sqsubseteq Y_{j_v}} \wedge {Y_{j_v} \in QS}\}}}{u({Y'}_{j_v}, j_v, QS)}$. Let $|S|$ be the length of $S$, the instance average utility at position $p$ is its utility divided by its length and is defined as $au(S, p, QS)$ = $\frac{\sum_{\forall{X_v \in S}}{u(X_v, j_v, QS)}}{|S|}$ = $\frac{\sum_{\forall{\{{Y'}_{j_v}|{ {Y'}_{j_v} \sim X_v} \wedge {X_v \sqsubseteq Y_{j_v}} \wedge {Y_{j_v} \in QS}\}}}{u({Y'}_{j_v}, j_v, QS)}}{|S|}$.

For example, an \textit{q}-itemset \{($a$, 1), ($b$, 4), ($e$, 5)\} $contains$ an itemset \{$a$, $b$\}. In Table \ref{table: database}, the \textit{q}-itemset \{($a$, 2), ($c$, 8)\} $matches$ an itemset \{$a$, $c$\}, or we could say the sequence $\langle$\{$a$, $c$\}$\rangle$ has an \textit{instance} in ${QS}_1$ at position $p$: $\langle$1$\rangle$, its average utility is $au$($\langle$\{$a$, $c$\}$\rangle$, 1, ${QS}_1$) = $\frac{(4+32)}{2}$ = 18. There are three \textit{instances} of sequence $\langle$\{$a$\}, \{$e$\}$\rangle$ in ${QS}_1$, and the positions are $p_1$: $\langle$1, 2$\rangle$, $p_2$: $\langle$1, 4$\rangle$ and $p_3$: $\langle$2, 4$\rangle$, and the average utility at these positions are $au$($\langle$\{$a$\}, \{$e$\}$\rangle$, $\langle$1, 2$\rangle$, ${QS}_1)$ = $\frac{(4+30)}{2}$ = 17, $au$($\langle$\{$a$\}, \{$e$\}$\rangle$, $\langle$1, 4$\rangle$, ${QS}_1)$ = $\frac{(4+6)}{2}$ = $5$ and $au$($\langle$\{$a$\}, \{$e$\}$\rangle$, $\langle$2, 4$\rangle$, ${QS}_1)$ = $\frac{(2+6)}{2}$ = 4.

\begin{definition}[sequence average utility] \label{def: SAU}
  \rm If the sequence $S$: $\langle$$X_1$, $X_2$, $\cdots$, $X_m$$\rangle$ appears at different positions in the \textit{q}-sequence $QS$: $\langle$$Y_1$, $Y_2$, $\cdots$, $Y_n$$\rangle$. Let $P(S, QS)$ denote the set of all the positions of $S$ in $QS$, the utility of the sequence $S$ in $QS$ is the maximum $u(S, p, QS)$, and is denoted as $u(S, QS)$ = $\max\limits_{p \in P(S, QS)}{u(S, p, QS)}$. The average utility of the sequence $S$ in $QS$ is defined as $au(S, QS)$ = $\frac{\max\limits_{p\in P(S, QS)}{u(S, p, QS)}}{|S|}$ = $\max\limits_{p \in P(S, QS)}{\frac{u(S, p, QS)}{|S|}}$.
\end{definition} 

For example, in Table \ref{table: database}, the sequence $s_1$: $\langle$\{$a$\}, \{$c$\}, \{$e, f$\}$\rangle$ has two \textit{instances} in ${QS}_1$, and their average utilities are $au(s_1, {QS}_1)$ = $\max${\{$au$($s_1$, $\langle$1, 3, 4$\rangle$, ${QS}_1$), $au$($s_1$, $\langle$2, 3, 4$\rangle$, ${QS}_1$)\}} = $\max${\{$\frac{19}{4}$, $\frac{17}{4}$\} } = $4.75$.

\begin{definition}[high average utility sequential pattern] \label{def: HAUSP}
  \rm If there exists a sequence $S$ in a quantitative sequential database $\mathcal{D}$ whose average utility value is higher than $\xi$ $\times$ $u(\mathcal{D})$, which is the minimal acceptable threshold so then we define the sequence $S$ is a high average utility sequence. And we denote the minimum average utility threshold parameter by $\xi$.
\end{definition} 

\textbf{Problem statement:} Given a QSDB $\mathcal{D}$ and the corresponding external utility list of all the distinct items, $u(\mathcal{D})$ is its overall utility. The $\xi$ is a user-specified parameter that ranges from $0$ to $1$, and $minau$ = $\xi$ $\times$ $u(\mathcal{D})$ is the minimum acceptable average utility. The high average utility sequential pattern mining (HAUSPM) problem \cite{Ref35} is to find all the high average utility sequential patterns (HAUSPs) whose average utility is higher than $\xi$ $\times$ $u(\mathcal{D})$ in $\mathcal{D}$, when we get a value of $\xi$.

The problem with HAUPM differs from the problem with HUPM. Not only does it add another important consideration, but it also reflects the unique characteristics of the process and result of mining associated with the problem.  HUSP is not necessarily a HAUSP. As mentioned in the previous section, if a HUSP contains a substantial number of lower utility items, it remains categorized as a HUSP but does not qualify as a HAUSP. However, in this study, the method derived from HUSPM proves to be an effective approach for HAUSPM, resulting in improved performance. Additionally, two extra pruning strategies are employed to enhance algorithm efficiency. In Ref. \cite{Ref3}, the mining of maximal or top-\textit{k} patterns is classified as a form of lossy compression, whereas mining a closed pattern is considered lossless compression. During lossy compression, the support threshold increases dynamically, leading to a noticeable reduction in the projection space. As the threshold rises, the pruning effect on the construction and reuse of projection structures becomes increasingly significant. Based on this observation, this work regards HAUSPM as a variant of HUSPM embedded with dynamic threshold adjustment.

\section{Algorithm} \label{sec: algorithm}

\begin{figure*}
    \centering
    \includegraphics[width=0.98\linewidth]{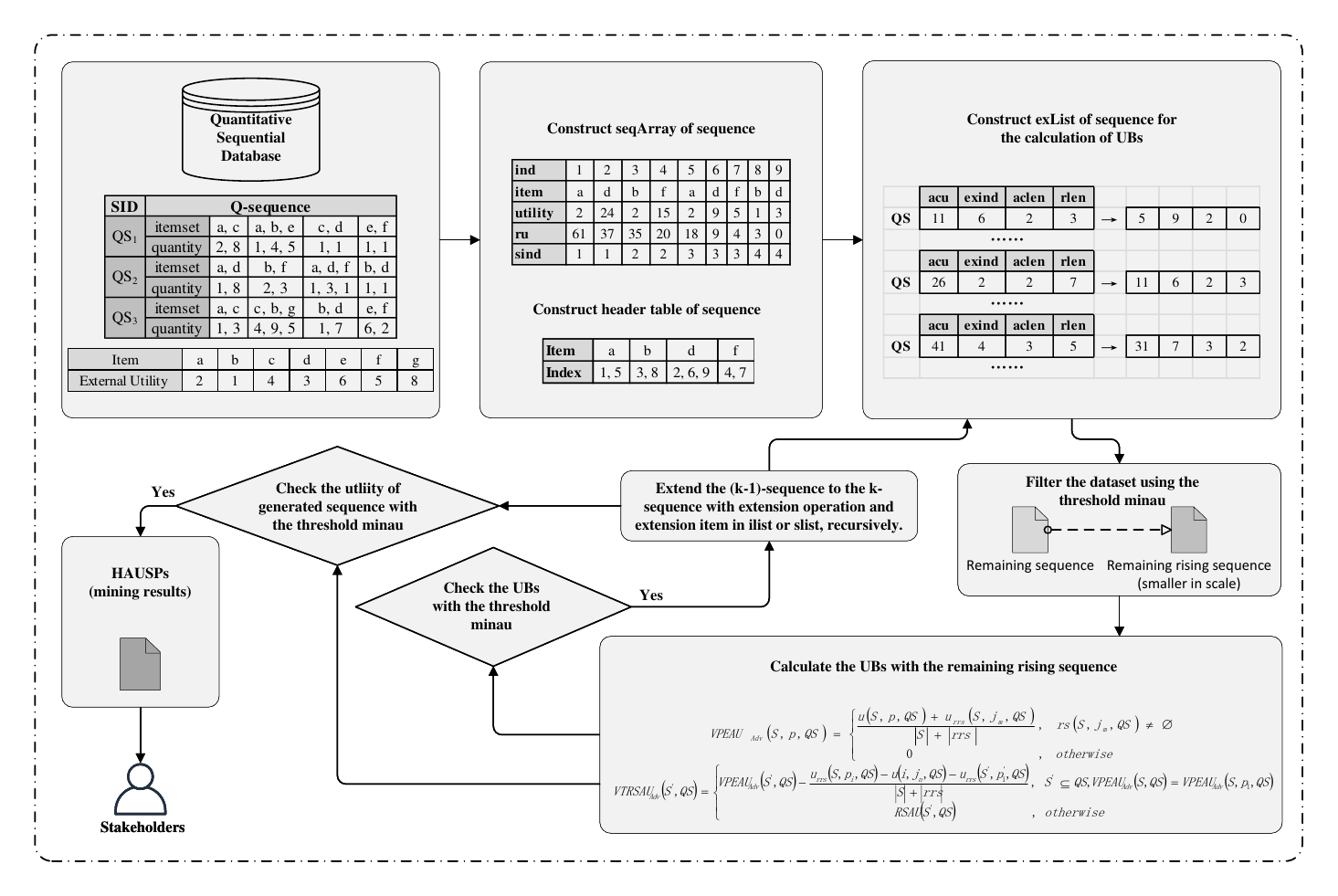}
    \caption{The overall process of the proposed HAUSP-PG.}
    \label{fig: 00}
\end{figure*}

Inspired by the algorithms for HUSPM and aiming for better performance, a novel algorithm named HAUSP-PG is designed for HAUSPM. Fig. \ref{fig: 00} presents the overall process of the proposed approach. In HAUSP-PG, we construct a projection database and adopt the pattern-growth method to avoid multiple scanning and combination explosions \cite{Ref62}. Based on the proposed strict upper bounds and their variants, the designed dual pruning method could effectively limit the search space in the algorithm. The following account will clearly illustrate the proposed algorithm.

\subsection{Pattern Generation and Search Space}
\label{sec: generation}

The proposed algorithm first obtains all the $1$-sequences, whose average utilities are equivalent to their utilities. Subsequently, more \textit{k}-sequences are recursively derived by extending their prefix \textit{(k-1)}-sequences. The collective set of sequences constitutes the entire search space for mining. The search space for HAUSPM is described as the lexicographically \textit{q}-sequence tree (LQS-tree) \cite{Ref27, Ref14}. In the LQS-tree, the root is null, and each child node represents a generated sequence whose parent node corresponds to the prefix of that sequence. The child nodes of a given node are not strictly ordered. Some of them are listed in lexicographical order. This is due to the different generation methods used. As defined in Ref. \cite{Ref63}, the \textit{extension} operation is appending an item to extend a sequence. The pattern-growth method \cite{Ref62} recursively calls the \textit{extension} operation, thereby incrementally increasing the length and utility of generated patterns in a step-wise manner. Analogous to how projection databases and pattern-growth methods are highly effective in HUSPM, they should likewise prove effective in HAUSPM.

\begin{definition}[{S}-Extension \, and \, \textit{I}-Extension \, \cite{Ref62, Ref14}]
\label{def: extension}
  \rm Consider the last itemset $X_k$: \{$i_1$, $i_2$, $\cdots$, $i_l$\} in the sequence $S$: $\langle$$X_1$, $X_2$, $\cdots$, $X_k$$\rangle$. Let $X_k$ be the position for the \textit{extension} operation. For an appending item $i$, if $i$ is appended to $X_k$ as $i_{l+1}$, the length of the sequence is increased by one, but the size of $S$ remains static. But if $i$ is appended to $S$ as $X_{k+1}$, both the size of $S$ and the length of the sequence are increased by one. The former case is defined as \textit{I}-Extension and is notated as $S$ $\oplus$ $i$. The latter case is defined as \textit{S}-Extension and is notated as $S$ $\otimes$ $i$.
\end{definition} 

\begin{definition}[extension item and extension position \cite{Ref27}] \label{def: EI&EP}
  \rm Consider the \textit{instances} of $S$: $\langle$$X_1$, $X_2$, $\cdots$, $X_m$$\rangle$ and a \textit{q}-sequence $QS$: $\langle$$Y_1$, $Y_2$, $\cdots$, $Y_n$$\rangle$, the \textit{instances} generally appear at several positions in $QS$. The set of positions is notated as $P(S,QS)$: \{$p_1$, $p_2$, $\cdots$, $p_w$\}. Let $p_k$: $\langle$$j_1$, $j_2$, $\cdots$, $j_m$$\rangle$ be one of positions, the \textit{extension position} $j_m$ is the sequence number of \textit{q}-itemset in $QS$ which contains $X_m$. The \textit{extension item} is the \textit{q}-item which corresponds to the last item within $X_m$. By definition \ref{def: SAU}, in a \textit{q}-sequence $QS$, the sequence average utility of $S$ is $au(S,QS)$ = $\max\limits_{p\in P}{\frac{u(S, p, QS)}{|S|}}$ = $\max${\{$\frac{u(S, p_1, QS)}{|S|}$, $\frac{u(S, p_2, QS)}{|S|}$, $\cdots$, $\frac{u(S, p_w, QS)}{|S|}$\}}.
\end{definition} 

\begin{definition}[remaining \textit{q}-sequence \cite{Ref27, Ref14}] \label{def: RS}
  \rm In the \textit{q}-sequence $QS$: $\langle$$Y_1$, $Y_2$, $\cdots$, $Y_n$$\rangle$, the \textit{extension position} of the \textit{instance} $S$ at positions $p_k$: $\langle$$j_1$, $j_2$, $\cdots$, $j_m$$\rangle$ is $j_m$. All the items that are behind the \textit{extension item} form a subsequence, we define the subsequence as the \textit{remaining sequence} of $QS$, denoted as $rs$. The utility of $rs$ is notated as $u_{rs}$$(S, j_m, QS)$.
\end{definition} 

For example, there are three \textit{instances} of $s_2$: $\langle$\{$d$\}, \{$b$\}$\rangle$ in ${QS}_2$ in Table \ref{table: database}, $au(s_2, {QS}_2)$ = $\max${\{$\frac{u(\langle\{d\}, \{b\}\rangle, \langle 1, 2\rangle, {QS}_2)}{2}$, $\frac{u(\langle\{d\}, \{b\}\rangle, \langle 1, 4\rangle, {QS}_2)}{2}$, $\frac{u(\langle\{d\}, \{b\}\rangle, \langle 3, 4\rangle, {QS}_2)}{2}$\}} = $\max${\{$\frac{26}{2}$, $\frac{25}{2}$, $\frac{10}{2}$\}} = 13. There is an \textit{instance} of $s_3$: $\langle$\{$b$, $f$\}$\rangle$ at position $\langle$2$\rangle$ in ${QS}_2$, then we have the remaining sequence utility $u_{rs}(s_3, 2, {QS}_2)$ = 20.

Obtaining the utility of a sequence $S$ within a \textit{q}-sequence $QS$ is time-consuming. We have to scan all the positions and calculate the utility of all its \textit{instance} ones repeatedly. Some existing data structures, such as \textit{utility-chain} \cite{Ref27} and \textit{seq-array} \cite{Ref46}, are designed to keep all the details and expedite the calculation process. In our work, a modified compact projection structure, \textit{proDB} is designed to store essential information about average utility. The projection structure includes \textit{sequence-array} (\textit{seqArray}), \textit{item-index} head table, and \textit{extension-list} (\textit{exList}).

\begin{table*}[ht]
    \renewcommand{\arraystretch}{1.35}
	\caption{Details of the \textit{proDB}.}  
	\label{table: projection}
	\centering
    \resizebox{0.76\linewidth}{!}{ 
	\begin{tabular}{lll}
		\toprule
	\textbf{Structure} & \textbf{Field}   & \textbf{Description}                                                           \\
	\midrule
		sequence-array     & \textit{ind}     & index of item                                                              \\
		                   & \textit{item}    & label of item                                                              \\
		                   & \textit{utility} & utility of item                                                          \\
		                   & \textit{ru}      & utility of \textit{remaining sequence} when the current item is an \textit{extension item} \\
		                   & \textit{sind}    & index of itemset containing the current item                             \\
		                   & 	        &                                                                                \\
		extension-list     & \textit{acu}     & utility of \textit{instance} with \textit{extension position}                                  \\
		                   & \textit{exind}   & index of \textit{extension item} at an \textit{extension position}                     \\
		                   & \textit{aclen}   & length of \textit{instance}                                                       \\
		                   & \textit{rlen}    & length of \textit{remaining sequence} when the current item is an \textit{extension item}  \\
	\bottomrule
	\end{tabular}}
\end{table*} 

The \textit{utility-chain} consists of a set of \textit{utility-lists}. It is less time-consuming, especially in cases where the \textit{q}-sequence has multiple \textit{instances}. Information stored in the \textit{utility-list} is used to calculate the utility of \textit{instance} quickly. Two fields \textit{aclen} and \textit{rlen} are expanded to provide essential information for the calculation of the average utility.

The refinement of the projection database structure, \textit{seqArray}, ensures the absence of null values in the head table \cite{Ref46}. This \textit{exList} speeds up the pattern generation and average utility calculation. The sequence average utility in the \textit{q}-sequence is the maximum average utility across all its \textit{instances}, while the sequence average utility in $\mathcal{D}$ is the sum of maximum average utilities in each \textit{q}-sequence. The scheme, which combines the pattern-growth and the database projection methods, has a significant improvement in computation complexity and time efficiency.

\begin{figure}
	\centering
			\includegraphics[width=0.79\linewidth]{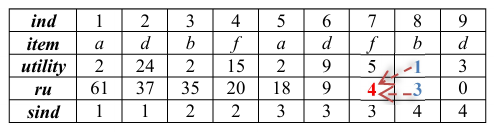} \\   
                \textnormal{(a) the \textit{seqArray} in \textit{proDB}} \\
			\includegraphics[width=0.79\linewidth]{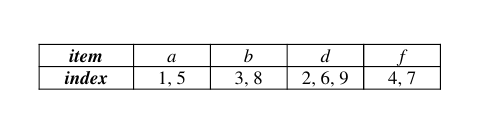} \\   
                \textnormal{(b) the \textit{item-index} head table in \textit{proDB}}
	\caption{An illustration of projection data structure of ${QS}_2$.}
	\label{fig: 01}
\end{figure}

For example, the \textit{seqArray} of the \textit{q}-sequence ${QS}_2$ is shown in Fig. \ref{fig: 01}(a). Each item is indexed according to its order of appearance. In the penultimate column, the utility of item $b$ is 1. When $b$ serves as the \textit{extension item}, its remaining sequence utility is 3. In this array, the remaining sequence utility of the current item is defined as the sum of its own utility and the remaining sequence utility of the subsequent item. For instance, in Fig. \ref{fig: 01}(a), the remaining sequence utility of $f$ (in the antepenultimate column) equals the utility of $b$ plus the remaining sequence utility of $b$ (in the penultimate column). The symbol \textit{sind} denotes the index of the itemset containing the current item. \textbf{Remark.} In the implementation stage, \textit{sind} is represented by the index of the first item in the corresponding itemset, which serves as its unique identifier. This identifier preserves the ordering of itemsets but is not necessarily continuous. For illustration in this paper, the continuous index $j$ defined in Definition \ref{def: q-Item} is still used. These two indexing schemes differ only in representation, while the algorithmic logic remains identical. In Fig. \ref{fig: 01}(b), for each distinct item within ${QS}_2$, their indices are recorded in the \textit{item-index} head table of the \textit{q}-sequence.

\begin{figure}
    \centering
    \includegraphics[width=1\linewidth]{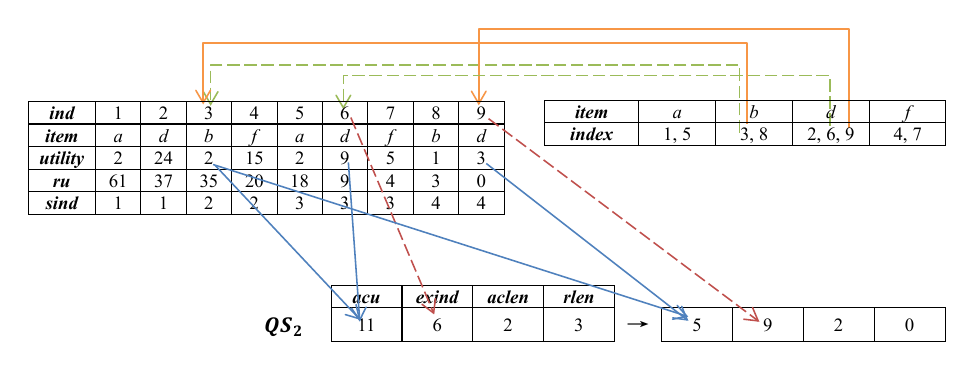}
    \caption{An illustration of \textit{exList} of sequence $s_4$ in ${QS}_2$.}
    \label{fig: 02}
\end{figure}

For example, there are two \textit{instances} of the sequence $s_4$: $\langle$\{$b$\},\{$d$\}$\rangle$ at positions \{$\langle$2, 3$\rangle$, $\langle$2, 4$\rangle$\} within the \textit{q}-sequence ${QS}_2$ in Table \ref{table: database}. The \textit{exList} of $s_4$ in ${QS}_2$ is shown in Fig. \ref{fig: 02}. The \textit{acu} of $s_4$ at position $\langle$2, 3$\rangle$ is 2+9 = 11, and the \textit{exind} of the \textit{extension item} $d$ at position $\langle$2, 3$\rangle$ is the index of $d$ in the $3^{rd}$ itemset, which is 6.

\begin{figure}
	\centering
	\includegraphics[width=1\linewidth]{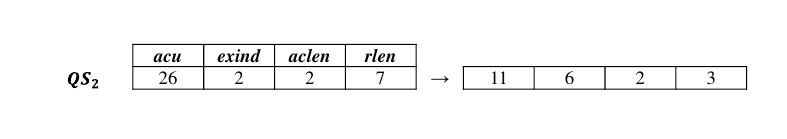}   
    \centerline{(a) the \textit{extension list} of sequence $\langle$\{$a$,$d$\}$\rangle$} 
	\includegraphics[width=1\linewidth]{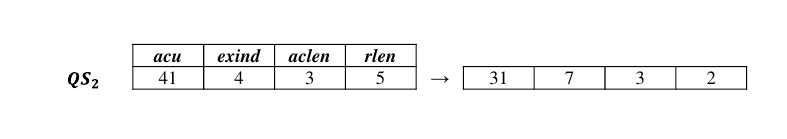}    
    \centerline{\textnormal{(b) the \textit{extension list} of sequence $\langle$\{$a$, $d$\}, \{$f$\}$\rangle$}}
	\caption{An illustration of \textit{exList} in one \textit{extension} operation.}
	\label{fig: 03}
\end{figure}

For example, consider the sequence $\langle$\{$a$, $d$\}, \{$f$\}$\rangle$ in Table \ref{table: database}, its utility calculation process could be succinctly described below. First, the sequence $\langle$\{$a$\}$\rangle$ as a prefix is contained in ${QS}_1$ and ${QS}_2$, but only ${QS}_2$ needs to be considered since it contains the sequence $\langle$\{$a$, $d$\}$\rangle$. Although there are two \textit{instances} of $\langle$\{$a$, $d$\}$\rangle$ at positions \{$\langle$1, 1$\rangle$, $\langle$3, 3$\rangle$\}, only at the position \{$\langle$1, 1$\rangle$\} could the sequence $\langle$\{$a$, $d$\}$\rangle$ be extended to sequence $\langle$\{$a$, $d$\}, \{$f$\}$\rangle$ through a \textit{S}-Extension. Based on the projected database, the utility of $\langle$\{$a$, $d$\}$\rangle$ can be efficiently calculated. Next, the sequence $\langle$\{$a$, $d$\}, \{$f$\}$\rangle$ could be generated by appending item $f$ to \textit{extension positions} $\langle 2 \rangle$ and $\langle 3 \rangle$ respectively. Fig. \ref{fig: 03} illustrates the \textit{exList} of $\langle$\{$a$, $d$\}$\rangle$ in ${QS}_2$. The \textit{exList} will be rapidly constructed by the projection structure \textit{proDB}. The \textit{acu} of the sequence $\langle$\{$a$, $d$\}, \{$f$\}$\rangle$ is the \textit{acu} of the sequence $\langle$\{$a$, $d$\}$\rangle$ and the utility of its appending item $f$ which has a larger \textit{ind} and \textit{sind} than the \textit{exind} of the sequence $\langle$\{$a$, $d$\}$\rangle$. The actual average utility of the corresponding generated sequence is $au$($\langle$\{$a$, $d$\}, \{$f$\}$\rangle$, ${QS}_2$) = $\max${\{$au$($\langle$\{$a$, $d$\}, \{$f$\}$\rangle$, $\langle$1, 2$\rangle$, ${QS}_2)$, $au$($\langle$\{$a$, $d$\}, \{$f$\}$\rangle$, $\langle$1, 3$\rangle$, ${QS}_2$)\}} = $max${\{$\frac{41}{3}$, $\frac{31}{3}$\}} = $\frac{41}{3}$, we could get the average utility of sequence $\langle$\{$a$, $d$\}, \{$f$\}$\rangle$ equals the average utility of the \textit{instance} at position $\langle$1, 2$\rangle$. Moreover, the \textit{exList} stores the necessary information required for intermediate calculations, which further accelerates the subsequent construction of average upper bounds.

\subsection{Search Space Pruning}
\label{sec: pruning}

Some HAUPM algorithms use the \textit{auub} model to overestimate the pattern average utility \cite{Ref30, Ref47}. It means that the average utility of a pattern is not greater than the maximum item utility in the transaction. Although the $auub$ model provides a correct and complete algorithm for HAUIM, it is loose and the algorithm consumes large amounts of execution time and memory. After that, the looser upper-bound utility (\textit{lub}) for mining HAUIs was proposed by Lin et al. \cite{Ref52} and defined as \textit{lub} = $\sum_{{X \sqsubseteq T_q} \wedge {T_q \in \mathcal{D}}}{\textit{lub}(X, T_q)}$ = $\sum_{{X \sqsubseteq T_q} \wedge {T_q \in \mathcal{D}}} {\frac{{u(X, T_q)+ |X| \times remu(X, T_q)}}{|X|}}$. The \textit{lub} model supposes that all the utility of items in \textit{remaining sequence} are equal to $remu$, the maximal item utility in the \textit{remaining sequence}. There is proof of the anti-monotonicity property of \textit{lub} to ensure completeness and correctness in Ref. \cite{Ref52}.

Sequence-weighted utility (SWU) has been utilized to eliminate unpromising sequences early in some existing algorithms \cite{Ref27}. However, the pruning effect of SWU-based algorithms is severely reduced because the value of SWU is much larger than the actual utility of the pattern or its supersequence, especially in a database with numerous long sequences. Wang et al. \cite{Ref27} proposed two utility UBs, PEU and RSU. Zhang et al. \cite{Ref46} compared the main utility upper bounds, adopted PEU and RSU, and proposed the tighter reduced sequence utility (TRSU). A theorem proposing that the utility of any extension sequence of sequence $s$ is not greater than the PEU of $s$ was proposed in Ref. \cite{Ref46}. So, if the corresponding PEU of a sequence is lower than the threshold, denoted by $\eta$ $\times$ $u(\mathcal{D})$, neither this sequence nor any of its extension sequences are HUSPs.

HAUSPM is an attempt to overcome the limitations of HUSPM. One potential attempt is to combine the utility upper bounds of HUSPM with the definition of sequence average utility. The fundamental modification involves incorporating the length of the sequence in the projection database to support the calculation of average utility. Therefore, we could have two primary UBs, prefix extension average utility (\textit{PEAU}) and reduced sequence average utility (\textit{RSAU}). Researchers proposed PEU and RSU instead of SWU to improve performance in databases with numerous long sequences. Furthermore, the proposed \textit{PEAU} and \textit{RSAU} are more suitable for long sequences containing lots of low utility items.

\begin{definition}[\textit{PEAU} Upper Bound] \label{def: PEAU}
  \rm Consider the \textit{q}-sequence $QS$, there is an \textit{instance} of the sequence $S$ at position $p$: $\langle$$j$, $j_2$, $\cdots$, $j_m$$\rangle$. Its \textit{extension position} is $j_m$, and the corresponding \textit{remaining sequence} is $rs$, which is the rest after $p$ to the end. The prefix extension average utility of the sequence $S$ at position $p$ is denoted as ${\textit{PEAU}}_{\textit{Ori}}(S,p,QS)$, is defined as
	\begin{displaymath}
	\begin{split}
	&{\textit{PEAU}}_{\textit{Ori}}(S, p, QS) \\
    &= \left\{\begin{array}{rcl} \frac{u(S, p, QS)+u_{rs}(S, j_m, QS)}{|S|}, & {rs}(S, j_m, QS) \neq \emptyset \\ 0, & otherwise \end{array}\right. \nonumber
	\end{split}
	\end{displaymath}
 
Then, let $p_i$ be one position of $S$ in $QS$, and $\textit{PEAU}(S, QS)$ = $\max{\{\textit{PEAU}(S, p_i, QS)\}}$ is the \textit{PEAU} of $S$ in the \textit{q}-sequence $QS$.

Moreover, $\textit{PEAU}(S)$ = $\sum_{{S \sqsubseteq QS} \wedge {QS \in \mathcal{D}}}{\textit{PEAU}(S, QS)}$ is defined as the \textit{PEAU} value of $S$ in database $\mathcal{D}$.
\end{definition} 

For example, in Table \ref{table: database}, consider the sequence $s_5$: $\langle$\{$a$, $c$\}, \{$b$\}$\rangle$, there are two \textit{instances} in ${QS}_3$, $u$($s_5$, $\langle$1, 2$\rangle$, ${QS}_3$) = 14+9 = 23 and $u$($s_5$, $\langle$1, 3$\rangle$, ${QS}_3$) = 14 + 1 = 15. The corresponding \textit{remaining sequences} are $\langle$\{$g$\}, \{$b$, $d$\}, \{$e$, $f$\}$\rangle$ and $\langle$\{$d$\}, \{$e$, $f$\}$\rangle$ respectively, $u_{rs}$($s_5$, $\langle$1, 2$\rangle$, ${QS}_3$) = 40 + 1 + 21 + 36 + 10 = 108 $>$ 0 and $u_{rs}$($s_5$, $\langle$1, 3$\rangle$, ${QS}_3$) = 21 + 36 + 10 = 67 > 0. ${\textit{PEAU}}_{\textit{Ori}}$($s_5$, ${QS}_3$) = $\max${\{${\textit{PEAU}}_{\textit{Ori}}$($s_5$, $\langle$1, 2$\rangle$, ${QS}_3$), ${\textit{PEAU}}_{\textit{Ori}}$($s_5$, $\langle$1, 3$\rangle$, ${QS}_3$)\}} = $\max$\{$\frac{23+108}{3}$, $\frac{15+67}{3}$\} = $\frac{131}{3}$. In ${QS}_1$, the sequence $s_5$ has an \textit{instance} and ${\textit{PEAU}}_{\textit{Ori}}$($s_5$, ${QS}_1$) = $\frac{4+32+4+30+4+3+6+5}{3}$ = $\frac{88}{3}$. In addition, there is no \textit{instance} in ${QS}_2$. we have ${\textit{PEAU}}_{\textit{Ori}}$($s_5$, ${QS}_2$) is equal to 0. Finally, the ${\textit{PEAU}}_{\textit{Ori}}$ of sequence $s_5$ is calculated as follows: ${\textit{PEAU}}_{\textit{Ori}}$($s_5$) = ${\textit{PEAU}}_{\textit{Ori}}$($s_5$, ${QS}_1$) + ${\textit{PEAU}}_{\textit{Ori}}$($s_5$, ${QS}_2$) + ${\textit{PEAU}}_{\textit{Ori}}$($s_5$, ${QS}_3$) = $\frac{88}{3}$+$0$+$\frac{131}{3}$ = $\frac{219}{3}$ = 73.

\textbf{Theorem:} For any sequence $S$ $\neq$ $\langle$$\rangle$ and its extension $S'$ in database $\mathcal{D}$, if $\textit{PEAU}$($S$) $\leqslant$ $\xi$ $\times$ $u(\mathcal{D})$ then $au$($S'$) $\leqslant$ $\xi$ $\times$ $u(\mathcal{D})$.

\textbf{Proof:} Let sequence $S$ be a prefix of a sequence $S'$, and the extension sequence $s$ is a subsequence of \textit{remaining sequence} $rs$. The utility of $rs$ must be larger than one of $s$. For any sequence $S'\sqsubseteq QS$ we obtain its average utility
 \begin{displaymath}
	\begin{split}
	au(S', QS) &= \frac{u(S', QS)}{|S'|} = \frac{\max{\{u(S', p', QS)\}}}{|S|+|s|} \\
	&= \frac{\max{\{u(S, p, QS)+u(s, p_i, QS)\}}}{|S|+|s|} \\
	&\leqslant \frac{\max{\{u(S, p, QS)+u_{rs}(S, j_m, QS)\}}}{|S|+|s|} \\
    &\leqslant \frac{\max{\{u(S, p, QS)+u_{rs}(S, j_m, QS)\}}}{|S|} \\
    &= \textit{PEAU}(S, QS).\nonumber
	\end{split}
	\end{displaymath}
where $j_m$ is the last item in $p$, and consider $j_n$ is the last item in $p'$, for any item $j_k$ $\in$ $p_i$, we have $j_m$ $\leqslant$ $j_k$ $\leqslant$ $j_n$. Then, we see that $au$($S'$, $QS$) $\leqslant$ $\textit{PEAU}(S, QS)$, such that $au$($S'$) $\leqslant$ $\textit{PEAU}(S)$. 

Based on the above derivation, the average utilities of all sequences that could be extended from $S$ are less than ${\textit{PEAU}}_{\textit{Ori}}$$(S, p, QS)$. Then, the $\textit{PEAU}(S)$ is the sequence maximum average utility upper bound of descendants of $S$ in $QS$. Thus, if the value of $\textit{PEAU}(S)$ is less than the minimum acceptable average utility, all descendants of $S$ are not considered HAUSPs, and these descendants could be safely pruned.

In the process of mining, all sequences are generated by \textit{extension} operation, including \textit{I}-Extension and \textit{S}-Extension. With each operation, an item will be appended to the original sequence, and the length of the sequence is going to increase by 1. Now, we take into account the increase in the length of the sequence. The incremental version of ${\textit{PEAU}}_{\textit{Ori}}(S,p,QS)$, denoted as ${\textit{PEAU}}_{\textit{Inc}}(S, p, QS)$, is defined as
	\begin{displaymath}
	\begin{split}
	&{\textit{PEAU}}_{\textit{Inc}}(S, p, QS) \\
    &= \left\{\begin{array}{rcl} \frac{u(S, p, QS)+u_{rs}(S, j_m, QS)}{|S|+1}, & {rs}(S, j_m, QS) \neq \emptyset \\ 0, & otherwise \end{array}\right. \nonumber
	\end{split}
	\end{displaymath}

For example, we continue using sequence $s_5$: $\langle$\{$a$, $c$\}, \{$b$\}$\rangle$ from Table \ref{table: database}. ${\textit{PEAU}}_{\textit{Inc}}$($s_5$, ${QS}_3$) = $\max${\{${\textit{PEAU}}_{\textit{Inc}}$($s_5$, $\langle$1, 2$\rangle$, ${QS}_3$), ${\textit{PEAU}}_{\textit{Inc}}$($s_5$, $\langle$1, 3$\rangle$, ${QS}_3$)\}} = $\max$\{$\frac{23+108}{3+1}$, $\frac{15+67}{3+1}$\} = $\frac{131}{4}$. In ${QS}_1$, the sequence $s_5$ has an \textit{instance} and ${\textit{PEAU}}_{\textit{Inc}}$($s_5$, ${QS}_1$) = $\frac{4+32+4+30+4+3+6+5}{3+1}$ = 22. In addition, there is no \textit{instance} in ${QS}_2$. we have ${\textit{PEAU}}_{\textit{Inc}}$($s_5$, ${QS}_2$) is equal to 0. Finally, the ${\textit{PEAU}}_{\textit{Inc}}$ of sequence $s_5$ is calculated as follows: ${\textit{PEAU}}_{\textit{Inc}}$($s_5$) = ${\textit{PEAU}}_{\textit{Inc}}$($s_5$, ${QS}_1$) + ${\textit{PEAU}}_{\textit{Inc}}$($s_5$, ${QS}_2$) + ${\textit{PEAU}}_{\textit{Inc}}$($s_5$, ${QS}_3$) = $\frac{88}{4}$+$0$+$\frac{131}{4}$ = $\frac{219}{4}$ = 54.75.

For any appending item, if its utility is smaller than the average utility of the prefix sequence, the generated sequence does not experience an average utility increase. In the \textit{remaining sequence}, any item whose utility is smaller than the minimum acceptable average utility does not support the generation of a high-average-utility sequence from a non-high-average-utility sequence. Therefore, an evaluation index for the ability of \textit{remaining sequence} to keep the average utility of a generated sequence rising is proposed. It means how many extra parts of utility the items in the \textit{remaining sequence} can provide beyond the minimum utility threshold, at the very most.

\begin{definition}[Remaining Rising Sequence] \label{def: RRS}
  \rm Let sequence $S$ have an \textit{instance} in the \textit{q}-sequence $QS$ with the \textit{extension position} $p$: $\langle$$j_1$, $j_2$, $\cdots$, $j_m$$\rangle$. The \textit{remaining sequence} is the rest after $p$ to the end, denoted as $rs$. Its subsequence, consisting of items whose utility value is not lower than a specified minimum threshold, is the \textit{remaining rising sequence} of $rs$ for this threshold, denoted as $rrs$. Its utility is denoted as $u_{rrs}(S, j_m, QS)$.
\end{definition} 

The \textit{remaining rising sequence} and the prefix could help us to determine how the lower utility of the item could be achieved by the \textit{extension} operation to keep the average utility of the generated sequence not lower than the user-specified threshold. Let $rrs$ be a subsequence containing only items with greater utility value than $\xi$ $\times$ $u(\mathcal{D})$ in $rs$. The revised version of ${\textit{PEAU}}_{\textit{Inc}}(S,p,QS)$ is denoted as ${\textit{PEAU}}_{\textit{Rev}}(S, p, QS)$, is defined as
	\begin{displaymath}
	\begin{split}
	&{\textit{PEAU}}_{\textit{Rev}}(S, p, QS) \\
    &= \left\{\begin{array}{rcl} \frac{u(S, p, QS)+u_{rrs}(S, j_m, QS)}{|S|+1}, & {rs}(S, j_m, QS) \neq \emptyset \\ 0, & otherwise \end{array}\right. \nonumber
	\end{split}
	\end{displaymath}
where $u_{rrs}(S, j_m, QS)$ is the utility of the subsequence $rrs$.

For example, we continue using the sequence $s_5$: $\langle$\{$a$, $c$\}, \{$b$\}$\rangle$ from Table \ref{table: database}. The threshold is set as $\xi$ $\times$ $u(\mathcal{D})$=0.12 $\times$ 300=36. For $s_5$, we have $u(e)$=76>36, $u(g)$=40>36, so, in $rs$, only $e$ and $g$ belongs to $rrs$, so ${\textit{PEAU}}_{\textit{Rev}}$($s_5$, ${QS}_3$) = $\max${\{${\textit{PEAU}}_{\textit{Rev}}$($s_5$, $\langle$1, 2$\rangle$, ${QS}_3$), ${\textit{PEAU}}_{\textit{Rev}}$($s_5$, $\langle$1, 3$\rangle$, ${QS}_3$)\}} = $\max$\{$\frac{23+76}{3+1}$, $\frac{15+36}{3+1}$\} = $\frac{99}{4}$. In ${QS}_1$, the sequence $s_5$ has an \textit{instance} and ${\textit{PEAU}}_{\textit{Rev}}$($s_5$, ${QS}_1$) = $\frac{4+32+4+30+6}{3+1}$ = 19. In addition, there is no \textit{instance} in ${QS}_2$. we have ${\textit{PEAU}}_{\textit{Rev}}$($s_5$, ${QS}_2$) is equal to 0. Finally, the ${\textit{PEAU}}_{\textit{Rev}}$ of sequence $s_5$ is calculated as follows: ${\textit{PEAU}}_{\textit{Rev}}$($s_5$) = ${\textit{PEAU}}_{\textit{Rev}}$($s_5$, ${QS}_1$) + ${\textit{PEAU}}_{\textit{Rev}}$($s_5$, ${QS}_2$) + ${\textit{PEAU}}_{\textit{Rev}}$($s_5$, ${QS}_3$) = $\frac{76}{4}$+$0$+$\frac{99}{4}$ = $\frac{175}{4}$ = 43.75 > $\xi$ $\times$ $u(\mathcal{D})$ = 36.

Based on the mentioned UBs, we can define an effective pruning strategy. However, this pruning strategy may not be optimal in certain situations. Since considering the length of $rrs$ can lead to a better pruning effect, we propose a variant of ${\textit{PEAU}}_{\textit{Rev}}(S,p,QS)$, ${\textit{VPEAU}}_{\textit{Adv}}(S, p, QS)$. It is defined as
	\begin{displaymath}
	\begin{split}
	&{\textit{VPEAU}}_{\textit{Adv}}(S, p, QS) \\
    &= \left\{\begin{array}{rcl} \frac{u(S, p, QS)+u_{rrs}(S, j_m, QS)}{|S|+|rrs|_d}, & {rs}(S, j_m, QS) \neq \emptyset \\ 0, & otherwise \end{array}\right. \nonumber
	\end{split}
	\end{displaymath}
where $u_{rrs}(S,j_m,QS)$ is the utility of the subsequence $rrs$, in which the item utility value is not lower than $\xi$ $\times$ $u(\mathcal{D})$. $|rrs|_d$ denotes the count of distinct items within all the $rrs$ in the database. As the number of $QS$ in the database increases, the value of $|rrs|_d$ could also rise accordingly. A larger $|rrs|_d$ will serve to make the ${\textit{VPEAU}}_{\textit{Adv}}$ tighter. Note that $|rrs|_d$ does not denote the maximum count of distinct items in $rrs$ within a single \textit{q}-sequences.

For example, we continue using the sequence $s_5$: $\langle$\{$a$, $c$\}, \{$b$\}$\rangle$ from Table \ref{table: database}. The threshold is set as $\xi$ $\times$ $u(\mathcal{D})$=0.12 $\times$ 300=36. For $s_5$, we have $u(e)$=76>36, $u(g)$=40>36, so, in $rs$, only $e$ and $g$ belongs to $rrs$, so ${\textit{PEAU}}_{\textit{Adv}}$($s_5$, ${QS}_3$) = $\max${\{${\textit{PEAU}}_{\textit{Adv}}$($s_5$, $\langle$1, 2$\rangle$, ${QS}_3$), ${\textit{PEAU}}_{\textit{Adv}}$($s_5$, $\langle$1, 3$\rangle$, ${QS}_3$)\}} = $\max$\{$\frac{23+76}{3+2}$, $\frac{15+36}{3+2}$\} = $\frac{99}{5}$. In ${QS}_1$, the sequence $s_5$ has an \textit{instance} and ${\textit{PEAU}}_{\textit{Adv}}$($s_5$, ${QS}_1$) = $\frac{4+32+4+30+6}{3+2}$ = $\frac{76}{5}$. In addition, there is no \textit{instance} in ${QS}_2$. we have ${\textit{PEAU}}_{\textit{Adv}}$($s_5$, ${QS}_2$) is equal to 0. Finally, the ${\textit{PEAU}}_{\textit{Adv}}$ of sequence $s_5$ is calculated as follows: ${\textit{PEAU}}_{\textit{Adv}}$($s_5$) = ${\textit{PEAU}}_{\textit{Adv}}$($s_5$, ${QS}_1$) + ${\textit{PEAU}}_{\textit{Adv}}$($s_5$, ${QS}_2$) + ${\textit{PEAU}}_{\textit{Adv}}$($s_5$, ${QS}_3$) = $\frac{76}{5}$+$0$+$\frac{99}{5}$ = $\frac{175}{5}$ = 35 < $\xi$ $\times$ $u(\mathcal{D})$ = 36.

It is crucial to emphasize that ${\textit{VPEAU}_{\textit{Adv}}(S, p, QS)}$ is not strictly an upper bound because the average utility of the generated sequence is reduced if the utility of the extension item is less than the average utility of its prefix. However, as demonstrated in the subsequent proof, ${\textit{VPEAU}_{\textit{Adv}}(S, p, QS)}$ still supports the development of more efficient pruning strategies.

\textbf{Theorem:} For any sequence $S$ $\neq$ $\langle$$\rangle$ and its extension $S'$ in database $\mathcal{D}$, if $\textit{VPEAU}_\textit{Adv}$($S$) $\leqslant$ $\xi$ $\times$ $u(\mathcal{D})$ then $au$($S'$) $\leqslant$ $\xi$ $\times$ $u(\mathcal{D})$.

\textbf{Proof:} Let sequence $S$ be a prefix of $S'$, with $rs$ represents its \textit{remaining sequence}. The $s$ is the extension sequence and is a subsequence of $rs$. Denoting $exu$($S$) as the excess utility of $S$, it represents an excess part of the utilities of sequence $S$ which is larger than a threshold. Let this threshold is $\xi$ $\times$ $u(\mathcal{D})$, we obtain $exu$($S$) = $u$($S$)-($|S|$ $\times$ $\xi$ $\times$ $u$($\mathcal{D}$)). As per the earlier definition \ref{def: RRS}, $rrs$ is the \textit{remaining rising sequence} of $S$, and $u_{rrs}$($S$, $QS$) is the utility of $rrs$, so the excess utility of \textit{remaining rising sequence} is ${exu}_{rrs}$($S$) = $u_{rrs}$($S$)-($|rrs|$ $\times$ $\xi$ $\times$ $u$($\mathcal{D}$)). The excess part of utilities of sequence $s$ is ${exu}$($s$) = $u$($s$)-($|s|$ $\times$ $\xi$ $\times$ $u$($\mathcal{D}$)), and in $s$, any item whose utility greater than or equal to this threshold belongs to $rrs$. It is obviously that $exu$($s$, $QS$) $\leqslant$ ${exu}_{rrs}$($S$, $QS$). Then, we derive,
	\begin{displaymath}
	\begin{split}
	&au(S') = \sum{au(S', QS)} = \frac{\sum{u(S', QS)}}{|S'|} \\ 
	&=\frac{\sum{u(S', QS)}+ \sum{u(s, QS)}}{|S|+|s|} \\ 
	&=\frac{[|S| \times \xi \times u(\mathcal{D})+exu(S)]+[|s| \times \xi \times u(\mathcal{D})+exu(s)]}{|S|+|s|} \\ 
	&= \xi \times u(\mathcal{D})+\frac{exu(S)+exu(s)}{|S|+|s|} \\ 
	&\leqslant \xi \times u(\mathcal{D})+\frac{exu(S)+exu(s)}{|S|} \\ 
	&\leqslant \xi \times u(\mathcal{D})+\frac{exu(S)+{exu}_{rrs}(S)}{|S|} \\ 
	&= \frac{[|S| \times \xi \times u(\mathcal{D})+exu(S)]+{exu}_{rrs}(S)}{|S|} \\ 
	&= \frac{u(S)+{exu}_{rrs}(S)}{|S|}. \nonumber
	\end{split}
	\end{displaymath}

Thus, if $\frac{u(S)+{exu}_{rrs}(S)}{|S|}$ $\leqslant$ $\xi$ $\times$ $u(\mathcal{D})$, then we can derive that $au$($S'$) $\leqslant$ $\xi$ $\times$ $u(\mathcal{D})$. Assume that $\frac{u(S)+u_{rrs}(S)}{|S|+|rrs|_d}$ $\leqslant$ $\xi$ $\times$ $u(\mathcal{D})$. Then, we derive,
	\begin{displaymath}
	\begin{split}
	&u(S)+u_{rrs}(S) \leqslant \xi \times u(\mathcal{D}) \times (|S|+|rrs|_d) \\
	&u(S)+u_{rrs}(S)-\xi \times u(\mathcal{D}) \times |rrs|_d \leqslant \xi \times u(\mathcal{D}) \times |S| \\
	&u(S)+u_{rrs}(S)-\xi \times u(\mathcal{D}) \times |rrs| \leqslant \xi \times u(\mathcal{D}) \times |S| \\
		&u(S)+{exu}_{rrs}(S) \leqslant \xi \times u(\mathcal{D}) \times |S| \\
	&\frac{u(S)+{exu}_{rrs}(S)}{|S|} \leqslant \xi \times u(\mathcal{D}) \nonumber
	\end{split}
	\end{displaymath}

Thus, if $\frac{u(S)+u_{rrs}(S)}{|S|+|rrs|_d}$ $\leqslant$ $\xi$ $\times$ $u(\mathcal{D})$, then we can derive that $\frac{u(S)+{exu}_{rrs}(S)}{|S|}$ $\leqslant$ $\xi$ $\times$ $u(\mathcal{D})$. Then, if $\frac{u(S)+u_{rrs}(S)}{|S|+|rrs|_d}$ $\leqslant$ $\xi$ $\times$ $u(\mathcal{D})$, then we can derive that $au(S')$ $\leqslant$ $\xi$ $\times$ $u(\mathcal{D})$. It can be shown that the ${\textit{VPEAU}}_{\textit{Adv}}$$(S, p, QS)$ does not strictly serve as an upper bound. However, it can also be used for removing unpromising items in the \textit{remaining sequence}.

\begin{definition}[\textit{RSAU} Upper Bound] \label{def: RSAU}
  \rm In a \textit{q}-sequence $QS$, let \textit{PEAU}$(S, QS)$ be the prefix extension average utility of $S$. Through one \textit{extension} operation, the sequence $S$ is expanded to a sequence $S'$. That means letting a node in the LQS-tree represent $S$, with the node representing $S'$ being its child node. The $\textit{RSAU}(S', QS)$ is reduced sequence average utility of $S'$ in $QS$, and is defined as
	\begin{displaymath}
	\begin{split}
	&\textit{RSAU}(S', QS) \\
    &= \left\{\begin{array}{rcl} \textit{PEAU}(S, QS), & {S \sqsubseteq QS} \wedge {S' \sqsubseteq QS} \wedge {QS \in D} \\ 0, & otherwise \end{array}\right. \nonumber
	\end{split}
	\end{displaymath}
\end{definition} 

Then, $\textit{RSAU}$($S$) = $\sum_{{S\sqsubseteq QS} \wedge {QS \in D}}{\textit{RSAU}(S,QS)}$ is defined as the \textit{RSAU} of $S$ in database $\mathcal{D}$.

\textbf{Theorem:} For any sequence $S'$ $\neq$ $\langle$$\rangle$ in database $\mathcal{D}$, assume the sequence $S''$ = $S'$ or is extended from $S'$, if $\textit{RSAU}$($S'$) $\leqslant$ $\xi$ $\times$ $u(\mathcal{D})$ then $au$($S''$) $\leqslant$ $\xi$ $\times$ $u(\mathcal{D})$.

\textbf{Proof:} Consider sequences $S'$ and $S$, both of them are contained in the \textit{q}-sequence $QS$. We assume that $S'$ is extended from sequence $S$ through one \textit{extension} operation. Then, based on Definition \ref{def: RSAU}, we have $\textit{RSAU}$$(S', QS)$ = $\textit{PEAU}$$(S, QS)$. Consider any sequence $S''$ which is extended from $S'$ or $S''$ = $S'$, we have $S$ is also a prefix of $S''$. By comparison with Definition \ref{def: PEAU}, we see that $au$($S''$, $QS$) $\leqslant$ $\textit{PEAU}$($S$, $QS$). So that we also have $au$($S''$, $QS$) $\leqslant$ $\textit{RSAU}$($S'$). Therefore, the proposed reduced sequence average utility is one of UBs of sequence average utility.

Hence, any descendants of $S$ are not HAUSPs, if the value of $\textit{RSAU}$($S'$) is below the minimum acceptable average utility. All its descendants of $S$ could be pruned.

The RSU takes more account of the prefix sequence and neglects items between the extension item and the appending item. Although these neglected items are irrelevant to the generated sequence, they contribute to a tighter upper bound \cite{Ref22}. Therefore, the TRSU was designed by Zhang et al. \cite{Ref46} as a tighter utility upper bound for HUSP. This improvement is still effective for HAUSP.

\begin{definition}[\textit{TRSAU} Upper Bound] \label{def: TRSAU}
  \rm In a \textit{q}-sequence $QS$, a sequence $S$ is expanded to $S'$ through one \textit{extension} operation. Let $rs$ be the \textit{remaining sequence} which is the rest after the extension position to the end of $QS$. Let ${p'}_1$: $\langle$ ${j'}_1, {j'}_2$, $\cdots$, ${j'}_n$$\rangle$ be the position of the first \textit{instance} of the sequence $S'$, ${j'}_n$ is the \textit{extension position} of the first \textit{instance} of $S'$ and $i$ is the \textit{extension item}. The position of the first \textit{instance} of the sequence $S$ is $p_1$. Assume $j_m$ is the last \textit{extension position} of $S$ before ${j'}_n$, and $p_i$ is the position of the corresponding \textit{instance} of $S$. The \textit{TRSAU(S,QS)} is tighter reduced sequence average utility of $S$ in $QS$, and could be defined as
	\begin{displaymath}
	\begin{split}
	&\textit{TRSAU}(S', QS) \\
    &= \left\{\begin{array}{rcl} \textit{PEAU}(S, QS)-\frac{u_{rs}(S, j_m, QS)-u(i, {j'}_n, QS)-u_{rs}(S', {j'}_n, QS)}{|S|}\\,  S'\sqsubseteq QS,\textit{PEAU}(S, QS)=\textit{PEAU}(S, p_1, QS) \\ \textit{RSAU}(S', QS), otherwise \end{array}\right. \nonumber
	\end{split}
	\end{displaymath}
\end{definition} 

Then, $\textit{TRSAU(S)}$ = $\sum_{{S\sqsubseteq QS} \wedge {QS\in \mathcal{D}}}{\textit{TRSAU}(S, QS)}$ is defined as the \textit{TRSAU} of $S$ in a database $\mathcal{D}$.

\textbf{Theorem:} For any sequence $S'$ $\neq$ $\langle$$\rangle$ in a  database $\mathcal{D}$, assume the sequence $S''$ = $S'$ or is extended from $S'$, if $\textit{TRSAU}$($S'$) $\leqslant$ $\xi$ $\times$ $u(\mathcal{D})$ then $au$($S''$) $\leqslant$ $\xi$ $\times$ $u(\mathcal{D})$.

\textbf{Proof:} Consider the sequences $S'$ and $S$, both of them are contained in the \textit{q}-sequence $QS$. We assume that $S'$ is extended from sequence $S$ through one \textit{extension} operation. Let sequence $S''$ be extended from $S'$ or $S''$ = $S'$, we have $S$ is also a prefix of sequence $S''$. For $S$, the extension position is $j_m$. Let $rs$ be the \textit{remaining sequence} which is the rest after $j_m$ to the end of $QS$. Based on Definition \ref{def: PEAU}, we see that $au$($S''$, $QS$) $\leqslant$ $\textit{PEAU}$($S$, $QS$). By comparing this with Definition \ref{def: TRSAU}, the subsequence between extension positions, $j_m$ and ${j'}_n$, is irrelevant to the generated sequence $S'$ and sequence $S''$. That is, the item in this subsequence will not be contained in any generated sequence and could be removed. The corresponding subsequence utility is $u_{rs}$($S'$, $j_m$, $QS$)-$u_{rs}$($S'$, ${j'}_{n-1}$, $QS$), which equals $u_{rs}$($S$, $j_m$, $QS$)-$u_{rs}$($S'$, ${j'}_n$, $QS$)-$u$($i$, ${j'}_n$, $QS$). From Definition \ref{def: RS}, we get $u_{rs}$($S$, $p_1$, $QS$) $\geqslant$ $u_{rs}$($S$, $p_{max}$, $QS$). Then we can derive that $u$($S$, $p_1$, $QS$) $\leqslant$ $u$($S$, $p_{max}$, $QS$). Thus, when $\textit{PEAU}$($S$, $QS$) = $\textit{PEAU}$($S$, $p_1$, $QS$), we have $\textit{PEAU}$($S'$, $QS$) = $\max{\{\textit{PEAU}(S', p, QS)\}}$ = $\textit{PEAU}$($S'$, ${p'}_1$, $QS$). Therefore, we can obtain
	\begin{displaymath}
	\begin{split}
	&\textit{TRSAU}(S', QS) \\
	&= \textit{PEAU}(S, p_1, QS) \\
    & \quad -\frac{u_{rs}(S, j_m, QS)-u(i, {j'}_n, QS)-u_{rs}(S', {j'}_n, QS)}{|S|} \\ 
	&= \frac{u(S, p_1, QS)+u_{rs}(S, j_i, QS)}{|S|} \\
    & \quad -\frac{u_{rs}(S, j_m, QS)-u(i, {j'}_n, QS)-u_{rs}(S', {j'}_n, QS)}{|S|} \\ 
	&= \frac{u(S, p_1, QS)+u(i, {j'}_n, QS)}{|S|} \\
    & \quad +\frac{u_{rs}(S, j_i, QS)-u_{rs}(S, j_m, QS)+u_{rs}(S', {j'}_n, QS)}{|S|} \\ 
	&= \frac{u(S', {p'}_1, QS)+u_{rs}(S', {j'}_n, QS)}{|S|} \\
    & \quad +\frac{u_{rs}(S, j_i, QS)-u_{rs}(S, j_m, QS)}{|S|} \\ 
	&\geqslant \frac{u(S', {p'}_1, QS)+u_{rs}(S', {j'}_n, QS)}{|S|} \\
    &\geqslant \textit{PEAU}(S', {p'}_1, QS) = \textit{PEAU}(S', QS) \\ 
	&\geqslant au(S'', QS). \nonumber
	\end{split}
	\end{displaymath}

This means that we also have $au$($S''$, $QS$) $\leqslant$ $\textit{TRSAU}$($S'$). Hence, the proposed reduced sequence average utility is also one of UBs of sequence average utility. 

Hence, $S'$ and any descendants of $S'$ are not HAUSPs, if the value of \textit{TRSAU(S')} is below the minimum acceptable average utility. $S'$ and all its descendants could be pruned. For example, $s_6$: $\langle$\{$b$, $c$\}, \{$d$\}, \{$e$\}$\rangle$ is a sequence, we assume that $s_5$: $\langle$\{$b$, $c$\}, \{$d$\}$\rangle$ is its prefix. In Table \ref{table: database}, there are three instances of $s_5$ at position $\langle$1, 2$\rangle$, $\langle$1, 4$\rangle$ and $\langle$3, 4$\rangle$ in ${QS}_3$, but only one \textit{instances} of $s_6$ at position $\langle$1, 2, 3$\rangle$. $\textit{TRSAU}$($s_6$, ${QS}_3$) = $\textit{TRSAU}$($s_6$, $\langle$1, 2, 3$\rangle$, ${QS}_3$) = $\frac{(12+34-21)}{3}$ = $\frac{(12+7+6)}{3}$ = $\frac{25}{3}$.

The proposed variant of \textit{TRSAU} is based on the ${\textit{VPEAU}}_{\textit{Adv}}$ $(S,p,QS)$, and is defined as
\begin{displaymath}
	\begin{split}
	&{\textit{VTRSAU}}_{\textit{Adv}}(S', QS) \\
    &= \left\{\begin{array}{rcl} {\textit{VPEAU}}_{\textit{Adv}}(S, QS)-\frac{u_{rrs}(S, p_i, QS)-u(i, {j'}_n, QS)-u_{rrs}(S', {p'}_1, QS)}{|S|+|rrs|_d}\\, S'\sqsubseteq QS, {\textit{VPEAU}}_{\textit{Adv}}(S, QS)={\textit{VPEAU}}_{\textit{Adv}}(S, p_1, QS) \\ \textit{RSAU}(S', QS), otherwise \end{array}\right. \nonumber
	\end{split}
\end{displaymath}
where $u_{rrs}$ is the utility of the subsequence $rrs$. The subsequence $rrs$ is the \textit{remaining rising sequence} of $S'$. It's essential to note that, because ${\textit{VPEAU}_{\textit{Adv}}(S, p, QS)}$ is not strictly an upper bound, ${\textit{VTRSAU}_{\textit{Adv}}(S', QS)}$ does not strictly serve as an upper bound. It is also only used to facilitate a more efficient pruning strategy.

So far, we have designed two primary UBs, PEAU and RSAU. We also enhanced these two bounds by incorporating the pattern-growth method. In pursuit of higher pruning efficiency, we design new measurements and variants of these UBs, which are not upper bounds. Combinations of different strategies can be utilized as various versions of the algorithm HAUSP-PG, and in section \ref{sec: experiment}, we will evaluate the efficiency of three typical versions among them.

\subsection{An Illustration of the Proposed Method}
\label{sec: illustration}

\begin{figure*}
    \centering
    \includegraphics[width=1\linewidth]{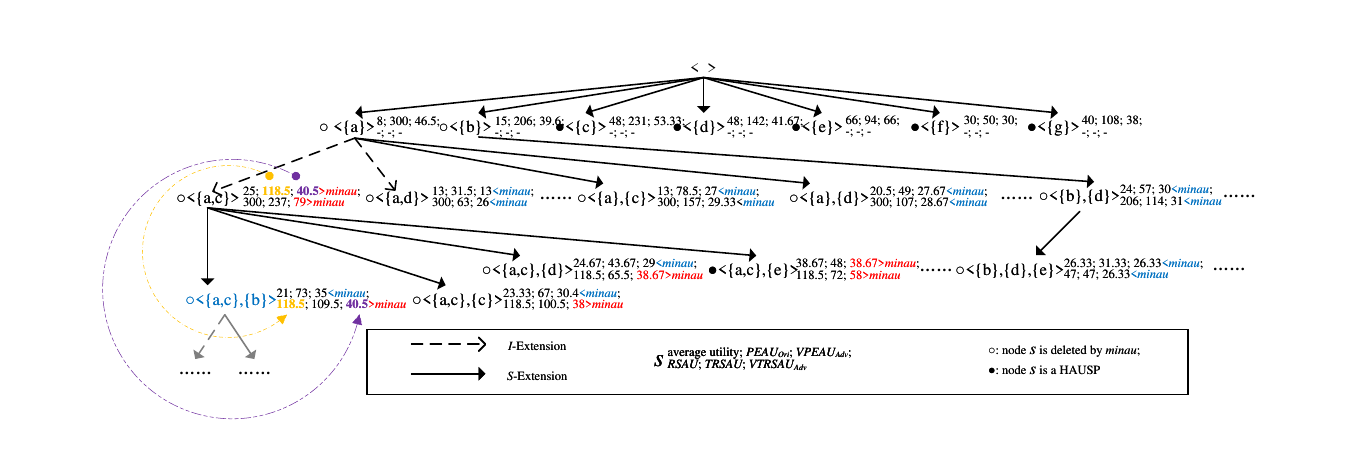}
	\caption{Process of an illustrative example of HAUSP-PG.}
    \label{fig: 04}
\end{figure*}

Fig. \ref{fig: 04} is an illustrative example of HAUSP-PG with $\xi$ = 12$\%$, such that the minimum acceptable average utility of $\mathcal{D}$ in Table \ref{table: database} is $\xi$ $\times$ $u(\mathcal{D})$ = 36. Each sequence in the search space corresponds to a node in the LQS-tree. All nodes, except the root node, are represented in the form of \textit{s}(\textit{average utility}; ${\textit{PEAU}}_{\textit{Ori}}$; ${\textit{VPEAU}}_{\textit{Adv}}$; \textit{RSAU}; \textit{TRSAU}; ${\textit{VTRSAU}}_{\textit{Adv}}$), and these notations will be discussed in Section \ref{sec: pruning}. Fig. \ref{fig: 04} mainly shows some of the nodes in two branches. Despite their average utility does not meet the minimum threshold, nodes $\langle$\{$a$\}$\rangle$ and $\langle$\{$b$\}$\rangle$ still are expected to generate new HAUSPs. Once the corresponding \textit{PEAU} could not reach this threshold, for example, the node $\langle$\{$a$, $d$\}$\rangle$, its descendants will be pruned. Furthermore, if the corresponding \textit{TRSAU} could not reach this threshold, for example, the node $\langle$\{$b$\}, \{$d$\}, \{$f$\}$\rangle$, the whole branch which includes the current node and all descendants will be pruned. In addition, we could see that even if the actual average utility of the prefix sequence cannot reach the minimum acceptable average utility, its generated sequence may still be a HAUSP. Furthermore, the value of ${\textit{VPEAU}}_{\textit{Adv}}$ is lower than ${\textit{PEAU}}_{\textit{Ori}}$, and \textit{TRSAU} is not higher than \textit{RSAU}, so we could distinguish unpromising pattern and prune them early. However, each upper bound in the parent node is an overestimate of the average utility of the generated sequence in the child node.

As in the earlier example of upper bound calculation in this section, consider sequence $s_5$: $\langle$\{$a$, $c$\}, \{$b$\}$\rangle$ from Table \ref{table: database}. Its corresponding upper bounds are given as follows: ${\textit{PEAU}}_{\textit{Ori}}$($s_5$) = 73 > $\xi$ $\times$ $u(\mathcal{D})$ = 36, ${\textit{PEAU}}_{\textit{Inc}}$($s_5$) = 54.75 > $\xi$ $\times$ $u(\mathcal{D})$ = 36, ${\textit{PEAU}}_{\textit{Rev}}$($s_5$) = 43.75 > $\xi$ $\times$ $u(\mathcal{D})$ = 36, ${\textit{PEAU}}_{\textit{Adv}}$($s_5$) = 35 < $\xi$ $\times$ $u(\mathcal{D})$ = 36, with detailed calculation procedures referring to the previous content. Based on the variant forms of the upper bound model, we judge that extending $s_5$ cannot yield new target patterns. In fact, the generated sequences that may potentially become target patterns by extending $s_5$ include: $\langle$\{$a$, $c$\}, \{$b$, $g$\}$\rangle$, $\langle$\{$a$, $c$\}, \{$b$, $g$\}$, \{$e$\}\rangle$, $\langle$\{$a$, $c$\}, \{$b$, $g$\}$, \{$d$\}, \{$e$\}\rangle$, $\langle$\{$a$, $c$\}, \{$b$\}$, \{$e$\}\rangle$ and $\langle$\{$a$, $c$\}, \{$b$, $e$\}$\rangle$. Their corresponding actual average utilities are 15.75, 19.8, 20, 26, and 17.5, respectively, which thus confirms the conclusion derived from the upper bound judgment.
 
\subsection{Proposed HAUSP Algorithm}
\label{sec: hausp-pg}

The proposed algorithm HAUSP-PG integrates the aforementioned data structure and mechanism into the mining method. It is divided into three parts as follows. The first part starts by scanning the quantitative sequential database $\mathcal{D}$ and constructing the \textit{proDB} (Lines 2 to 7). It is the first procedure and the main body of HAUSP-PG. The prefix sequence is stored in a projection database and is used in the next procedure PGrowth.

\begin{algorithm}[ht]
    \small
    \caption{The HAUSP-PG algorithm}
    \label{alg: 1}
    \LinesNumbered
    \KwIn{A quantitative sequential database $\mathcal{D}$; A minimum average utility threshold, $\xi$.} 
    \KwOut{HAUSPs in $\mathcal{D}$.}
	scan $\mathcal{D}$ and calculate $u$($\mathcal{D}$)\\
	\For {\rm \textbf{each} $QS$ $\in$ $\mathcal{D}$}{
		build the \textit{seqArray}\\
	}
	build the projection database \textit{proDB}\\
	\quad 1). \textit{proDB.seqArray} $\leftarrow$ \{\textit{seqArray} of $QS$ | $QS$ $\in$ $\mathcal{D}$\}\\
	\quad 2). \textit{proDB.exList} $\leftarrow$ \textit{NULL}\\
	initialize HAUSPs $\leftarrow$ $\phi$\\
    \For {\rm \textbf{each} $s$ $\in$ \textit{1}-sequence}{
	calculate the average utility and \textit{PEAU} of all \textit{1}-sequences\\
	\If {$au$($s$) $ \geqslant$ $\xi$ $\times$ $u$($\mathcal{D}$)}{
		HAUSPs $\leftarrow$ HAUSPs $\cup$ $s$\\
	}
	\If {\textit{PEAU}($s$) $\geqslant$ $\xi$ $\times$ $u$($\mathcal{D}$)}{
		call PGrowth($s$, \textit{proDB}($s$), HAUSPs)\\
	}
    }

 \textbf{return} HAUSPs
\end{algorithm}

\begin{algorithm}[ht]
    \small
    \caption{The PGrowth algorithm}
    \label{alg: 2}
    \LinesNumbered
    \KwIn{A projected database, \textit{proDB}; A prefix, $S$; HAUSPs.} 
    \KwOut{HAUSPs in $\mathcal{D}$.}
	scan \textit{proDB}($S$) to remove the unpromising item and the irrelevant item\\
	\quad 1). get the set of \textit{I}-Extension items of $S$, \textit{ilist}\\
	\quad 2). get the set of \textit{S}-Extension items of $S$, \textit{slist}\\
	\For {\rm \textbf{each} item $i$ $\in$ \textit{ilist}}{
	\If{\textit{TRSAU}($S$) < $\xi$ $\times$ $u(\mathcal{D})$}{
		mark the irrelevant item $i$\\
            continue
	}
		call AUCalcu($S$ $\oplus$ $i$, \textit{proDB}($S$), HAUSPs)\\
	}
	\For {\rm \textbf{each} item $i$ $\in$ \textit{slist}}{
	\If{\textit{TRSAU}($S$) < $\xi$ $\times$ $u(\mathcal{D})$}{
		mark the irrelevant item $i$\\
            continue
	}
		call AUCalcu($S$ $\otimes$ $i$, \textit{proDB}($S$), HAUSPs)\\
	}
\end{algorithm}

In the PGrowth procedure, firstly, it removes all the irrelevant items and the unpromising items from the set of appending items. These items are selected by upper bounders and corresponding pruning strategies. Then, from lines 4 to 17, sequences are generated by the \textit{I}-Extension or \textit{S}-Extension operations. The average utility and \textit{PEAU} are calculated by calling the procedure AUCalcu.

\begin{algorithm}[ht]
    \small
    \caption{The AUCalcu algorithm}
    \label{alg: 3}
    \LinesNumbered
    \KwIn{A projected database, \textit{proDB}; A sequence extended by the prefix $S$, $S'$; HAUSPs.} 
    \KwOut{HAUSPs in $\mathcal{D}$.}
	\textit{proDB}($S'$) $\leftarrow$ \{\textit{proDB} of $S'$ | {$S'$ $\sqsubseteq$ $QS$} $\wedge$ {$QS$ $\in$ \textit{proDB}($S$)}\}\\
	calculate $au$($S'$) and \textit{PEAU}($S'$)\\
	\If {$au$($S'$) $\geqslant$ $\xi$ $\times$ $u(\mathcal{D})$}{
		HAUSPs $\leftarrow$ HAUSPs $\cup$ $S'$\\
	}

	\If {\textit{PEAU}($S'$) $\geqslant$ $\xi$ $\times$ $u$($\mathcal{D}$)}{
		call PGrowth($S'$, \textit{proDB}($S'$), HAUSPs)\\
	}
\end{algorithm}

As shown in the AUCalcu procedure, based on the \textit{proDB} of prefix sequence, a new projected database for the sequence $S'$ is established in the first line. The average utility and \textit{PEAU} are then calculated. If the value of average utility is not less than $\xi$ $\times$ $u(\mathcal{D})$, the generated sequence $S'$ is a HAUSP (lines 3 to 5). Moreover, if the value of \textit{PEAU} is equal to or greater than $\xi$ $\times$ $u$($\mathcal{D}$), the generated sequence $S'$ could be a prefix of a HAUSP. Sequences are generated with the prefix $S$ by calling the PGrowth procedure. Eventually, when no new sequences are generated, the algorithm HAUSP-PG returns the set of HAUSPs and then terminates.

Although they do not address the same problem, the proposed HAUSP-PG algorithm has indeed drawn on the design ideas of the HUSP-SP algorithm and adopted a similar pattern-growth approach. This choice is based on two considerations: On the one hand, HAUPM can be regarded as a variant of HUPM that incorporates both sequence length and utility, making it more fair and practical for scenarios requiring balanced evaluation of utility and length. On the other hand, the pattern-growth method appears to yield greater efficiency gains in quantitative sequential databases compared to transactional databases. This is because the fixed prefix items and their ordered combinations in sequential data effectively restrict the size of the search space, reducing redundant computations. In addition, the design of HAUSP-PG offers several additional advantages: Firstly, it achieves higher pruning efficiency through threshold growth. Considering the general definition of average utility, for a specified threshold (e.g., \textit{minau}), the total utility threshold for a pattern, $u(s)$, dynamically increases with its length, $|s|$, (i.e., $u(s)$ = \textit{minau} $\times$ $|s|$). This dynamic adjustment creates a partial effect similar to the threshold elevation strategy in top-$k$ method, enhancing the efficiency of generated pattern filtering. Secondly, the search space is constrained by filtering $rs$, making the proposed method efficient. Appending an item with low utility will reduce the average utility of the generated sequence; thus, only items with utility not below a specific threshold are considered to maintain or improve the average utility. By filtering such items, HAUSP-PG reduces the number of items that need to be considered during utility estimation, directly improving efficiency. In summary, the proposed HAUSP-PG algorithm ensures efficiency through dual pruning: it not only utilizes the upper bound \textit{TRSAU} or its variants to remove irrelevant items from prefix sequences but also employs ${\textit{PEAU}}_{\textit{Adv}}$ and ${\textit{RSAU}}_{\textit{Adv}}$ to eliminate unpromising items with utility below the threshold in remaining sequences.

Overall, the proposed algorithm HAUSP-PG represents a novel attempt to discover HAUSPs. Inspired by HUSP-SP, it further refines its design by leveraging certain characteristics of average utility, thereby achieving favorable processing efficiency while responding to the inherent need to balance utility and sequence length.  Furthermore, the use of a consistent processing framework facilitates more effective integration between HUSPM and HAUSPM, opening up new possibilities for advancements in this field.

\subsection{Complexity Analysis}
\label{sec: complexity}

Suppose the quantitative sequential database is composed of |$\mathcal{D}$| \textit{q}-sequences. There are |$I$| distinct items in it. And the average number of items pre \textit{q}-sequence in the database is denoted by |$QS$|. First of all, starting with the first procedure of the proposed algorithm, the original database is scanned, so the first step takes $\mathcal{O}$(|$D$| $\times$ |$QS$|). The memory complexity is also $\mathcal{O}$(|$D$| $\times$ |$QS$|) to construct a \textit{sequence-array} and the corresponding \textit{item-index} head table. Then, the second function, PGrowth, is called recursively, and the set of HAUSPs is returned. Let |$L$| be the length of the longest generated sequence. Then, |$QS$| $\leqslant$ |$L$|. It should be noted that, in the worst case, the maximum depth and the number of times of recursively calling are |$L$| and $|I|^{|L|}$, respectively.

Consider the function PGrowth. In this function, all items in the projected database are read, and the unpromising items and the irrelevant items are marked first. It takes $\mathcal{O}$(|$D$| $\times$ |$QS$|) to calculate the utility of each extension item, and to mark the unpromising ones with low item utility and the irrelevant ones with low \textit{TRSAU} or \textit{RSAU}. The \textit{iList} and \textit{sList} are built with the other extension items. At this time, in the worst case, none of them could be removed. The corresponding time complexity is the sum of all the time complexities of the calling processing, and the memory complexity is $\mathcal{O}$(|$I$|). Then, each item in \textit{iList} and \textit{sList} is appended into the prefix, and the next function, AUCalcu, is called to calculate the actual average utility of the generated sequence. In the function AUCalcu, the actual average utility and the \textit{PEAU} of the generated sequence are calculated for each appending item. Thus, it takes $\mathcal{O}$(|$D$| $\times$ |$QS$|)+$\mathcal{O}$(|$D$| $\times$ |$QS$|), which equals $\mathcal{O}$(|$D$| $\times$ |$QS$|). In this step, its memory complexity is $\mathcal{O}$(1). Therefore, the time complexity and memory complexity of function PGrowth are $\mathcal{O}$($|D$| $\times$ |$QS$|+|$I$| $\times$ |$D$| $\times$ |$QS$|) and $\mathcal{O}$(|$D$| $\times$ |$QS$|+|$I$|).

Based on the above, the worst time complexity of HAUSP-PG is $\mathcal{O}$(|$D$| $\times$ |$QS$|)+$|I|^{|L|}$ $\mathcal{O}$(|$D$| $\times$ |$QS$|+|$I$| $\times$ |$D$| $\times$ |$QS$|), equivalent to $\mathcal{O}$($|I|^{|L|}$|$D$||$QS$|). The worst memory complexity of HAUSP-PG is $\mathcal{O}$(|$D$| $\times$ |$QS$|)+|$L$|$\mathcal{O}$(|$D$| $\times$ |$QS$|+|$I$|), equivalent to $\mathcal{O}$(|$L$||$D$||$QS$|+|$I$|). Since |$QS$| $\leqslant$ |$L$|, the maximum time and memory complexity respectively are $\mathcal{O}$($|I|^{|L|}$|$D$||$L$|) and $\mathcal{O}$($|L|^2$ |$D$|+|$I$|). In Ref. \cite{Ref35}, the complexity of the EHAUSM algorithm is given as $\mathcal{O}$($|M|^{|L|}$), where $M$ denotes the number of distinct items. Compared with the method proposed in this paper, both algorithms have exponential time complexity (related to the number of items and sequence length). In terms of space complexity, both utilize projected databases, while in practical scenarios, the efficiency of both depends on the strength of pruning strategies. However, the optimizations of HAUSP-PG in sorting-free computation and remaining sequence processing make it potentially more efficient in specific complex datasets, especially those with long sequences and multiple items.

\section{Experiments} \label{sec: experiment}

The performance of the HAUSP-PG algorithm is assessed with the results of the experiment in this section. All algorithms were implemented in Java. The experiments are conducted on a cloud virtual machine equipped with an AMD EPYC 7542 32-Core CPU and the Linux version 5.4.0-166-generic.x86\_64 operating system. The source code is available at https://github.com/HNUSCS-DMLab/HAUSP-PG.

\subsection{Experimental Settings}
\label{sec: setting}

The EHAUSM is considered the first algorithm for mining HAUSPs in the general case \cite{Ref35}. Based on EHAUSM, we developed SimEHAUSM as one of the baseline algorithms for comparison. This algorithm replicates the three Upper Bounds (UBs) and their corresponding pruning strategies originally proposed in EHAUSM. However, we did not adopt the AMUB model in the early stage of the algorithm; instead, we kept consistency with the method proposed in this paper. The reason is that for specific datasets (e.g., the Kosarak dataset), this early-stage pruning exhibits significant advantages, which would hinder the comparison of upper-bound models in the iterative stage of the contrast algorithms. HUSP-SP \cite{Ref46} adopted PEU and a pattern-growth method and is more efficient than USpan. In the proposed HAUSP-PG, the \textit{PEAU} is inspired by PEU, and likewise adopts the pattern-growth method. The first experiment assesses the performance of HAUSP-PG against SimEHAUSM and HUSP-SP. The second experiment evaluates the effects of various UBs and their variants, and the dual pruning strategies mentioned above. The performance is evaluated in terms of the runtime, space overhead, total number of sequences generated, and scalability against large datasets.

\subsection{Data Description}
\label{sec: data}

The datasets for experiments include two synthetic datasets and five real-world datasets. The detailed characteristics of datasets are given in Table \ref{table: datasets}. These six parameters are used in the description of characteristics. These datasets can be downloaded from SPMF (http://www.philippe-fournier-viger.com/spmf/). Note that $|D|$ and $|I|$ are respectively the number of \textit{q}-sequences and the number of distinct items in the original database. \textit{MaxLen} is the maximal length of \textit{q}-sequence in the database. \textit{AvgLen}, \textit{AvgSeqSize} and \textit{AvgSeqSize} are the average length of \textit{q}-sequence, the average number of \textit{q}-itemsets in one \textit{q}-sequence, and the average number of \textit{q}-items in one \textit{q}-itemset respectively.

\begin{table}[ht]
    \renewcommand{\arraystretch}{1.35}
	\caption{Features of datasets.}  
	\label{table: datasets}
	\centering
    \resizebox{0.76\linewidth}{!}{ 
	\begin{tabular}{@{}ccccccc@{}}
	\toprule
		Dataset		& $|D|$	& $|I|$	& \textit{AvgLen}	& \textit{MaxLen}	& \textit{AvgSeqSize}	& \textit{AvgSeqSize} \\
	\midrule
		Bible			& 36369	& 13905	& 21.64	& 100 & 17.85 & 1.0 \\
		Leviathan		& 5834 & 9025 & 33.81	& 100 & 26.34 & 1.0 \\
		Sign			& 730 & 267 & 52 & 94 & 51.99	& 1.0 \\
		Yoochoose		& 234300 & 16,004 & 2.25 & 112 & 1.14 & 1.97 \\
		{Kosarak}\_{10K}	& 10000 & 10094 & 8.14 & 608 & 8.14 & 1.0 \\
		{SynDataset}\_{80K}	& 79718 & 7584 & 26.80 & 213 & 6.20 & 4.32 \\
		{SynDataset}\_{10K}	& 10000 & 7313 & 27.11 & 213 & 6.23 & 4.35 \\
	\bottomrule
	\end{tabular}}
\end{table} 

The sentence in the conversion dataset Bible or Leviathan is considered as a sequence, and each word was replaced by a digital item. The length of sequences is medium. The Sign is the sign language utterance dataset in a new format, and most of the sequences are long. Both Yoochoose and Kosarak are click-stream datasets; the former is collected in e-commerce, and the latter comes from the Hungarian online news portal. Yoochoose includes single-itemset sequences and multi-itemset sequences. The length of the sequence in Kosarak is extremely long. {SynDataset}\_{80K} and {SynDataset}\_{10K} are two different synthetic datasets. There are more than 80000 and 1000 sequences, respectively. {SynDataset}\_{160K} and other synthetic datasets could be generated by repeatedly duplicating the duplication of dataset {SynDataset}\_{80K}.

To delve into the utility distribution characteristics of different datasets and enable more precise analysis of how data features influence experimental performance when discussing in conjunction with experimental results, we have compiled key statistical indicators for several common datasets. These indicators include sample size, extreme values, central tendency, dispersion, and distribution shape parameters, all of which clearly reflect the distribution characteristics of each dataset. The specific statistical results are presented in Table \ref{table: dataset_distribution_utility}.

\begin{table*}[htbp]
    \centering
    \caption{Statistical Characteristics of Utility Distribution Across Datasets}
    \resizebox{\textwidth}{!}{
    \begin{tabular}{lccccccccccc}
    \hline
    \textbf{Dataset} & \textbf{Sample Size} & \textbf{Minimum Value} & \textbf{Maximum Value} & \textbf{Average Value} & \textbf{Median} & \textbf{Standard Deviation} & \textbf{CV} & \textbf{Skewness} & \textbf{Kurtosis} & \textbf{Distribution Shape} \\
    \hline
    BIBLE & 36369 & 46 & 2139 & 352.43 & 299.00 & 208.54 & 0.5917 & 1.903 & 8.522 & Significant long-tailed distribution \\
    \hline
    Leviathan & 5834 & 22 & 685 & 205.55 & 180.00 & 118.19 & 0.5750 & 1.039 & 3.800 & Intermediate form \\
    \hline
    SIGN & 730 & 176 & 1581 & 868.95 & 853.00 & 233.36 & 0.2686 & 0.348 & 3.135 & Approximate uniform distribution \\
    \hline
    Yoochoose & 234300 & 93 & 802253 & 8235.12 & 4655.00 & 12598.23 & 1.5298 & 9.839 & 253.801 & Significant long-tailed distribution \\
    \hline
    Kosarak10k & 10000 & 1 & 10270 & 139.63 & 54.00 & 384.93 & 2.7568 & 10.848 & 182.832 & Significant long-tailed distribution \\
    \hline
    Scalability\_160K & 159436 & 1 & 1286 & 165.90 & 141.00 & 113.35 & 0.6832 & 1.269 & 5.346 & Intermediate form \\
    \hline
    \end{tabular}}
    \label{table: dataset_distribution_utility}
\end{table*}

\subsection{Speed Efficiency Analysis}
\label{sec: efficiency}

\begin{figure*}
	\centering
	\begin{minipage}{0.98\textwidth}
			\centering
			\includegraphics[width=0.31\linewidth]{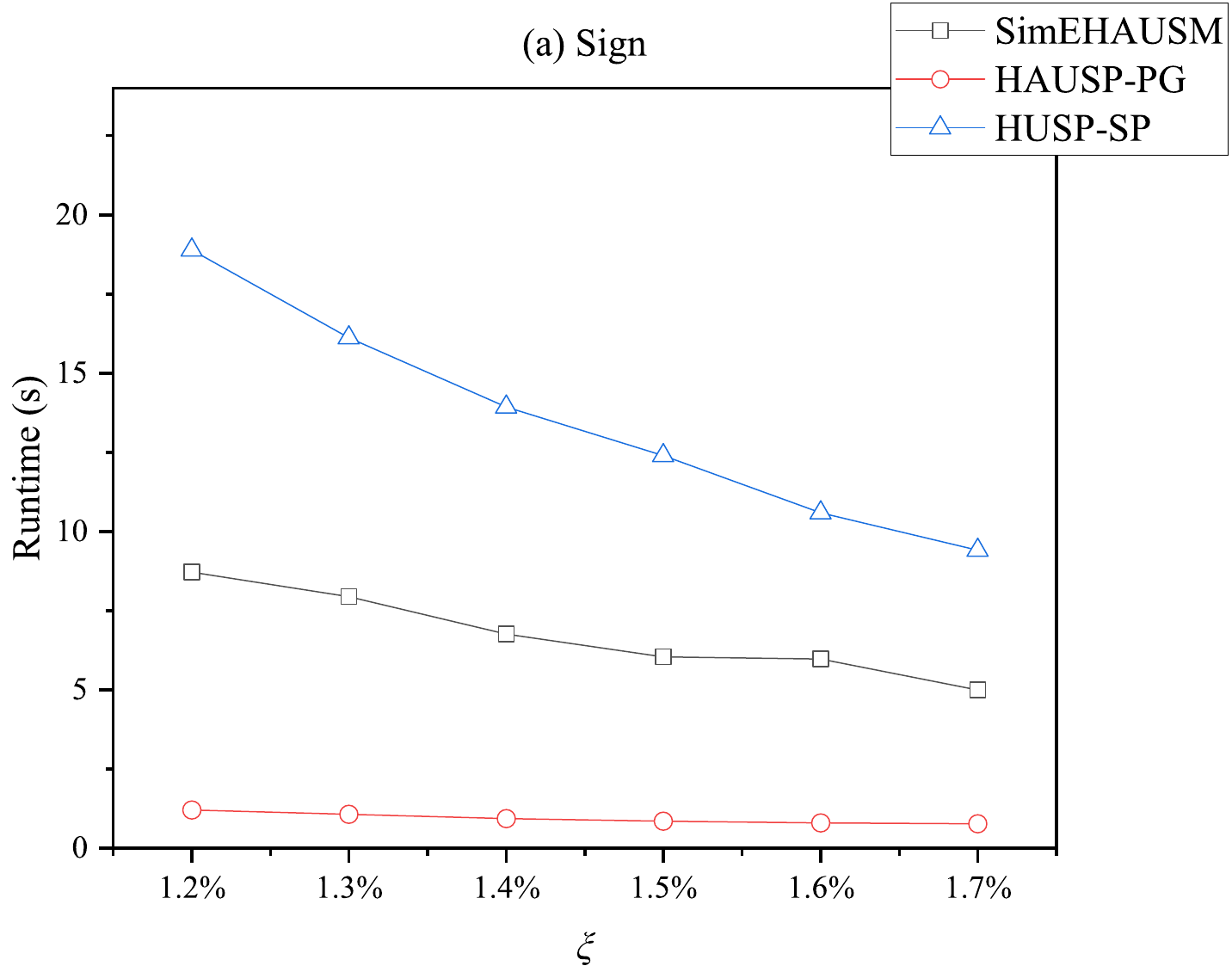}
			\label{fig: 06a}
			\includegraphics[width=0.31\linewidth]{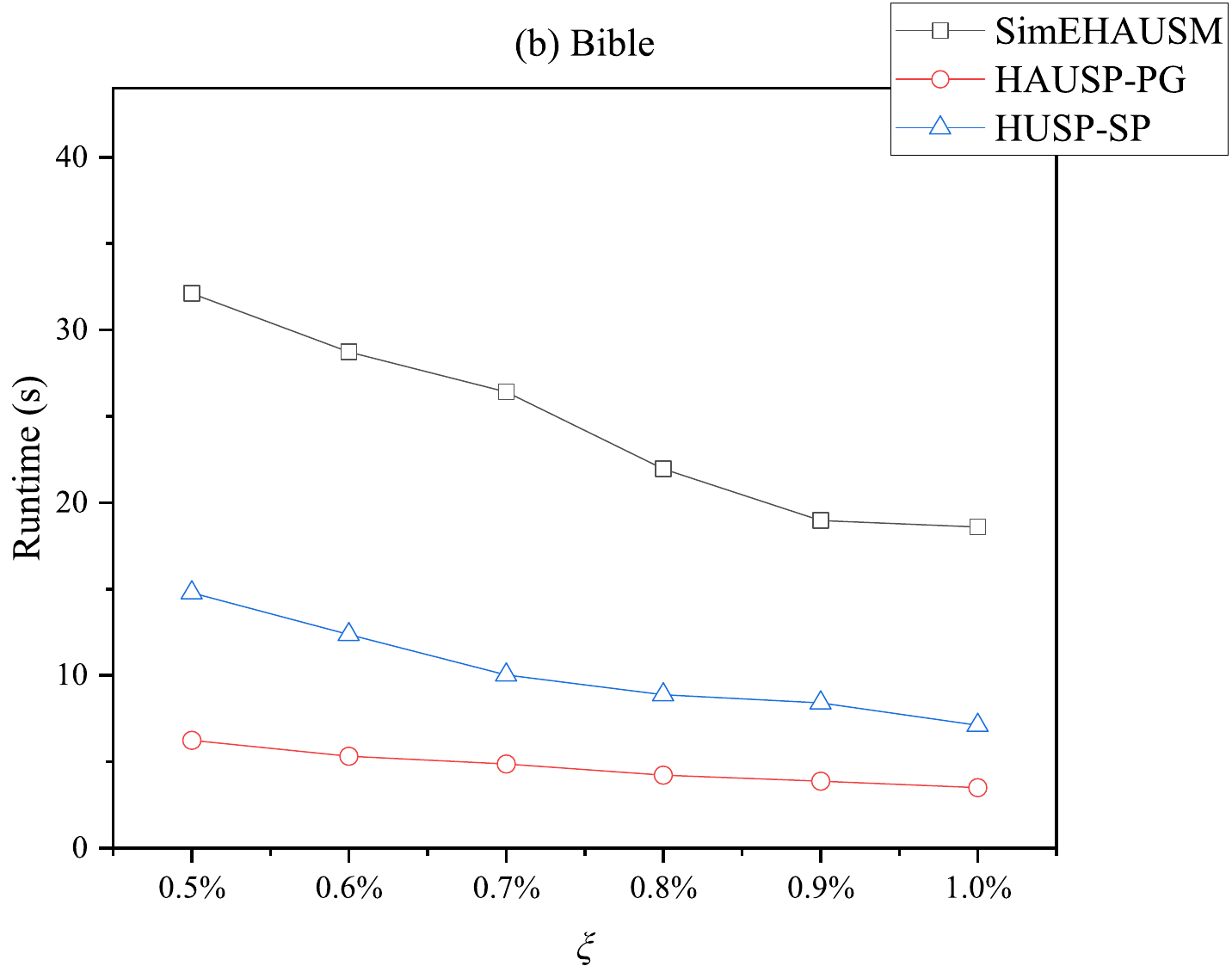}
			\label{fig: 06b}
			\includegraphics[width=0.31\linewidth]{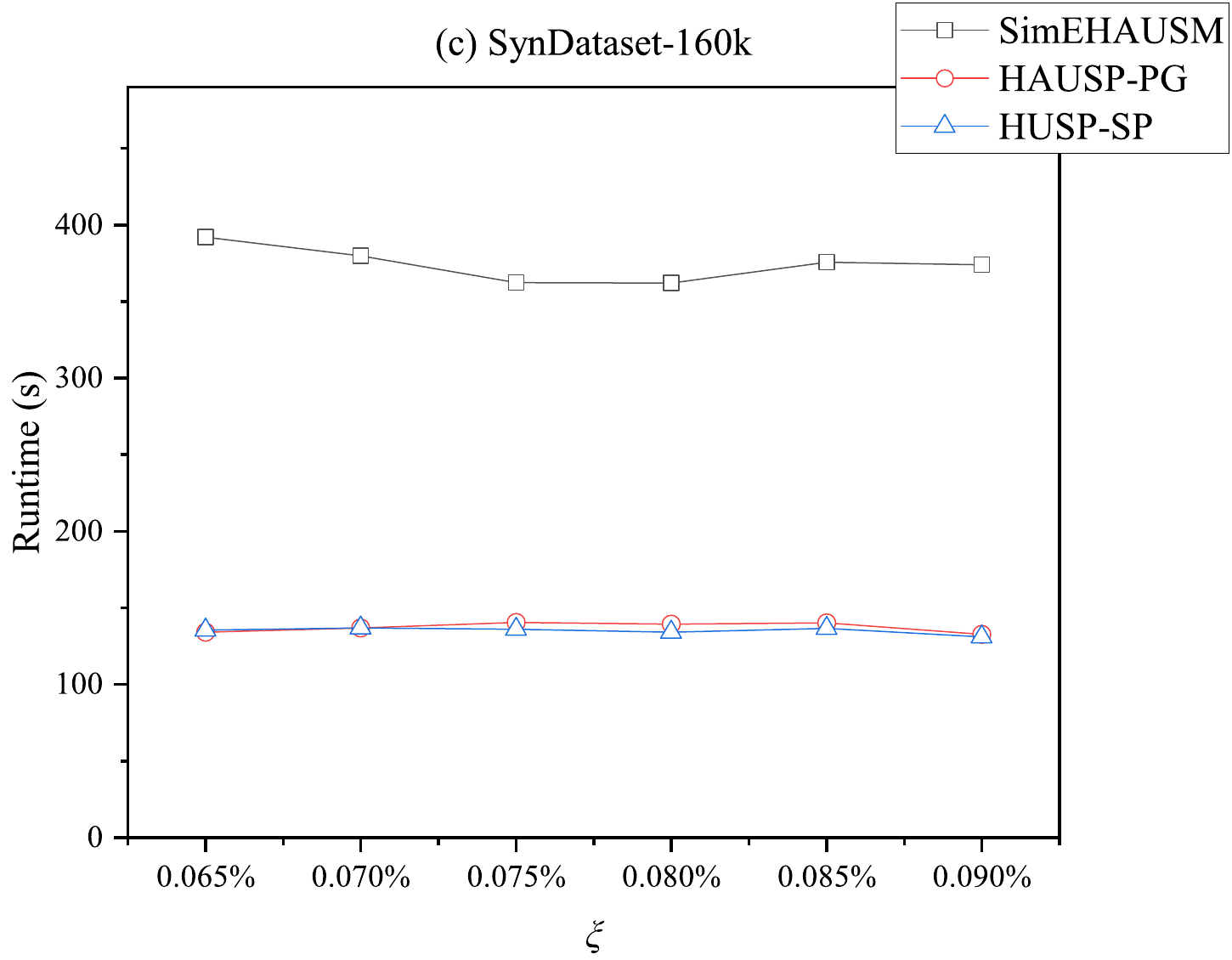}
			\label{fig: 06c}
	\end{minipage}
	\begin{minipage}{0.98\textwidth}
			\centering
			\includegraphics[width=0.31\linewidth]{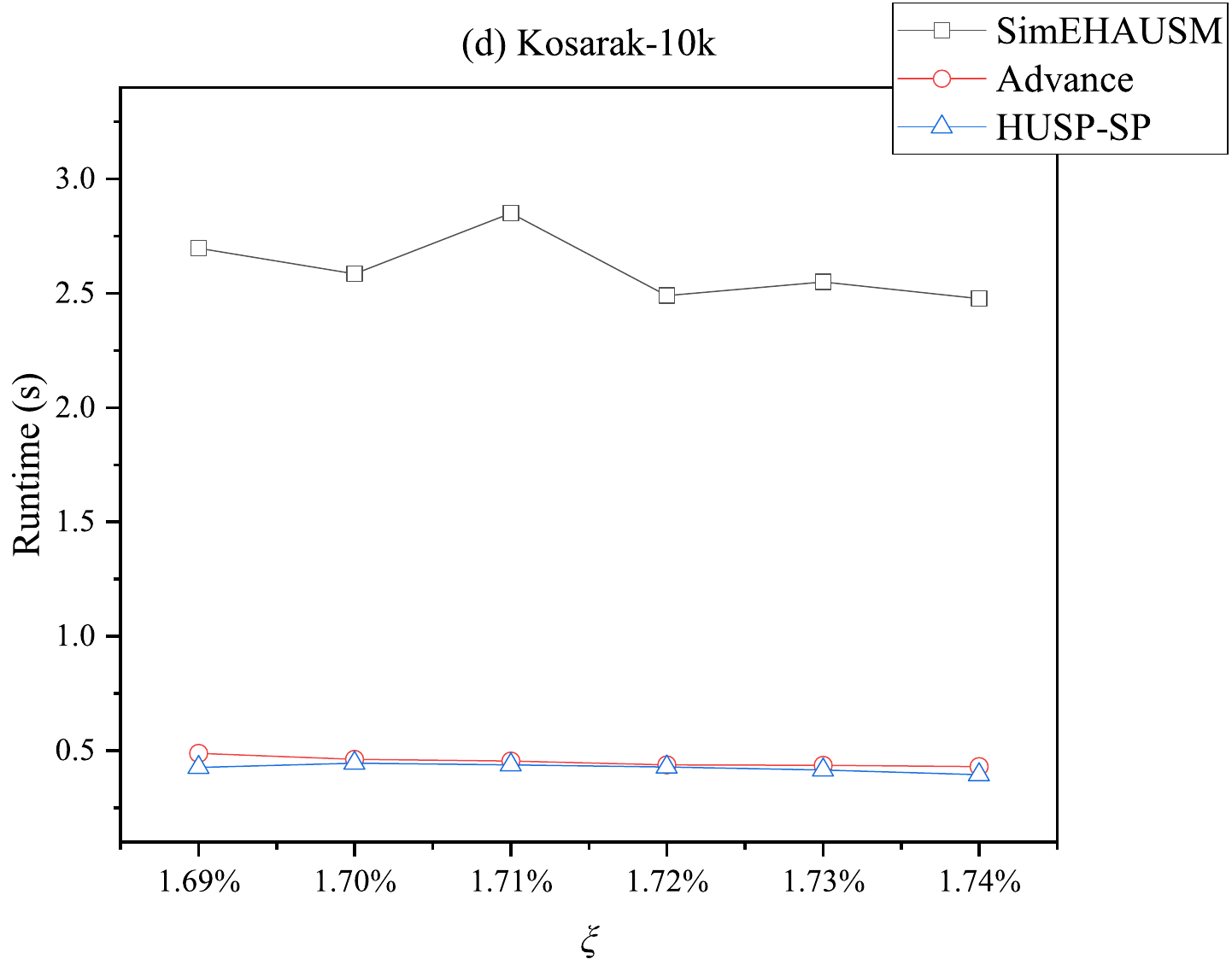}
			\label{fig: 06d}
			\includegraphics[width=0.31\linewidth]{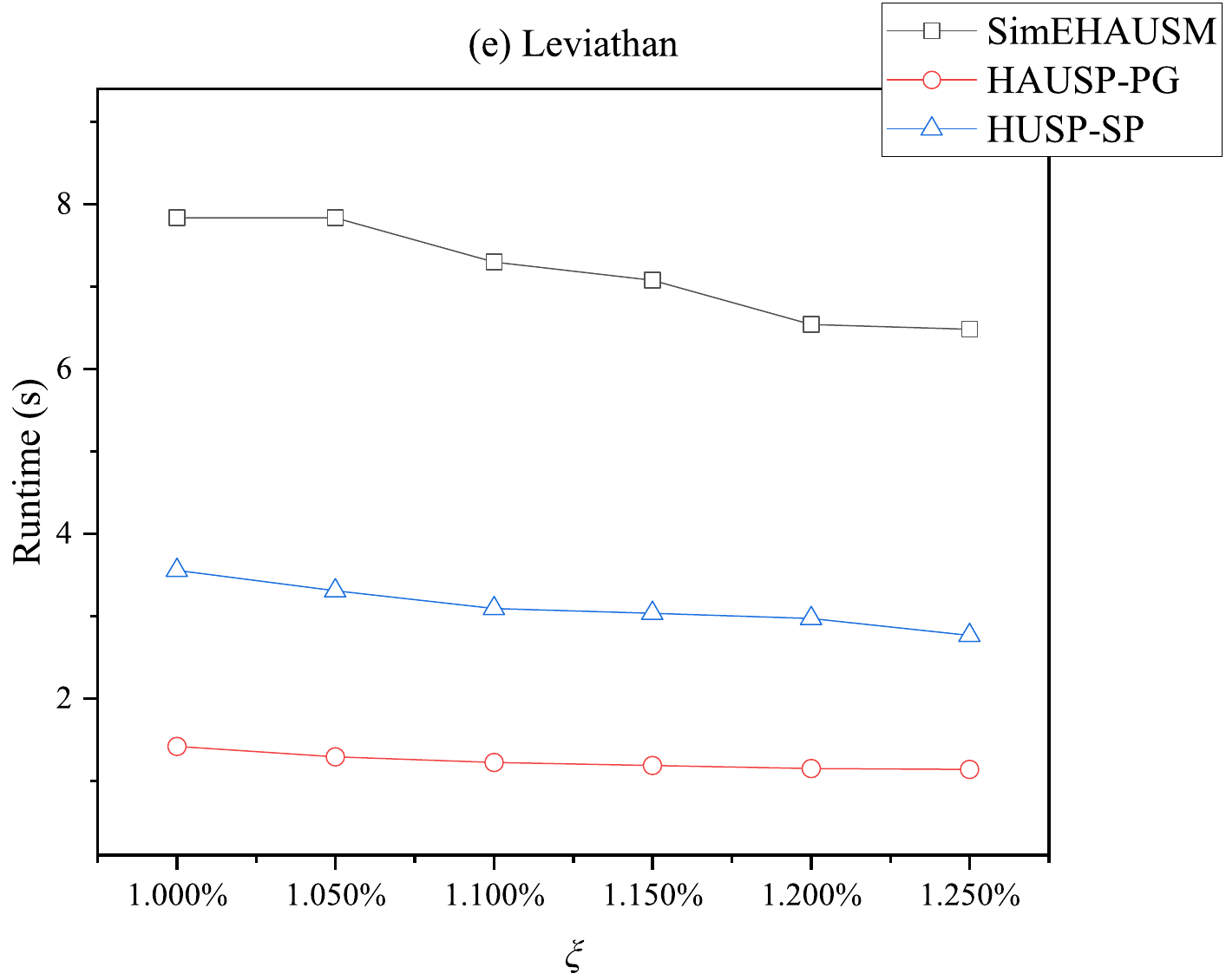}
			\label{fig: 06e}
			\includegraphics[width=0.31\linewidth]{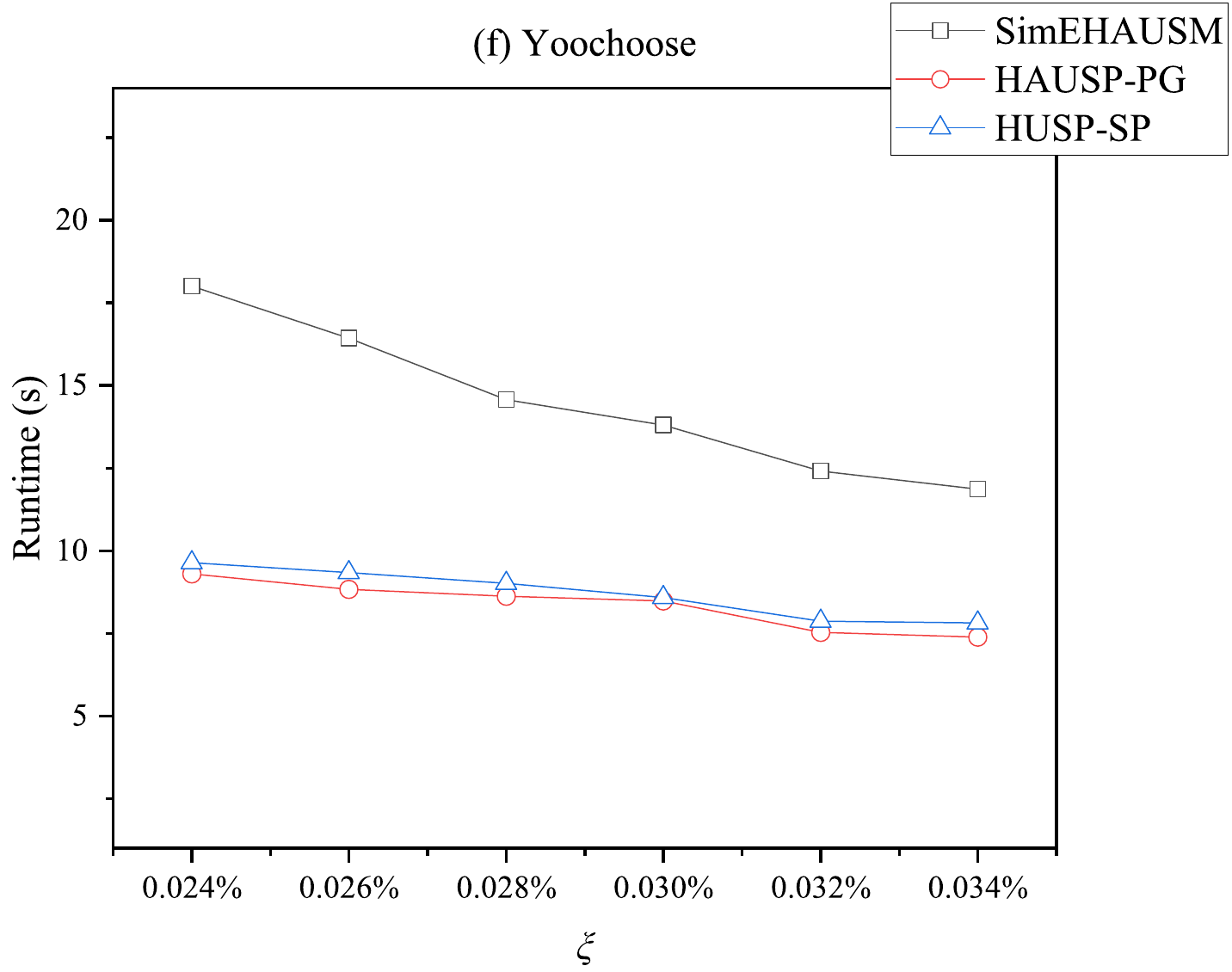}
			\label{fig: 06f}
	\end{minipage}
	\caption{Runtime for various minimum average utility thresholds.}
	\label{fig: 06}
\end{figure*}

Fig. \ref{fig: 06} is the variation in the runtime of the proposed algorithm with different values of parameter $\xi$ on each dataset. The overall runtime comprises the file access time and execution time of the algorithm. The minimum acceptable average utility increases with higher $\xi$ values. Consequently, fewer generated sequences meet the threshold, resulting in reduced overall execution times. Comparative results shown in Ref.\ cite{Ref46} demonstrate the superiority of HUSP-SP over state-of-the-art algorithms such as HUSP-ULL, ProUM, and USpan. Inspired by the algorithm HUSP-SP, our proposed HAUSP-PG, also employs the pattern-growth method to design pruning strategies. However, it is important to note that the average utility does not monotonically increase with pattern expansion. Utilizing this characteristic, we have adjusted the pruning strategy, resulting in a more efficient performance from HAUSP-PG. Indeed, the minimum acceptable total utility will increase progressively in HAUSPM owing to lengthening pattern sizes.

In Fig. \ref{fig: 06}(a), the runtime of HAUSP-PG even exceeds that of SimEHAUSM and HUSP-SP. The core reason lies in the approximately uniform distribution of the SIGN dataset: both the pruning strategies and upper-bound estimation models of HUSP-SP are designed for HAUSPM tasks. However, in a uniform distribution where utility values are concentrated and without significant long tails, its optimization logic fails to take effect, and even slows down performance due to redundant computations. As shown in Fig. \ref{fig: 06}(b), although HAUSP-PG, inspired by HUSP-SP, demonstrates improved advantages in datasets with long-tailed characteristics, in large-sample or highly unbalanced distribution scenarios, the added base overhead for adapting to long tails offsets part of the performance gains, as shown in Fig. \ref{fig: 06}(f), narrowing the efficiency improvement margin. Moreover, even when accounting for differences in tasks, the proposed algorithm does not require more runtime than SimEHAUSM on the experimental datasets.
 
\subsection{Number of Generated Sequences and Effectiveness Analysis}
\label{sec: effectiveness}

\begin{figure*}
	\centering
	\begin{minipage}{0.98\textwidth}
			\centering
			\includegraphics[width=0.31\linewidth]{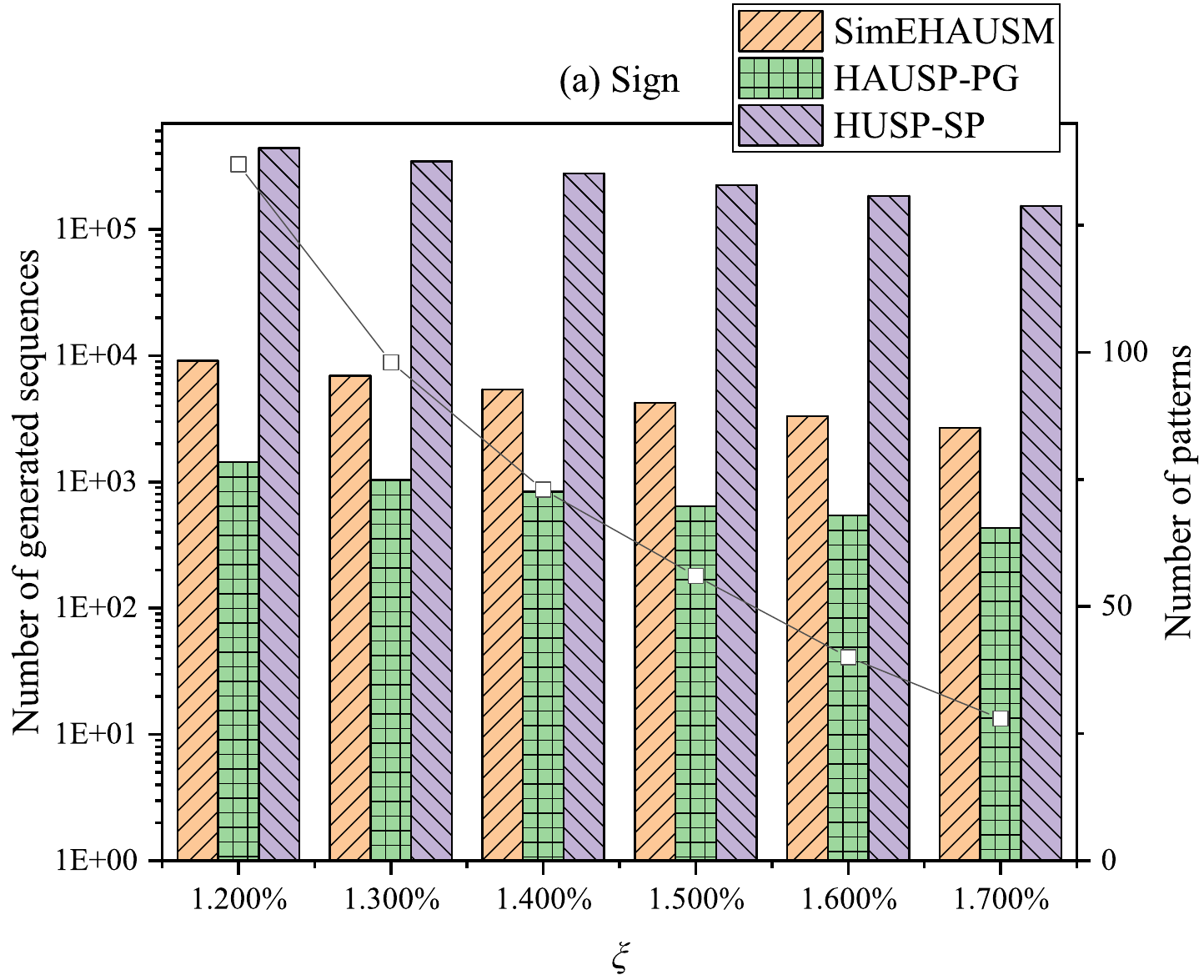}
			\label{fig: 07a}
			\includegraphics[width=0.31\linewidth]{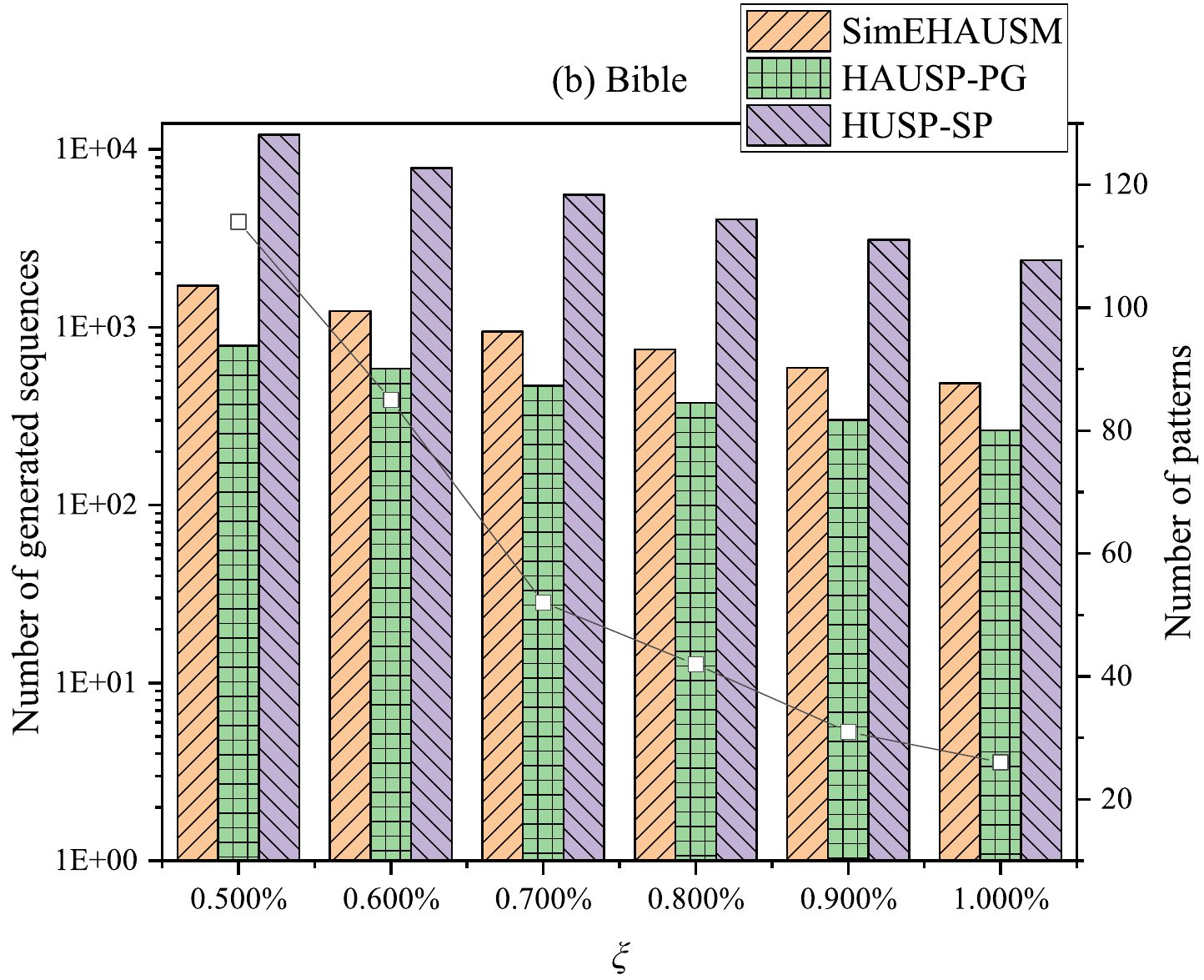}
			\label{fig: 07b}
			\includegraphics[width=0.31\linewidth]{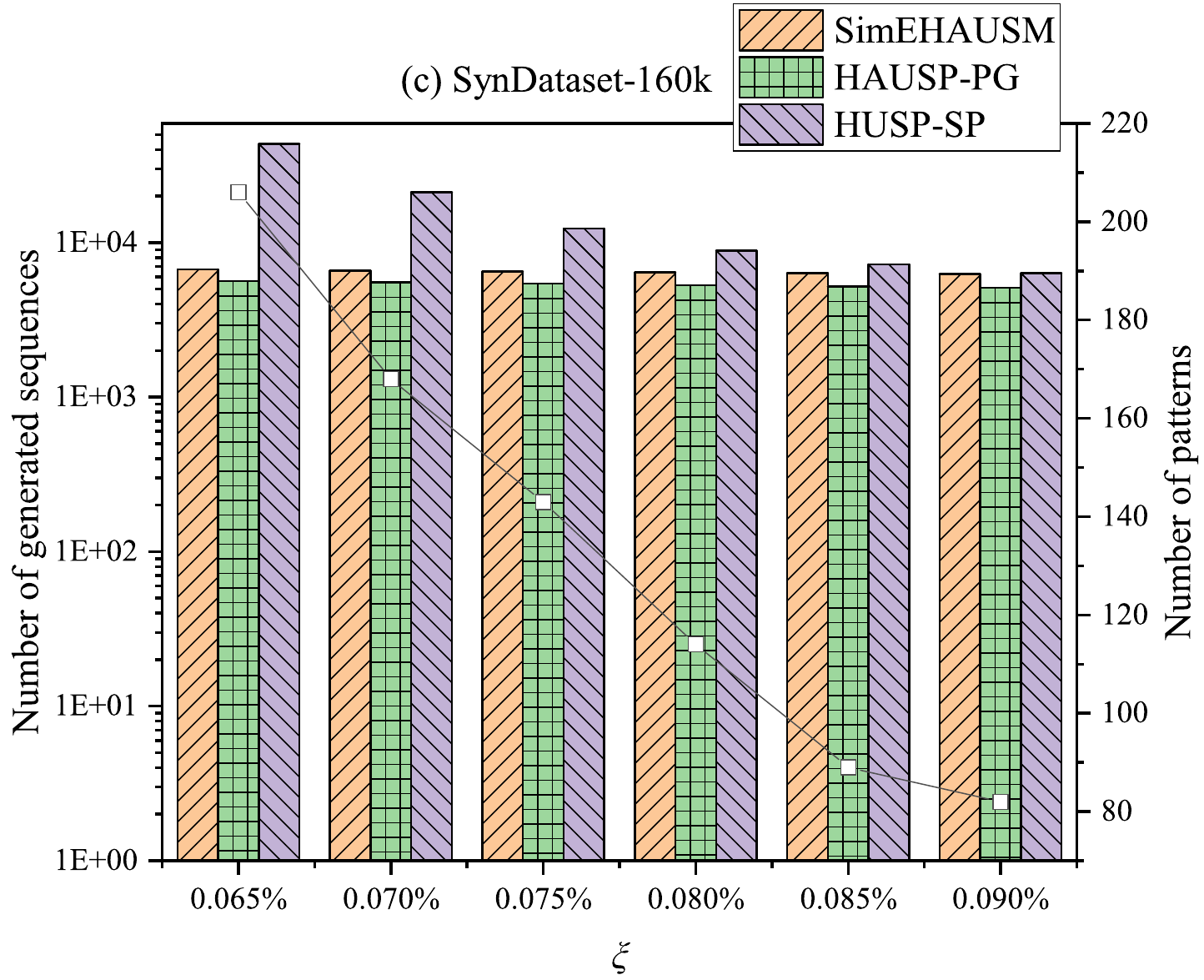}
			\label{fig: 07c}
	\end{minipage}
	\begin{minipage}{0.98\textwidth}
			\centering
			\includegraphics[width=0.31\linewidth]{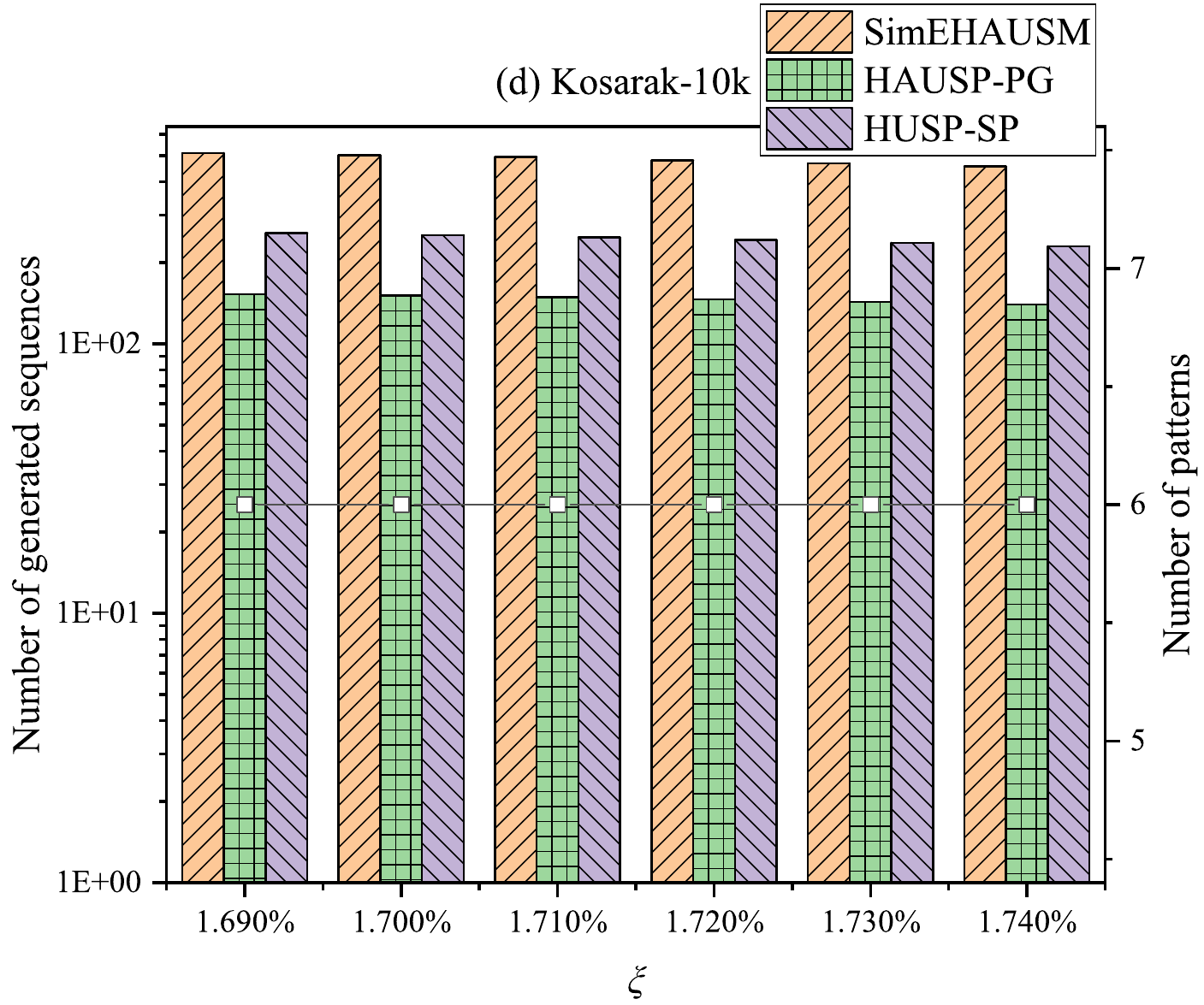}
			\label{fig: 07d}
			\includegraphics[width=0.31\linewidth]{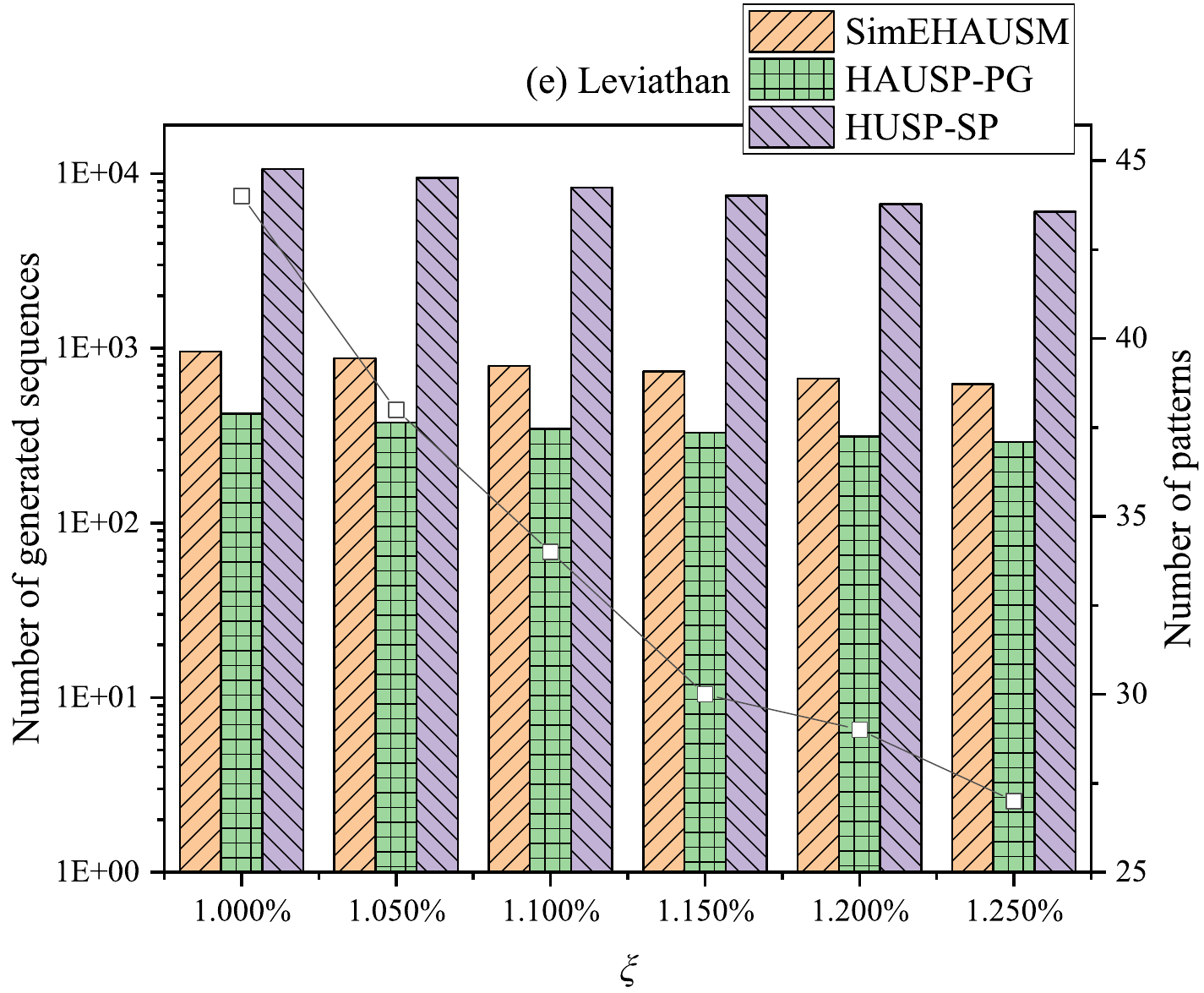}
			\label{fig: 07e}
			\includegraphics[width=0.31\linewidth]{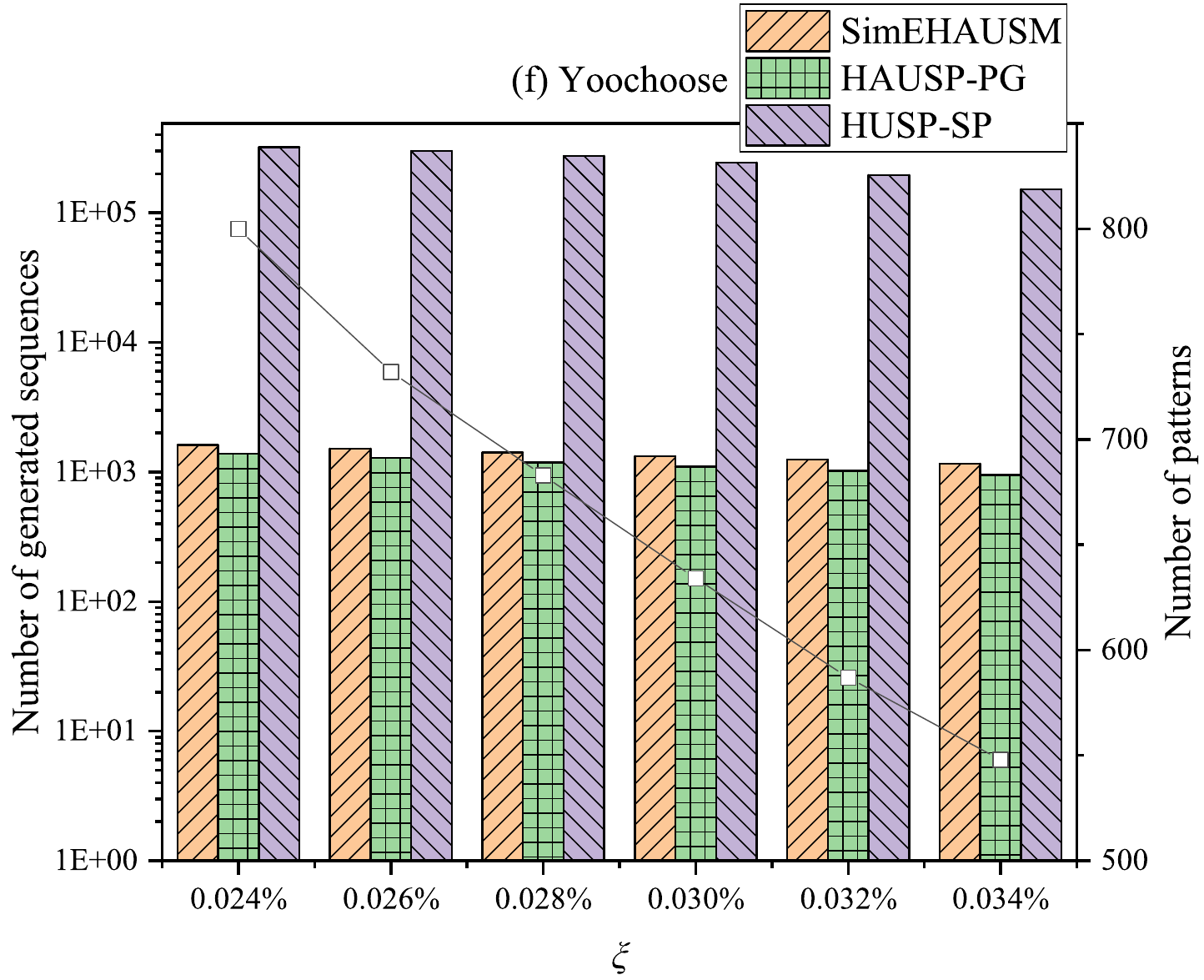}
			\label{fig: 07f}
	\end{minipage}
	\caption{Generated sequences for various minimum average utility thresholds.}
	\label{fig: 07}
\end{figure*}

The number of candidate sequences is used to measure the size of the search space. The experimental results in Ref. \cite{Ref46} show that HUSP-SP, which adopts the IIP strategy \cite{Ref22}, generates fewer candidates than other methods. Therefore, in the proposed algorithm, this strategy is used to remove irrelevant items. Additionally, unpromising items are eliminated based on the variant of the upper bound \textit{PEAU}, ${\textit{VPEAU}}_{\textit{Adv}}$, and its associated pruning strategy. Thus, through this improved method, HAUSP-PG can avoid generating a large number of candidate sequences, thereby reducing the search space. In Fig. \ref{fig: 07}(a), due to the uniform distribution of utility values in Sign dataset, any increase in length may lead to significant changes in pattern utility; however, this uniform distribution has relatively little impact on the HAUPM task. 

Furthermore, long sequences in the dataset result in the generation of more intermediate sequences. For instance, in Fig. \ref{fig: 07}(c), all three algorithms generate a relatively large number of intermediate sequences. Nevertheless, for algorithms that consider the average utility of pattern lengths, when the utility values in the dataset are more dispersed (i.e., with a high CV), methods oriented toward the HAUSPM task can better control the number of generated sequences. Hence, Fig. \ref{fig: 07}(f) shows that HAUSP-PG and SimEHAUSM generate fewer sequences.

Except for Fig. \ref{fig: 07}(d), on all other selected datasets, the two algorithms designed for the HAUSPM task generate fewer intermediate sequences than those designed for the HUSPM task. This is because HUSP-SP retains a large number of intermediate sequences with "high total utility but low average utility". However, as can be seen from Fig. \ref{fig: 07}(d), the difference in quantity is relatively small. A probable reason is that although the Kosarak dataset has a strong long-tail characteristic, its average utility value is low, and the dataset size is small, so the impact of the uneven distribution is limited. Nevertheless, because the SimEHAUSM algorithm's strategies and models are more sensitive to the maximum utility items in sequences, it generates a relatively larger number of sequences.

Moreover, by combining Fig. \ref{fig: 06}, Fig. \ref{fig: 07}, and Fig. \ref{fig: 08}, it can be observed that the runtime and memory usage of the algorithms do not solely depend on the number of generated candidate sequences. Both HAUSP-PG and SimEHAUSM require additional computational effort to process generated sequences. The smaller the difference in the number of generated sequences, the more noticeable the additional time and memory consumption caused by this extra computation, as shown in Fig. \ref{fig: 06}(e), Fig. \ref{fig: 06}(f) and Fig. \ref{fig: 08}(e), Fig. \ref{fig: 08}(f).

\subsection{Memory Overhead and Usage Evaluation}
\label{sec: memory}

\begin{figure*}
	\centering
	\begin{minipage}{0.98\textwidth}
			\centering
			\includegraphics[width=0.31\linewidth]{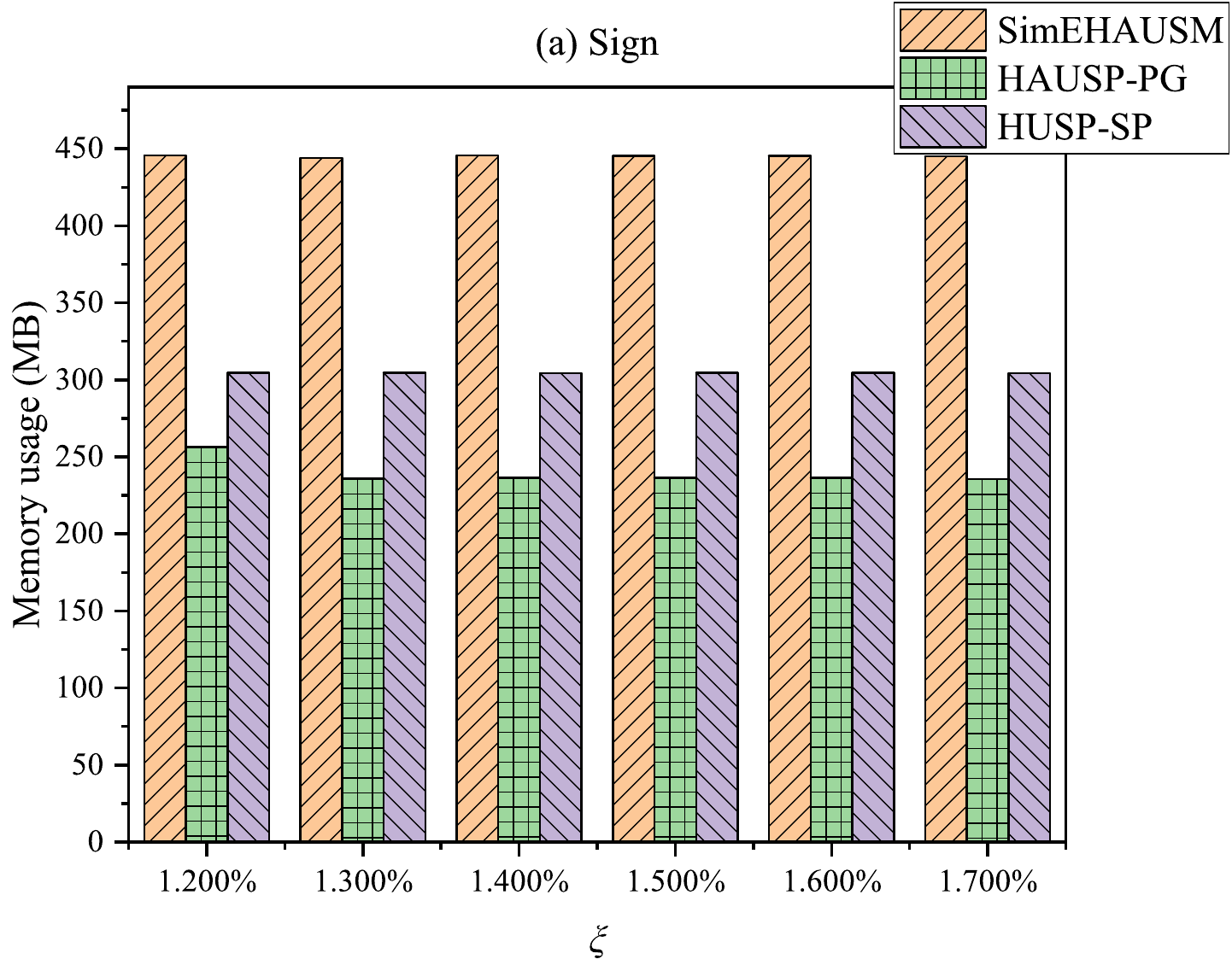}
			\label{fig: 08a}
			\includegraphics[width=0.31\linewidth]{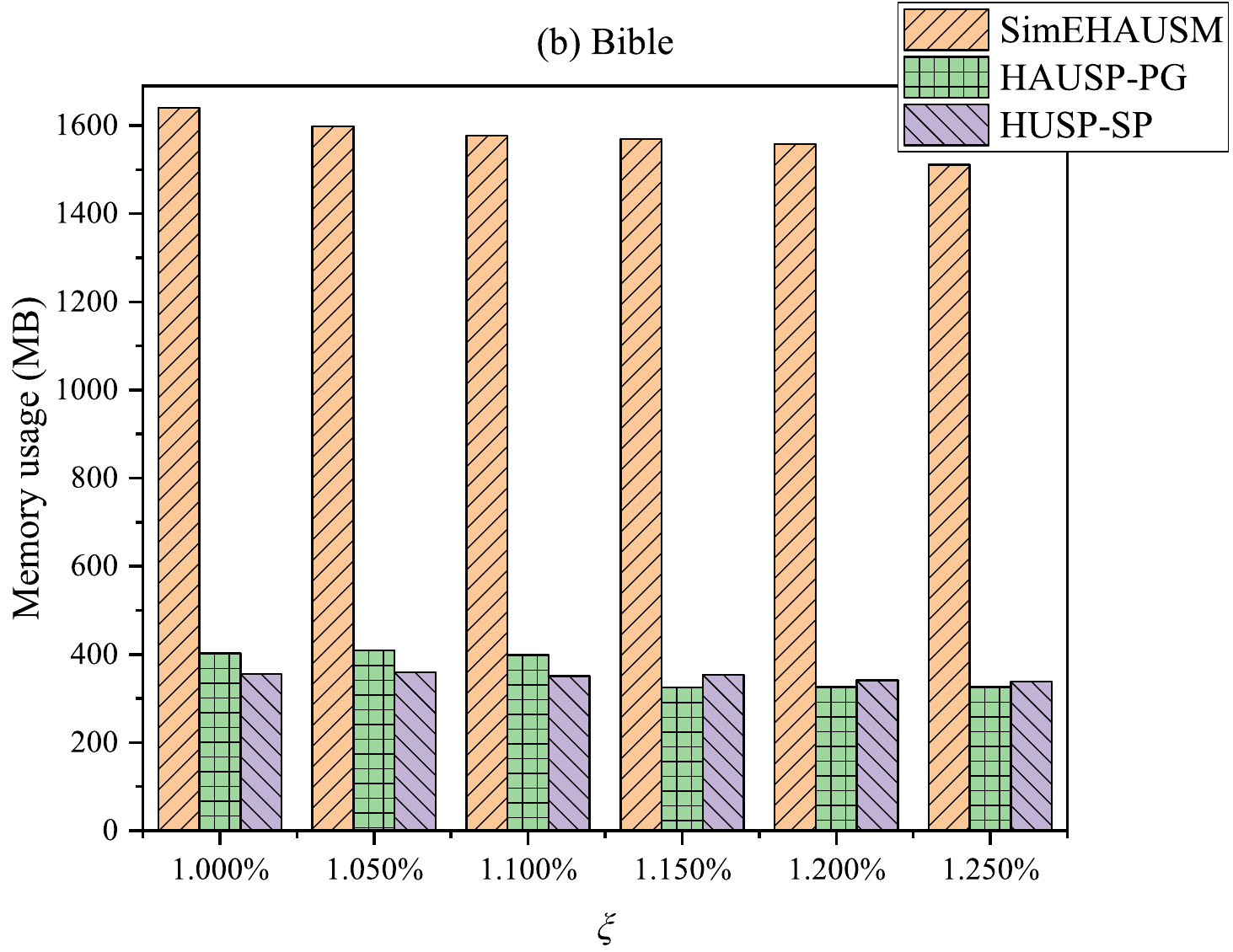}
			\label{fig: 08b}
			\includegraphics[width=0.31\linewidth]{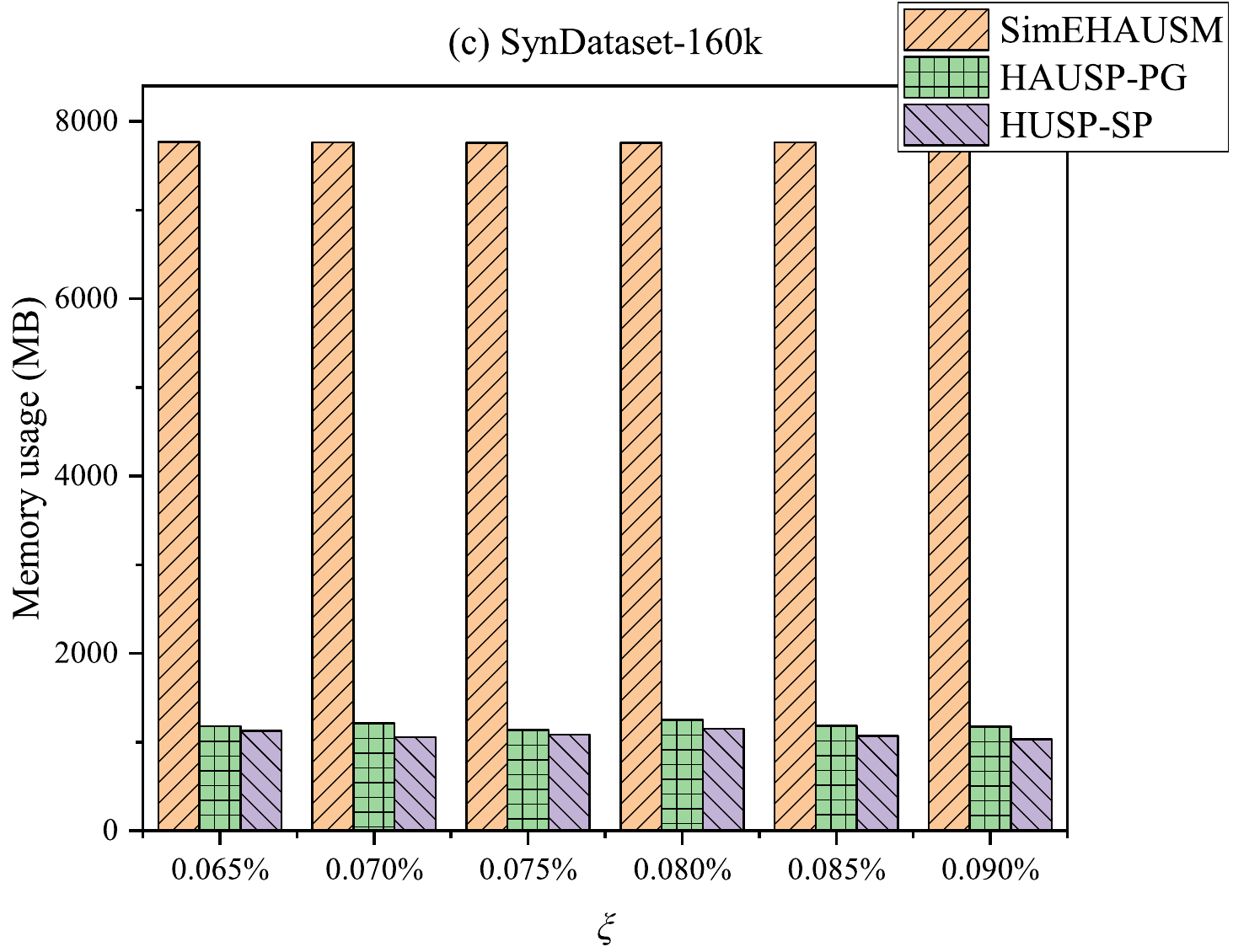}
			\label{fig: 08c}
	\end{minipage}
	\begin{minipage}{0.98\textwidth}
			\centering
			\includegraphics[width=0.31\linewidth]{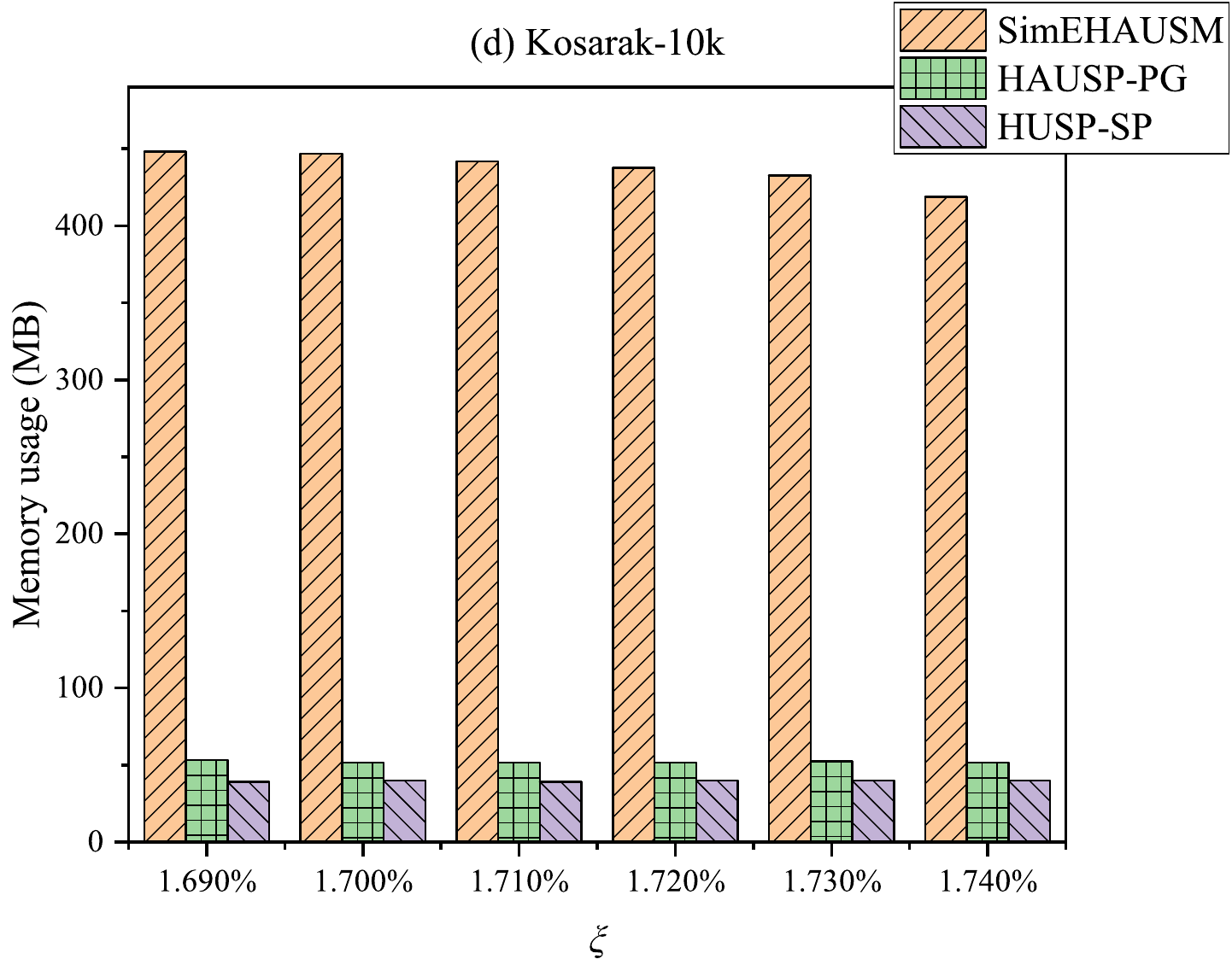}
			\label{fig: 08d}
			\includegraphics[width=0.31\linewidth]{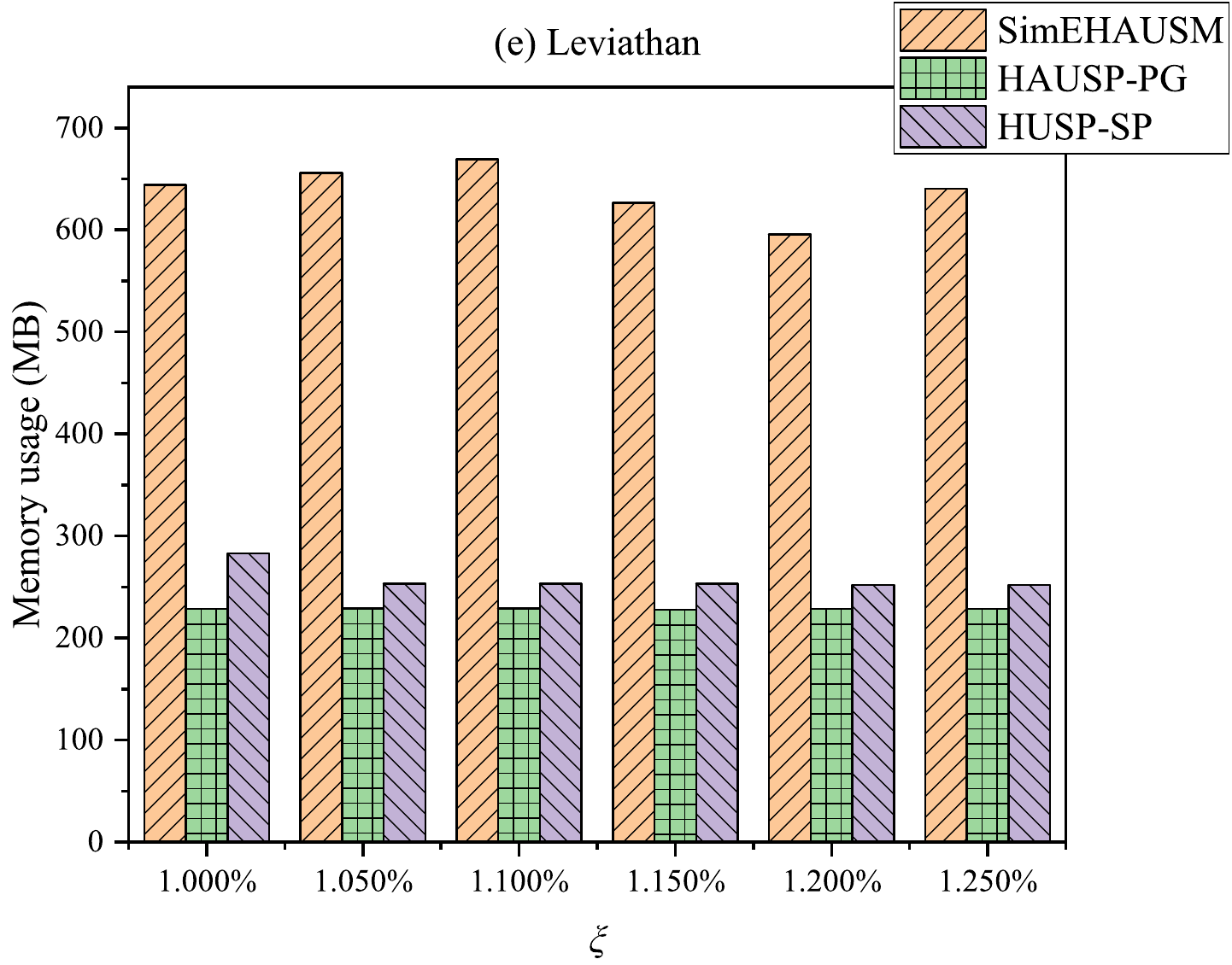}
			\label{fig: 08e}
			\centering
			\includegraphics[width=0.31\linewidth]{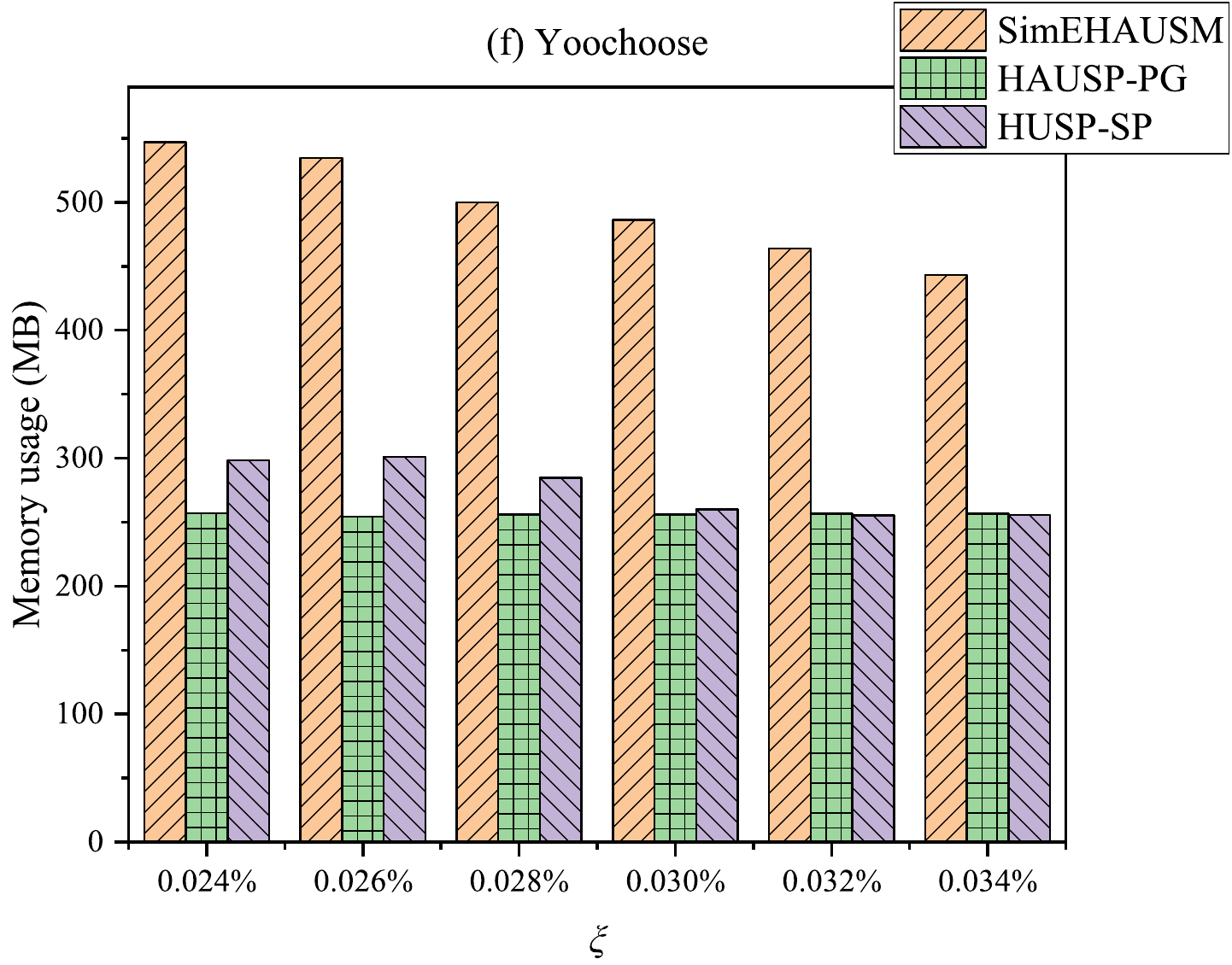}
			\label{fig: 08f}
	\end{minipage}
	\caption{Memory usage for various minimum average utility thresholds.}
	\label{fig: 08}
\end{figure*}

The HUSP-SP adopted a compact and effective projected structure, and the proposed algorithm for HAUSPM does likewise. Fig. \ref{fig: 08} clearly illustrates that a greater number of generated sequences necessitate higher memory usage.

Fig. \ref{fig: 08} reveals a stark contrast, in terms of memory usage with the algorithm HAUSP-PG. Despite the potential unfairness of comparing it with HUSPM and HAUSPM, Fig. \ref{fig: 08}(c) and \ref{fig: 08}(d) show similar memory usage between HUSP-SP and HAUSP-PG. The former requires more memory to handle more generated sequences, while the latter incurs overhead due to extra computations. In Ref. \cite{Ref63}, HUSP-SP has a better space performance due to its compact data structure and efficient strategies. Fig. \ref{fig: 08} demonstrates that HAUSP-PG requires lower memory usage than HUSP-SP in most cases, signifying meaningful enhancement. Although both of them adopt similar upper bounds and strategies to avoid generating more sequences, the proposed algorithm optimizes memory usage more effectively.

The memory consumption of HAUSP-PG is lower than SimEHAUSM, despite generating more generated sequences in some cases. In most HAUSPM algorithms, descending-sorted itemsets are typically required for UBs computation. In contrast to the filtering operations used in our proposed method, the memory complexity of such sorting operations is higher, regardless of whether the database is transaction or quantitative sequential databases. Furthermore, to enable fast updates and computations, the complex data structures used have also increased the memory usage of the algorithm.

\subsection{Strategy Performance and Efficiency Analysis}
\label{sec: strategyEfficiency}

\begin{figure*}
	\centering
	\begin{minipage}{0.98\textwidth}
			\centering
			\includegraphics[width=0.31\linewidth]{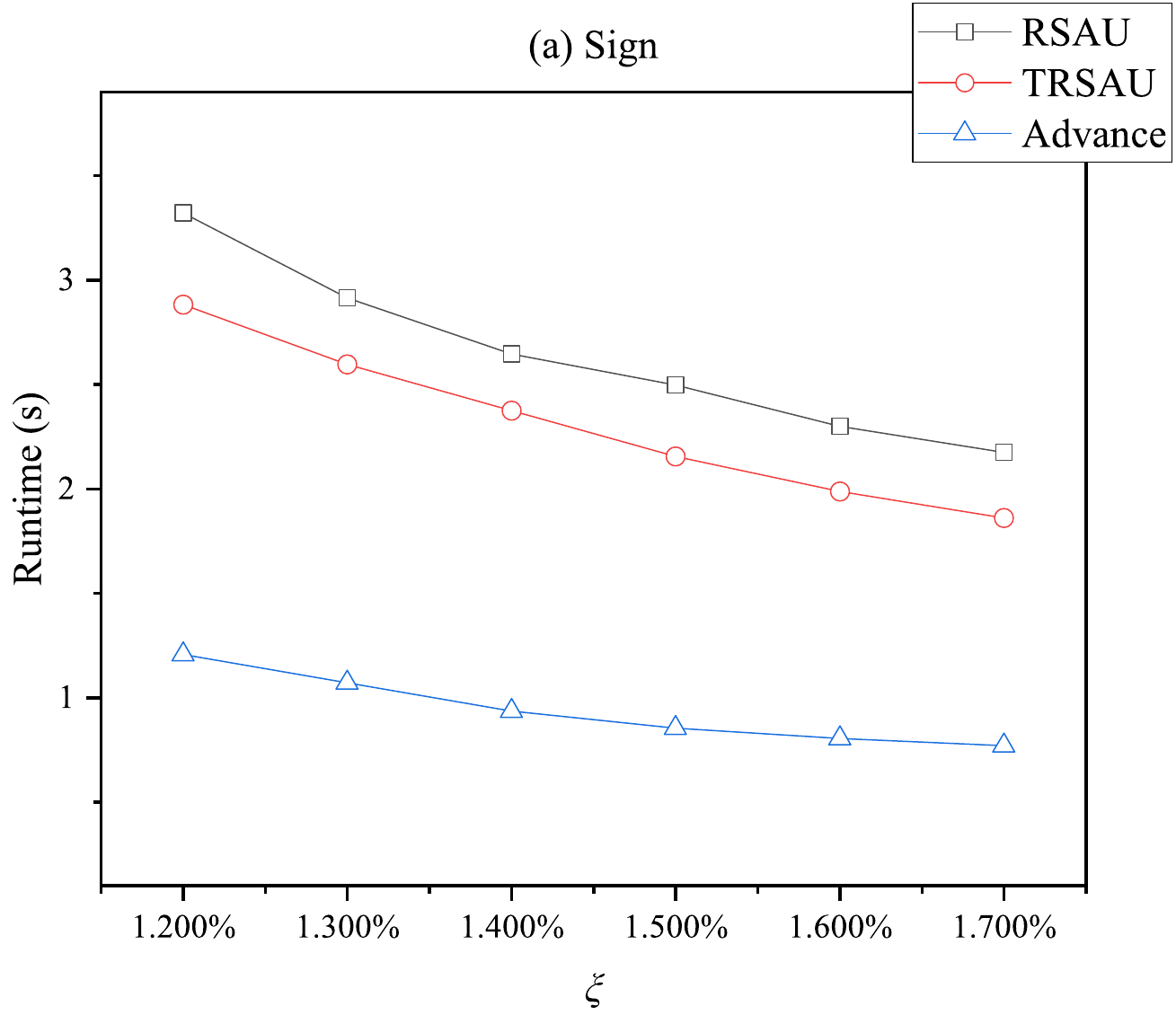}
			\label{fig: 09a}
			\includegraphics[width=0.31\linewidth]{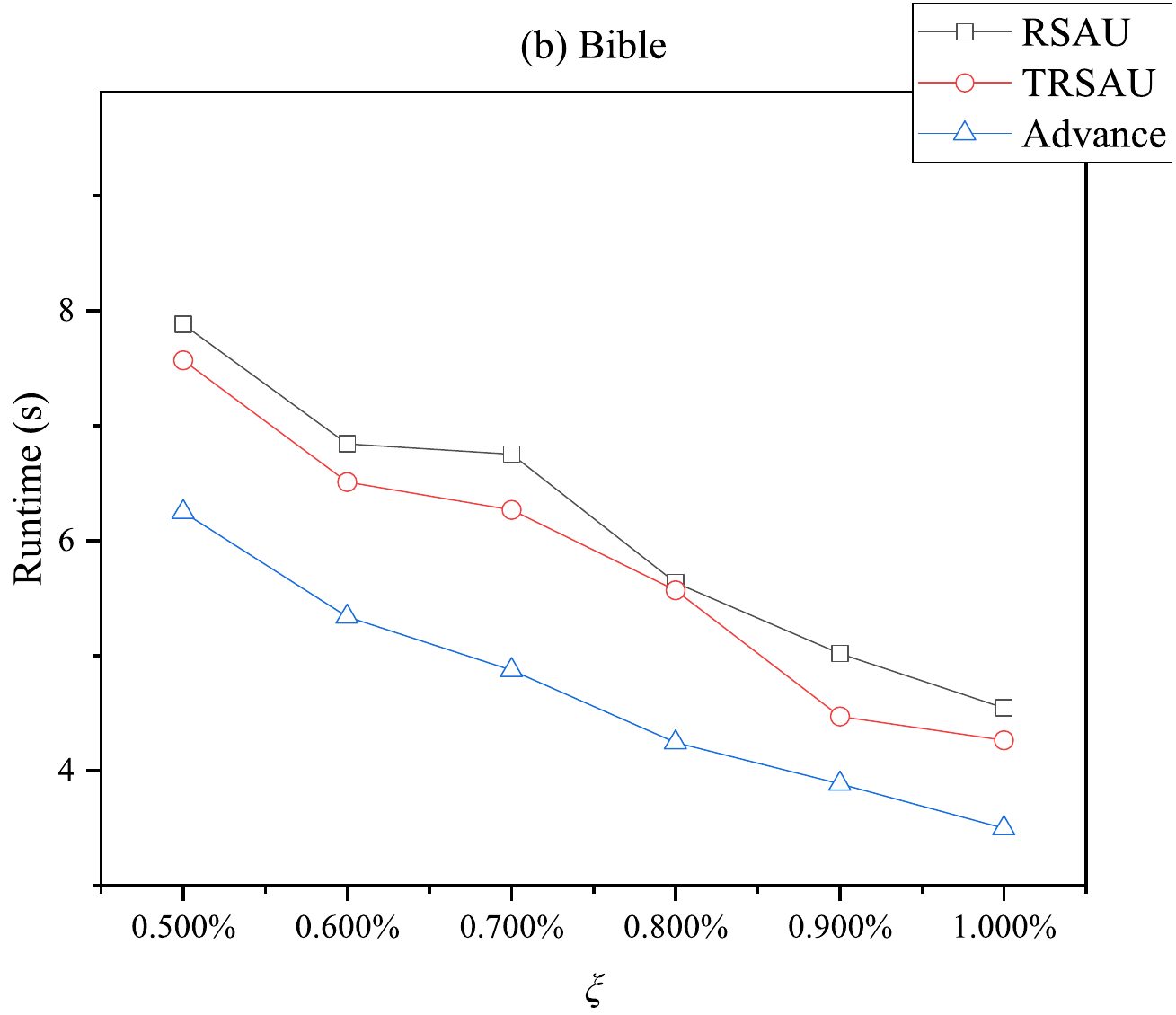}
			\label{fig: 09b}
			\includegraphics[width=0.31\linewidth]{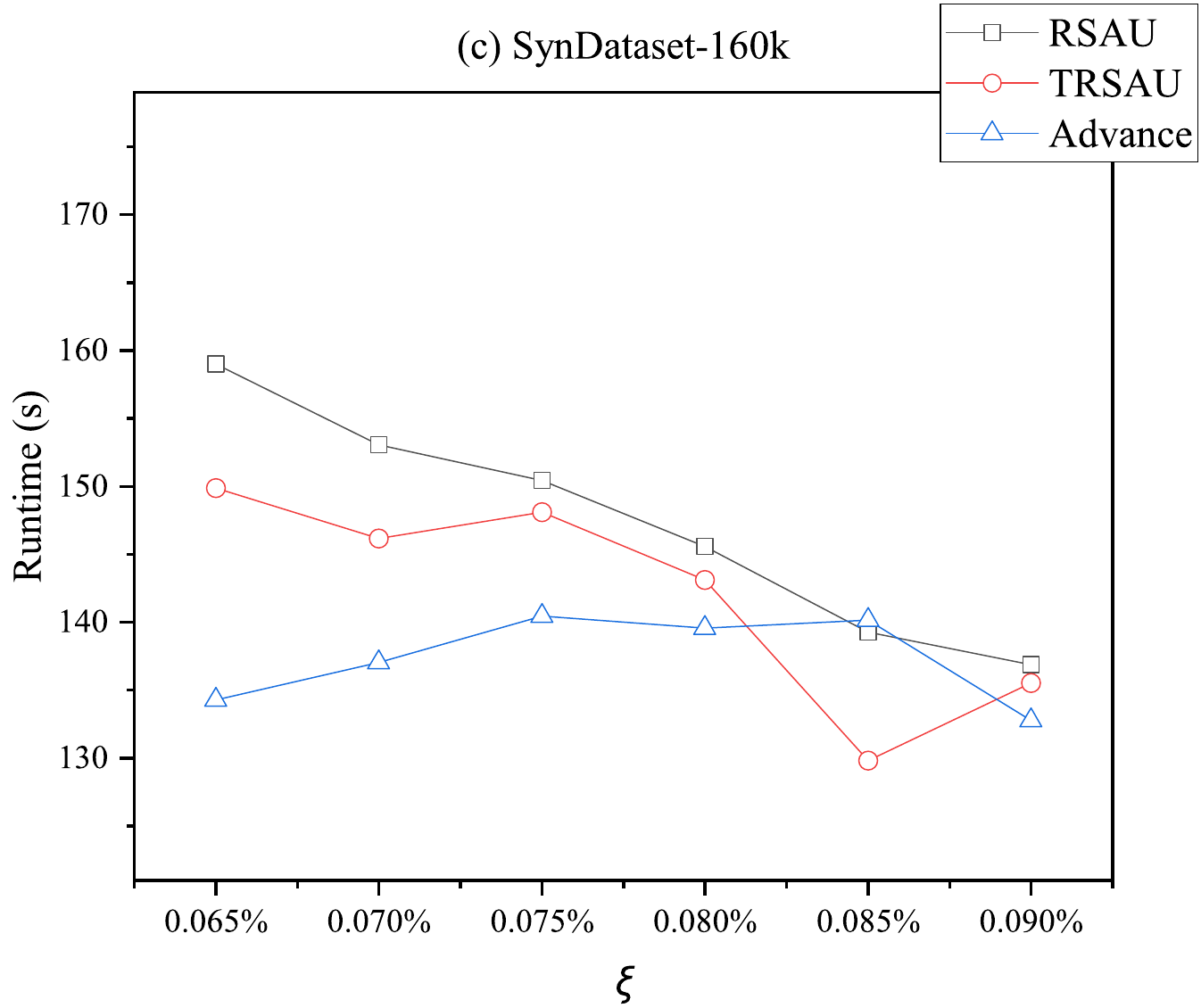}
			\label{fig: 09c}
	\end{minipage}
	\begin{minipage}{0.98\textwidth}
			\centering
			\includegraphics[width=0.31\linewidth]{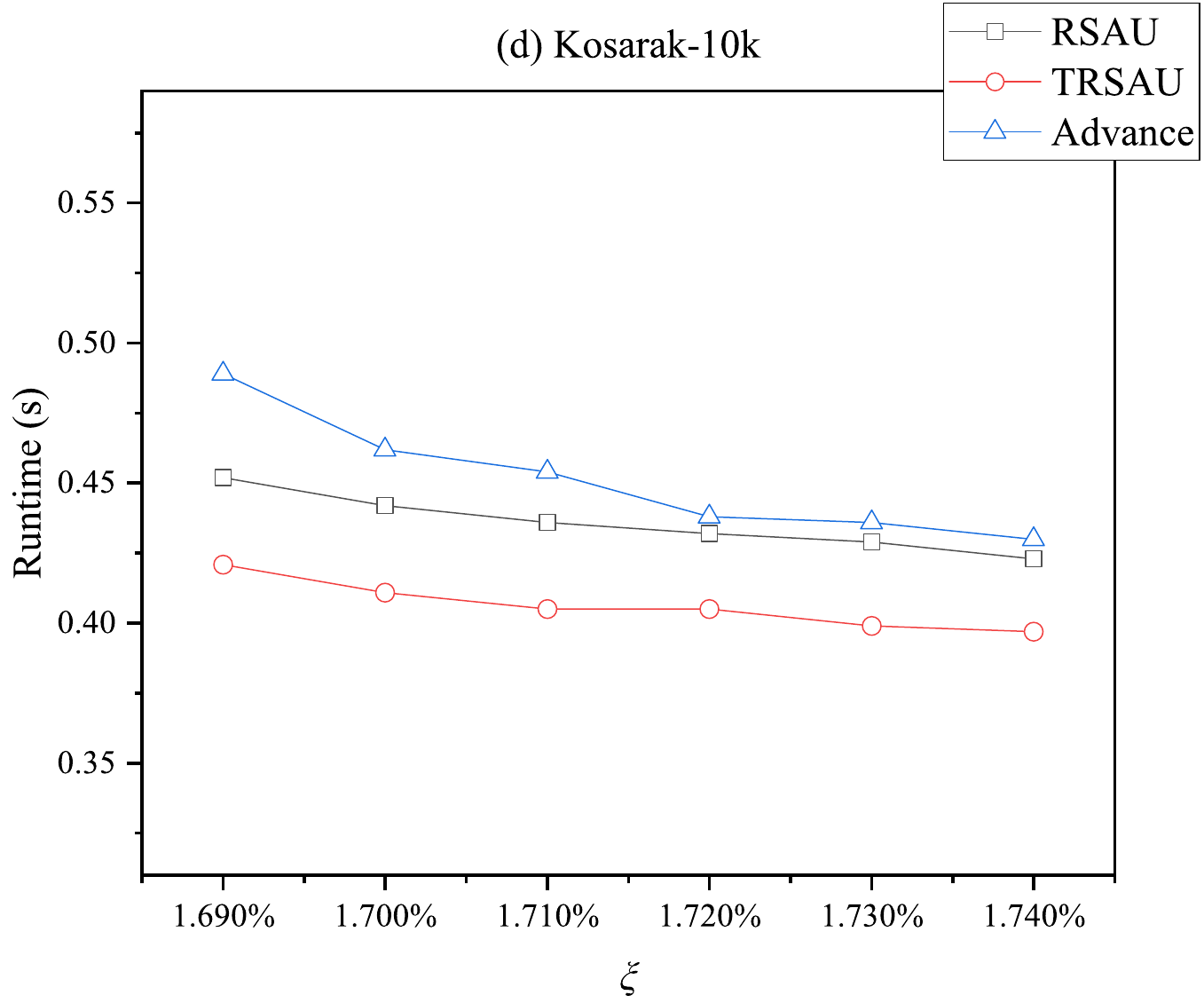}
			\label{fig: 09d}
			\includegraphics[width=0.31\linewidth]{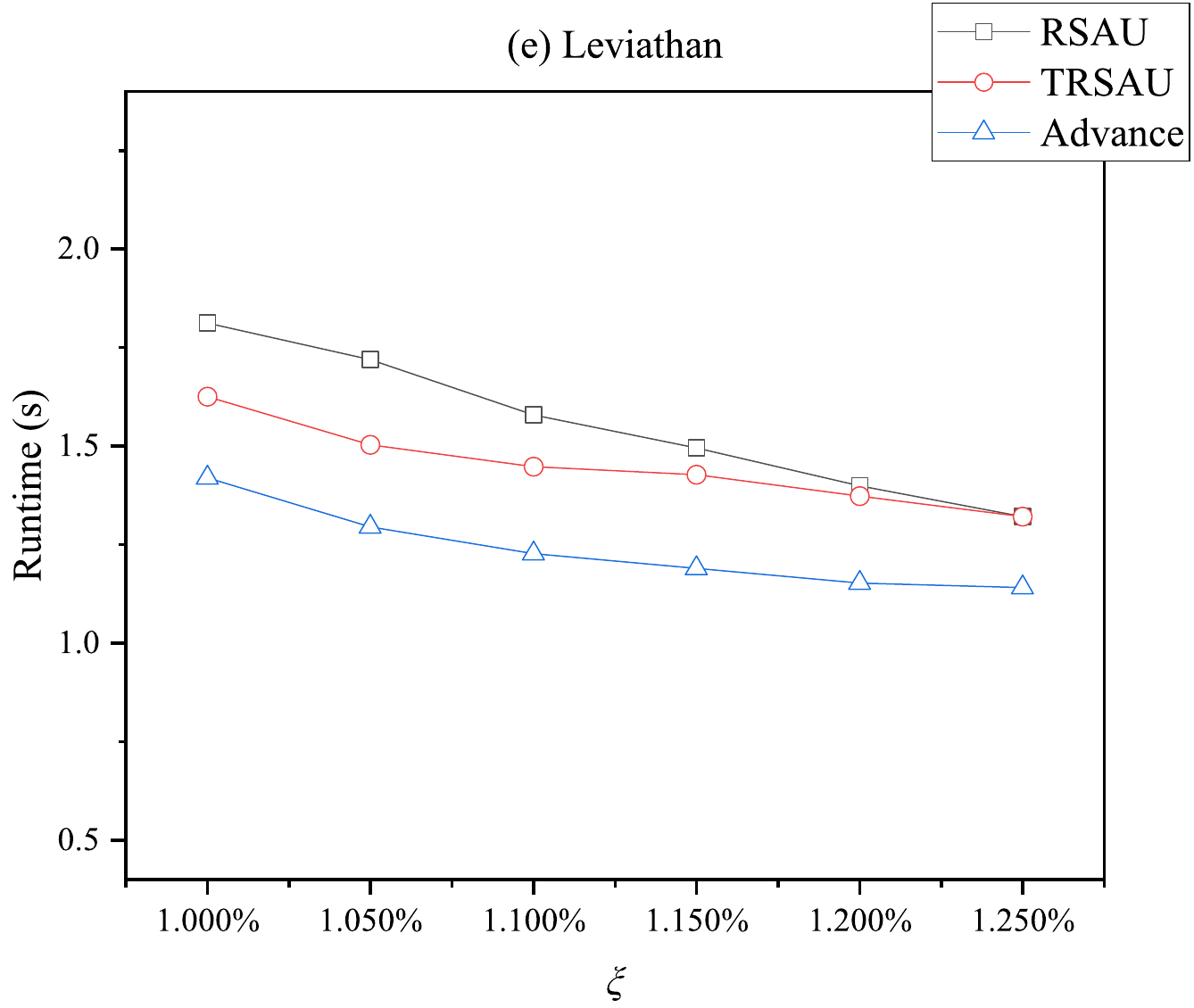}
			\label{fig: 09e}
			\includegraphics[width=0.31\linewidth]{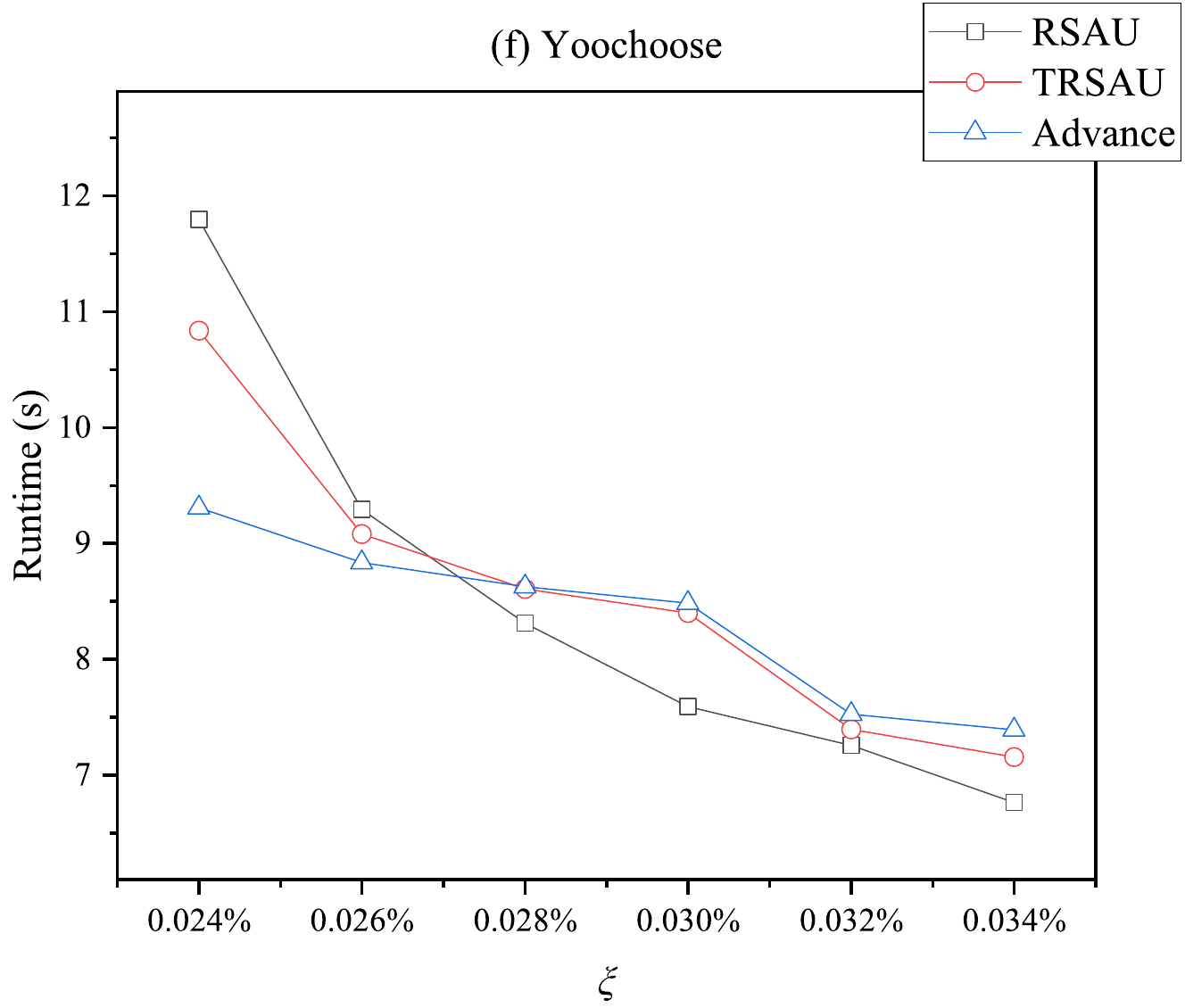}
			\label{fig: 09f}
	\end{minipage}
	\caption{Runtime for various versions of upper bounds.}
	\label{fig: 09}
\end{figure*}

The comparative results of runtime are shown in Fig. \ref{fig: 09}. RSAU denotes a version of the algorithm that combines ${\textit{PEAU}}_{\textit{Ori}}$ and \textit{RSAU} while adopting the corresponding pruning strategies. TRSAU refers to the version where \textit{TRSAU} is used instead of \textit{RSAU} as the upper bound. Advance denotes a version of the algorithm with two upper bounds: ${\textit{VPEAU}}_{\textit{Adv}}$ and ${\textit{VTRSAU}}_{\textit{Adv}}$. The results demonstrate that TRSAU is faster than RSAU. This is primarily because the strategy corresponding to \textit{TRSAU} removes irrelevant items from the set of appending items, thereby reducing execution time. Obviously, ${\textit{VPEAU}}_{\textit{Adv}}$ is not a strict upper bound; however, it remains tighter than other versions of upper bounds. These findings convincingly confirm the effectiveness of the proposed upper bounds and their variants in HAUSPM. Tighter UBs and their variants, along with associated pruning strategies, can accelerate the mining process of HAUSPs.

However, it can be observed from Fig. \ref{fig: 09}(d) that Advance is not highly efficient. This is mainly due to the characteristics of the Kosarak dataset. As observed in Section \ref{sec: efficiency}, length-oriented designs fail to show obvious advantages in this dataset; instead, the small dataset size and high data dispersion lead to relatively higher base consumption during processing, resulting in poor efficiency. In Fig. \ref{fig: 09}(f), the validity of this analysis is verified. In large-scale non-uniformly distributed data, a larger mean, higher CV, and significant positive skewness all indicate that the distribution has a significant impact on pattern utility in this dataset. As the threshold decreases, the number of generated patterns that need to be considered increases, and the advantages brought by the specialized designs will become increasingly evident.

\subsection{Strategy Performance and Effectiveness Analysis}
\label{sec: strategyEffectiveness}

\begin{figure*}
	\centering
	\begin{minipage}{0.98\textwidth}
			\centering
			\includegraphics[width=0.31\linewidth]{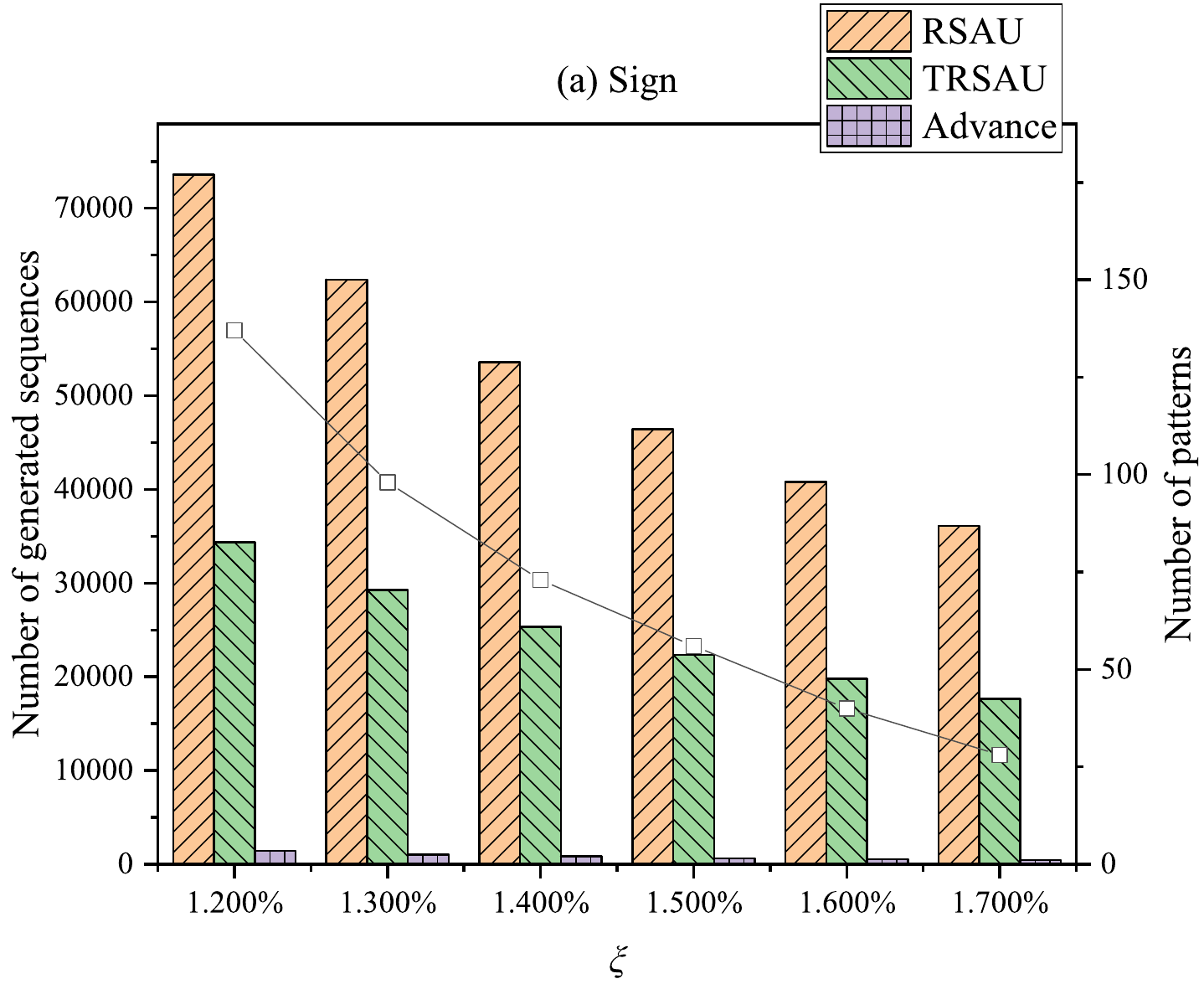}
			\label{fig: 10a}
			\includegraphics[width=0.31\linewidth]{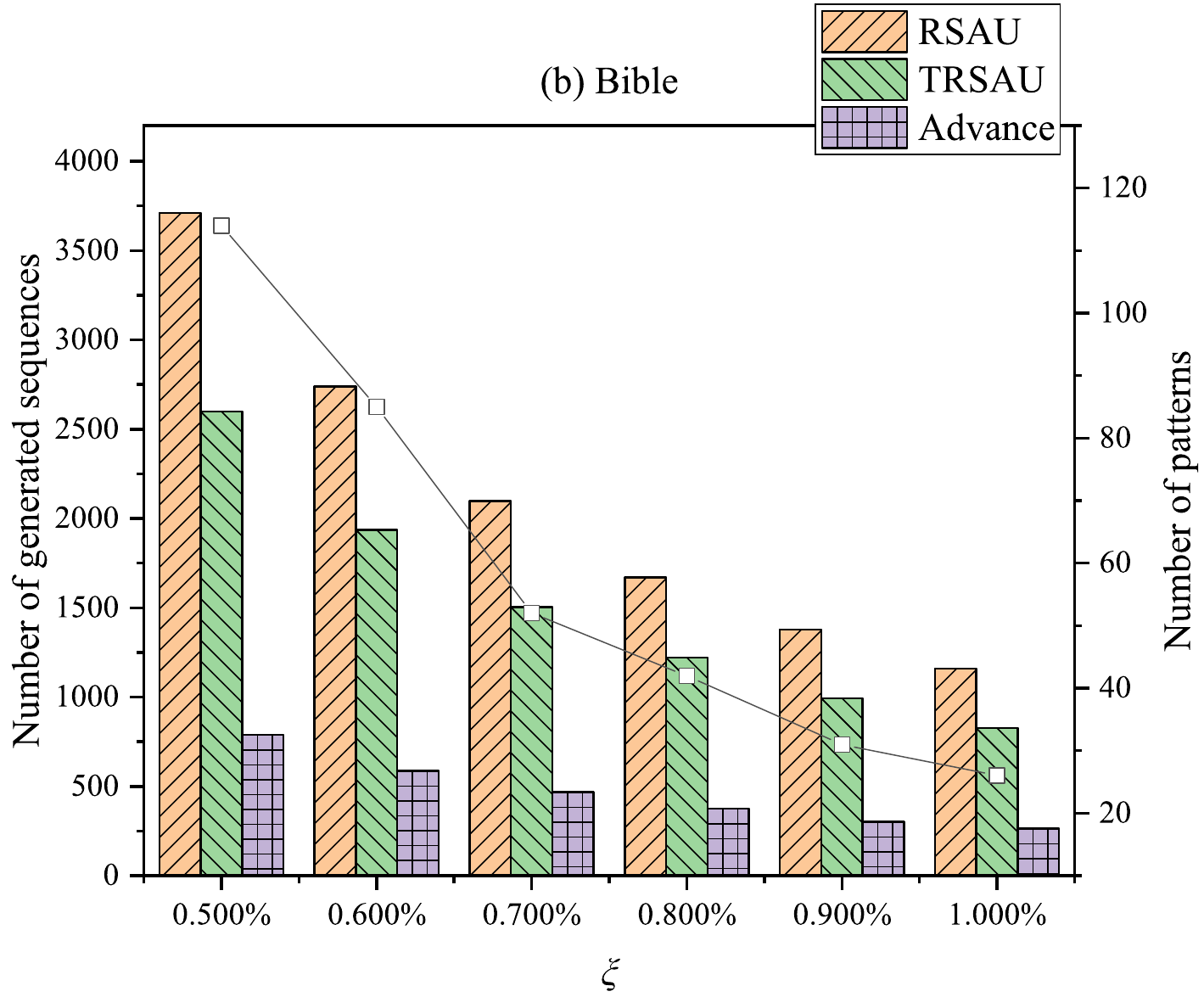}
			\label{fig: 10b}
			\includegraphics[width=0.31\linewidth]{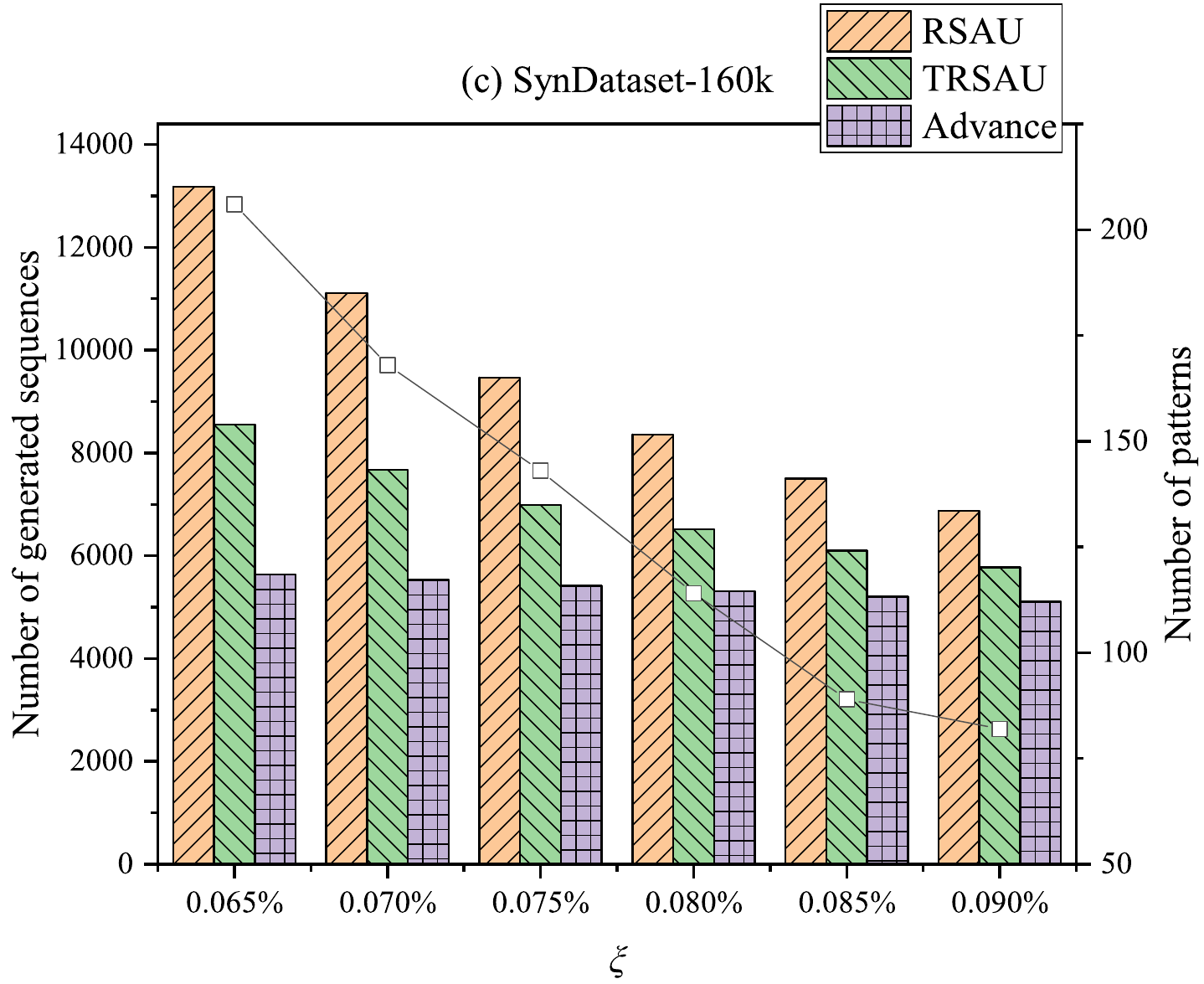}
			\label{fig: 10c}
	\end{minipage}
	\begin{minipage}{0.98\textwidth}
			\centering
			\includegraphics[width=0.31\linewidth]{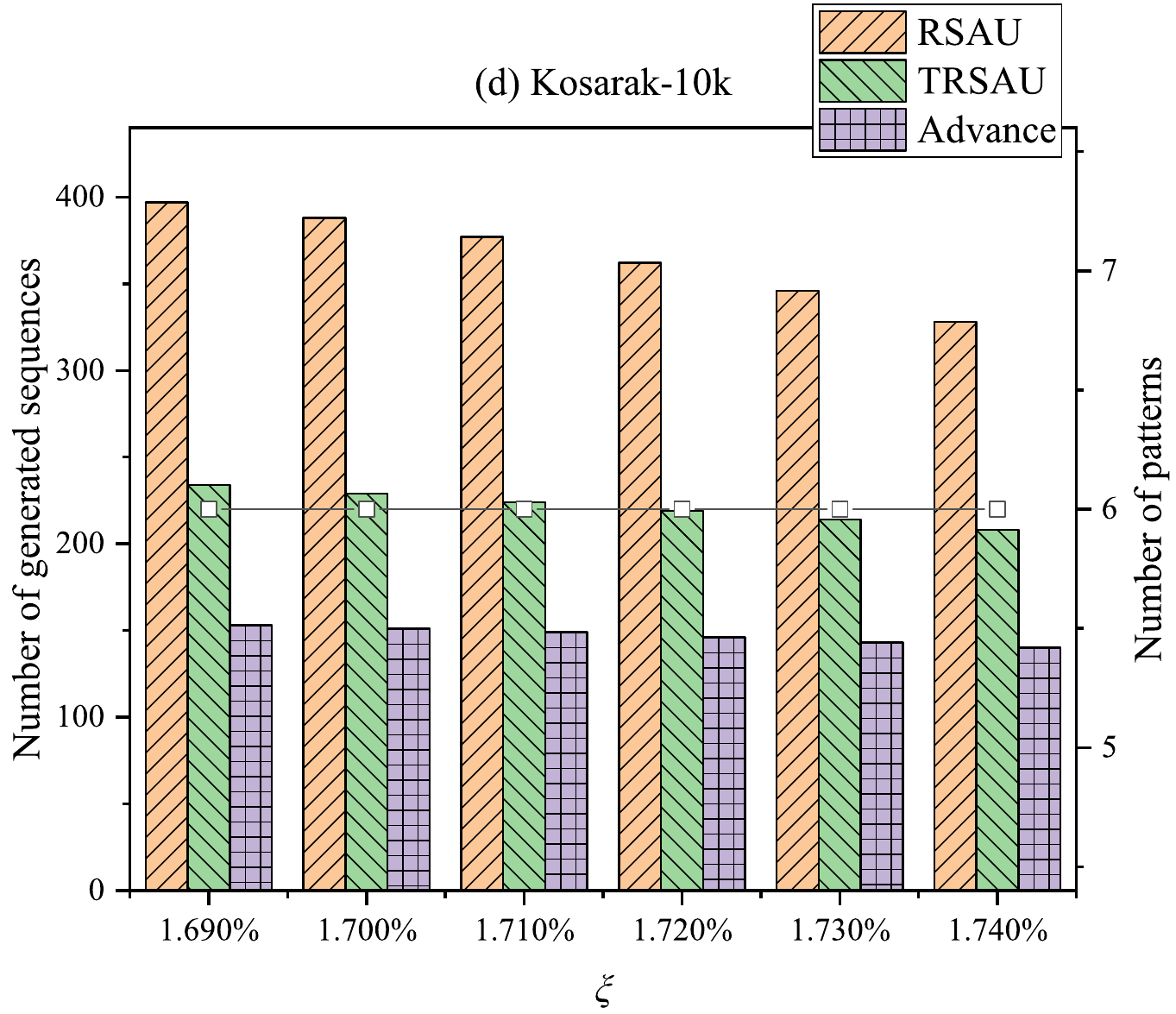}
			\label{fig: 10d}
			\includegraphics[width=0.31\linewidth]{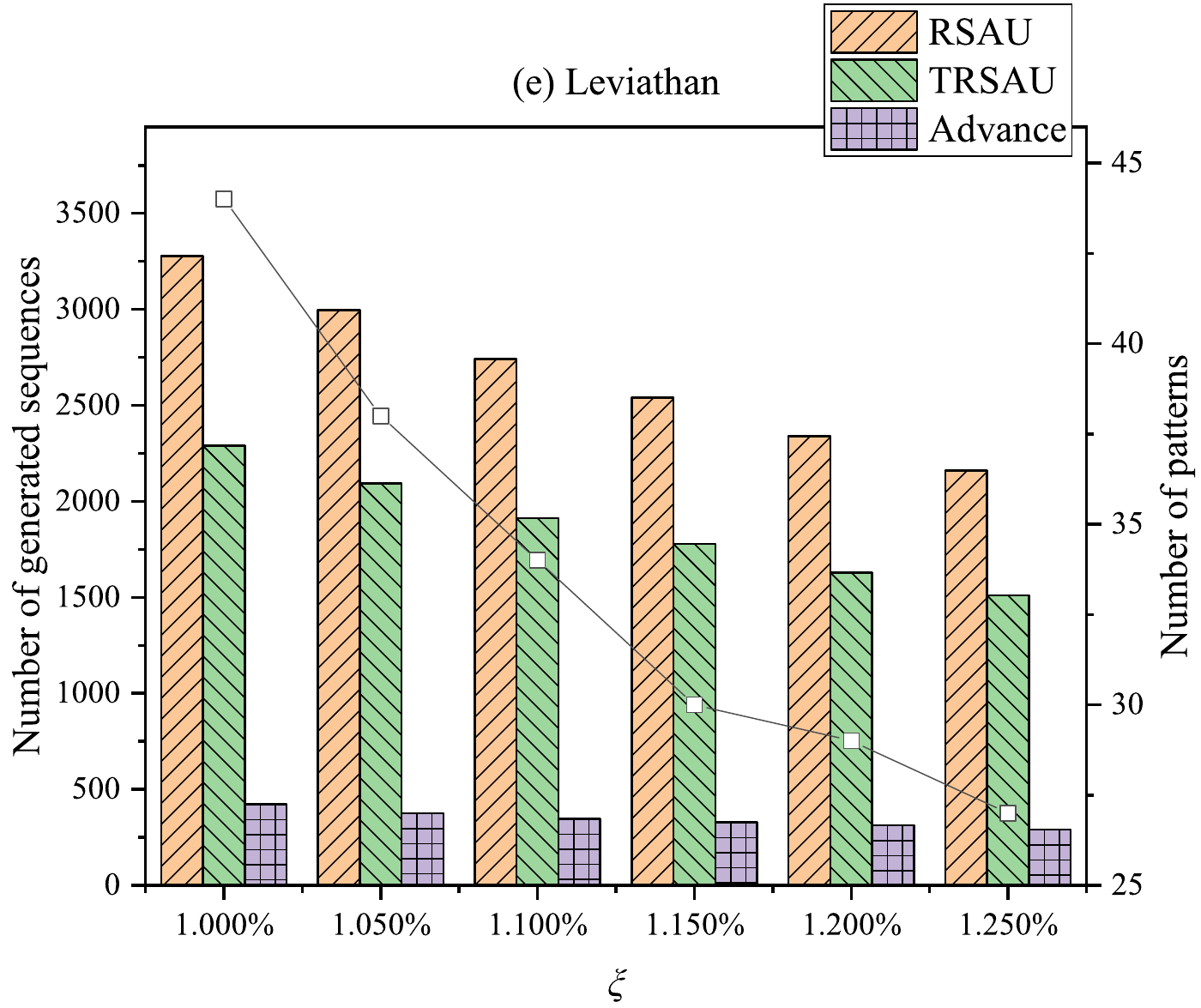}
			\label{fig: 10e}
			\includegraphics[width=0.31\linewidth]{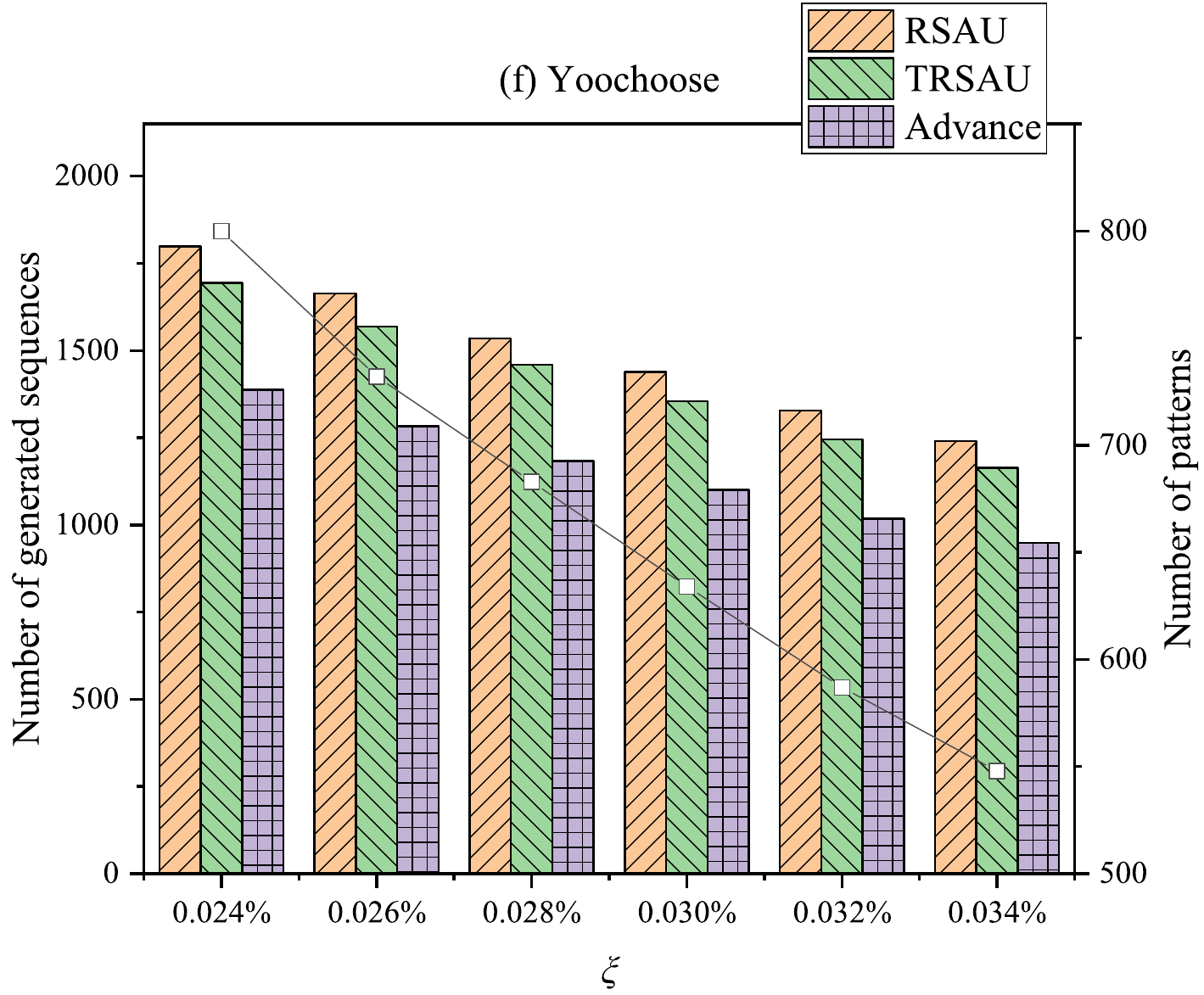}
			\label{fig: 10f}
	\end{minipage}
	\caption{Generated sequences for various versions of upper bounds.}
	\label{fig: 10}
\end{figure*}

As mentioned, the proposed algorithm could adopt \textit{TRSAU} to eliminate irrelevant items in the prefix sequence and prunes unpromising items whose utility does not surpass a threshold in the remaining sequence using the upper bound ${\textit{PEAU}}_{\textit{Rev}}$. The smaller number of generated sequences generated by Advance compared to TRSAU and RSAU in Fig. \ref{fig: 10} is chiefly responsible for the differentiation in results. The strategies could have different effects on diverse datasets. However, both approaches remain effective. In different datasets, the larger the dataset size and the more imbalanced the data distribution, the more sequences the novel algorithm will generate. For example, evident in Fig. \ref{fig: 10}(c) and Fig. \ref{fig: 10}(f).

\subsection{Strategy Performance and Memory Usage Evaluation} \label{sec: strategyMemory}

\begin{figure*}
	\centering
	\begin{minipage}{0.98\textwidth}
			\centering
			\includegraphics[width=0.31\linewidth]{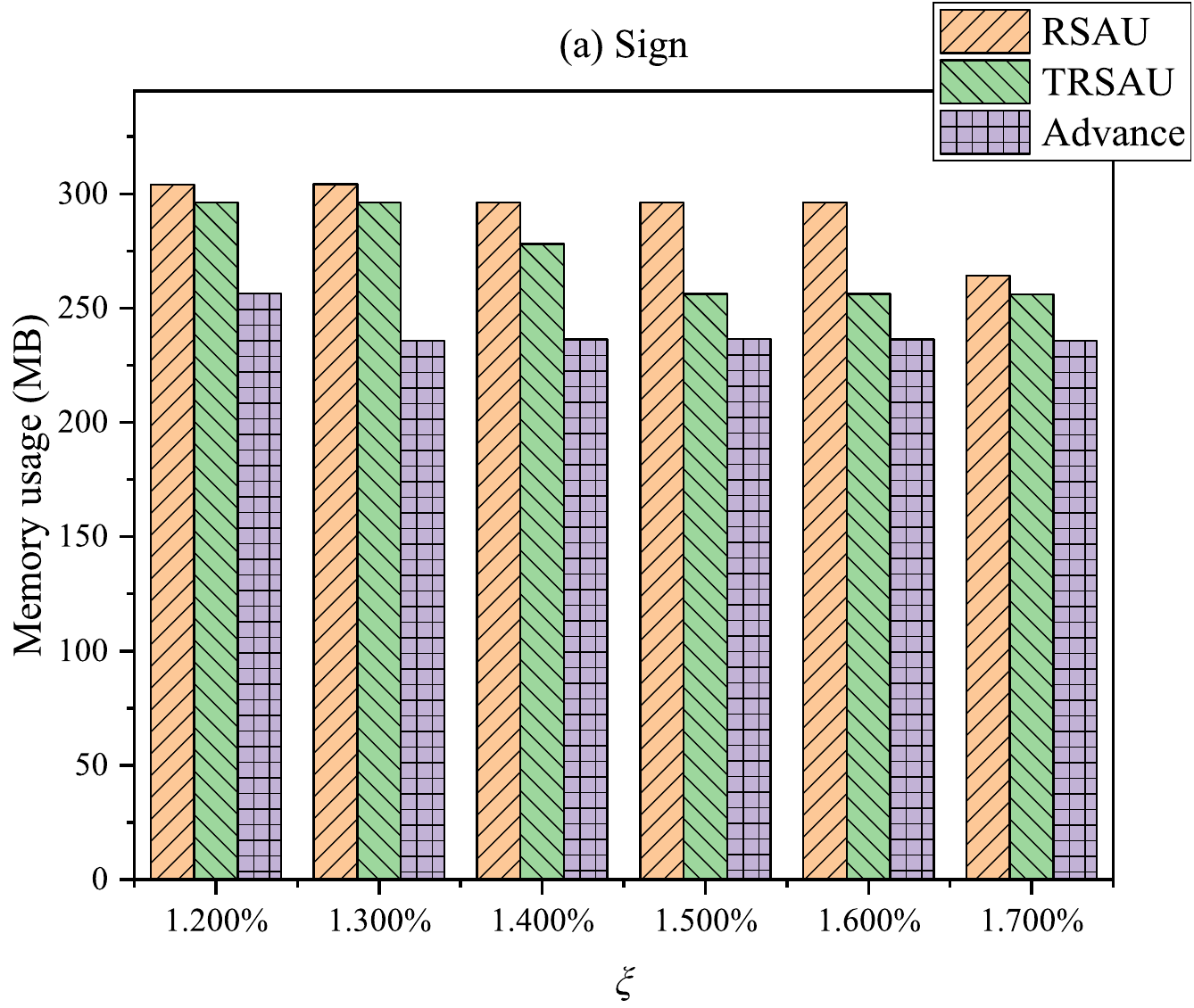}
			\label{fig: 11a}
			\includegraphics[width=0.31\linewidth]{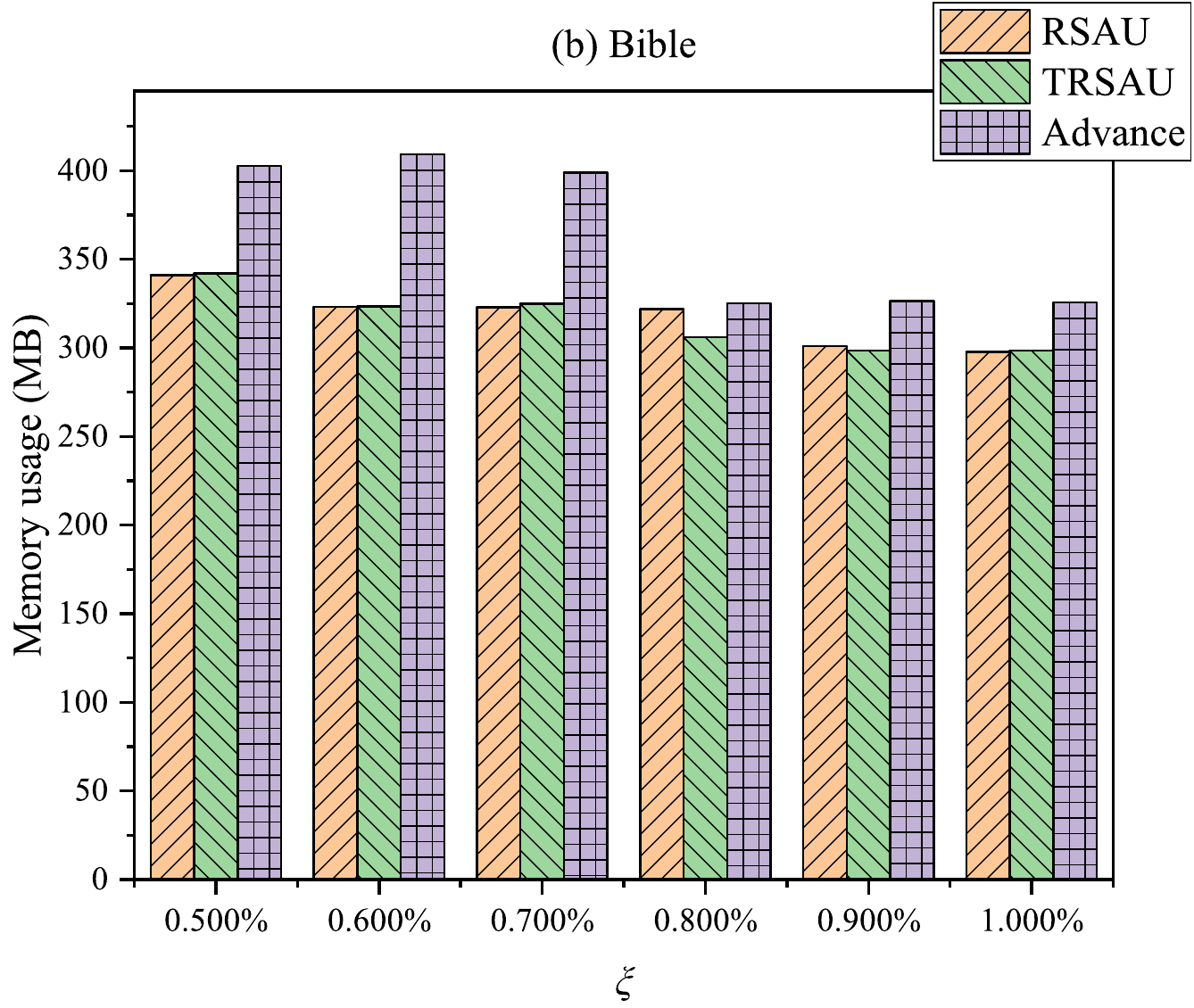}
			\label{fig: 11b}
			\includegraphics[width=0.31\linewidth]{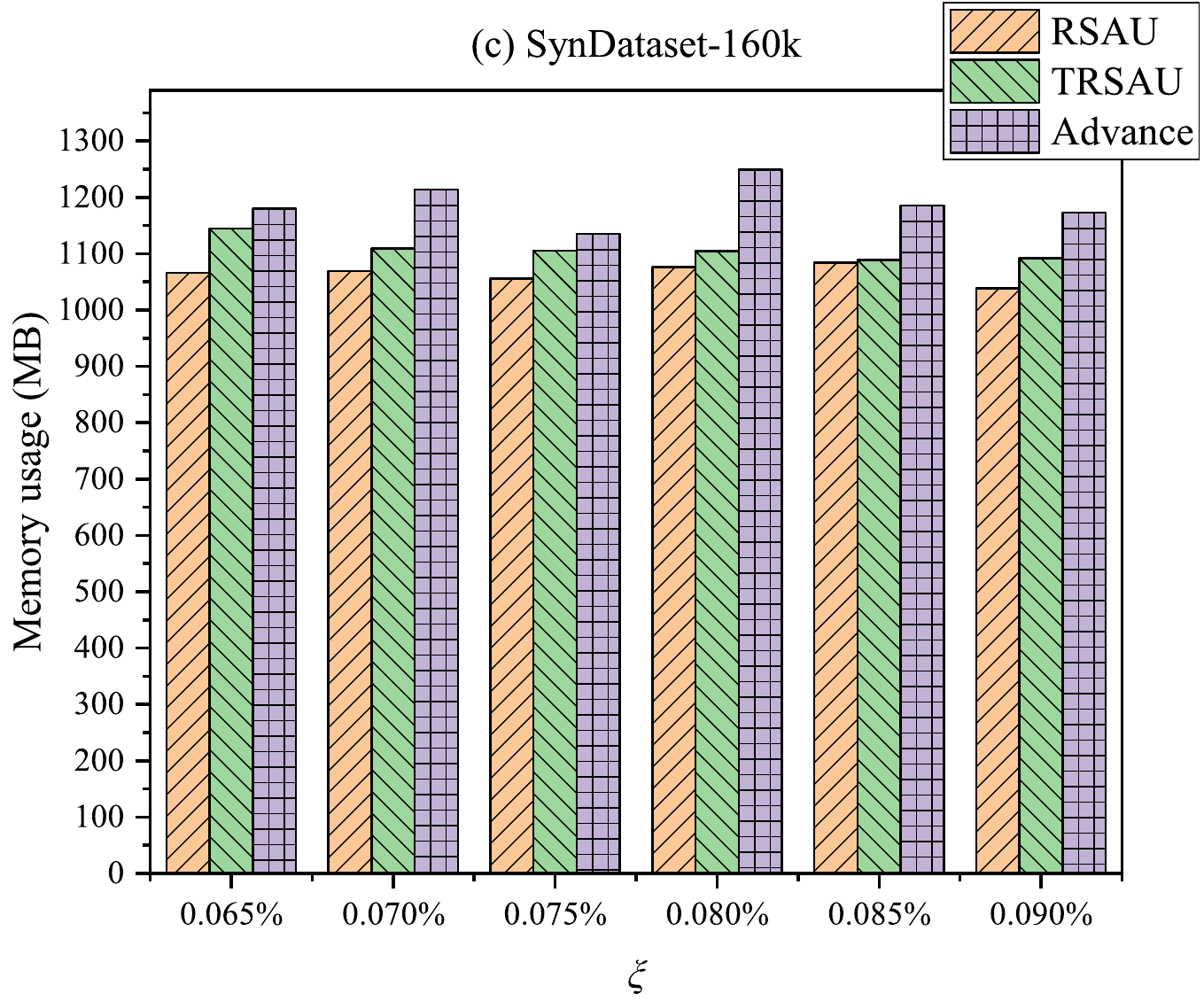}
			\label{fig: 11c}
	\end{minipage}
	\begin{minipage}{0.98\textwidth}
			\centering
			\includegraphics[width=0.31\linewidth]{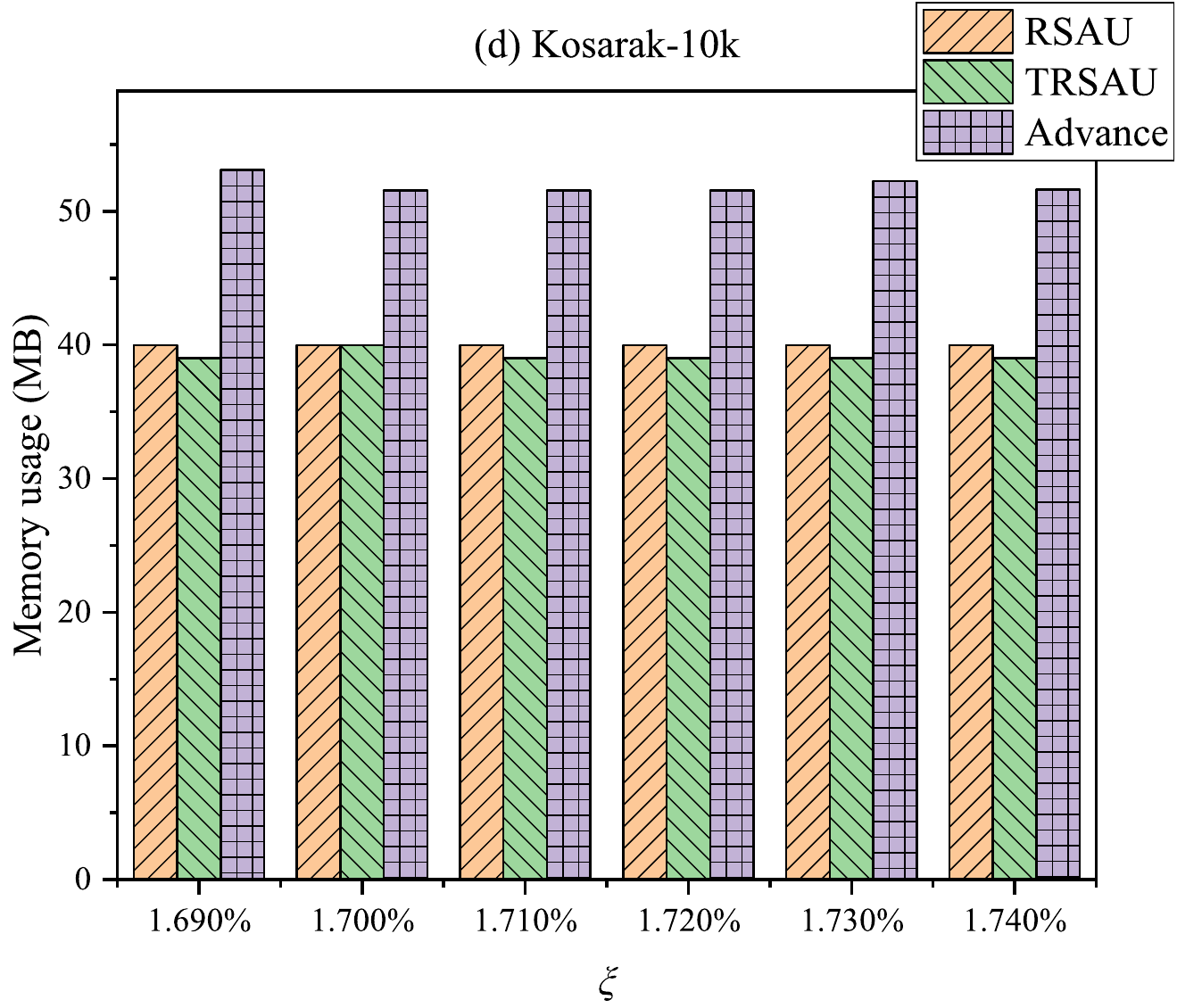}
			\label{fig: 11d}
			\includegraphics[width=0.31\linewidth]{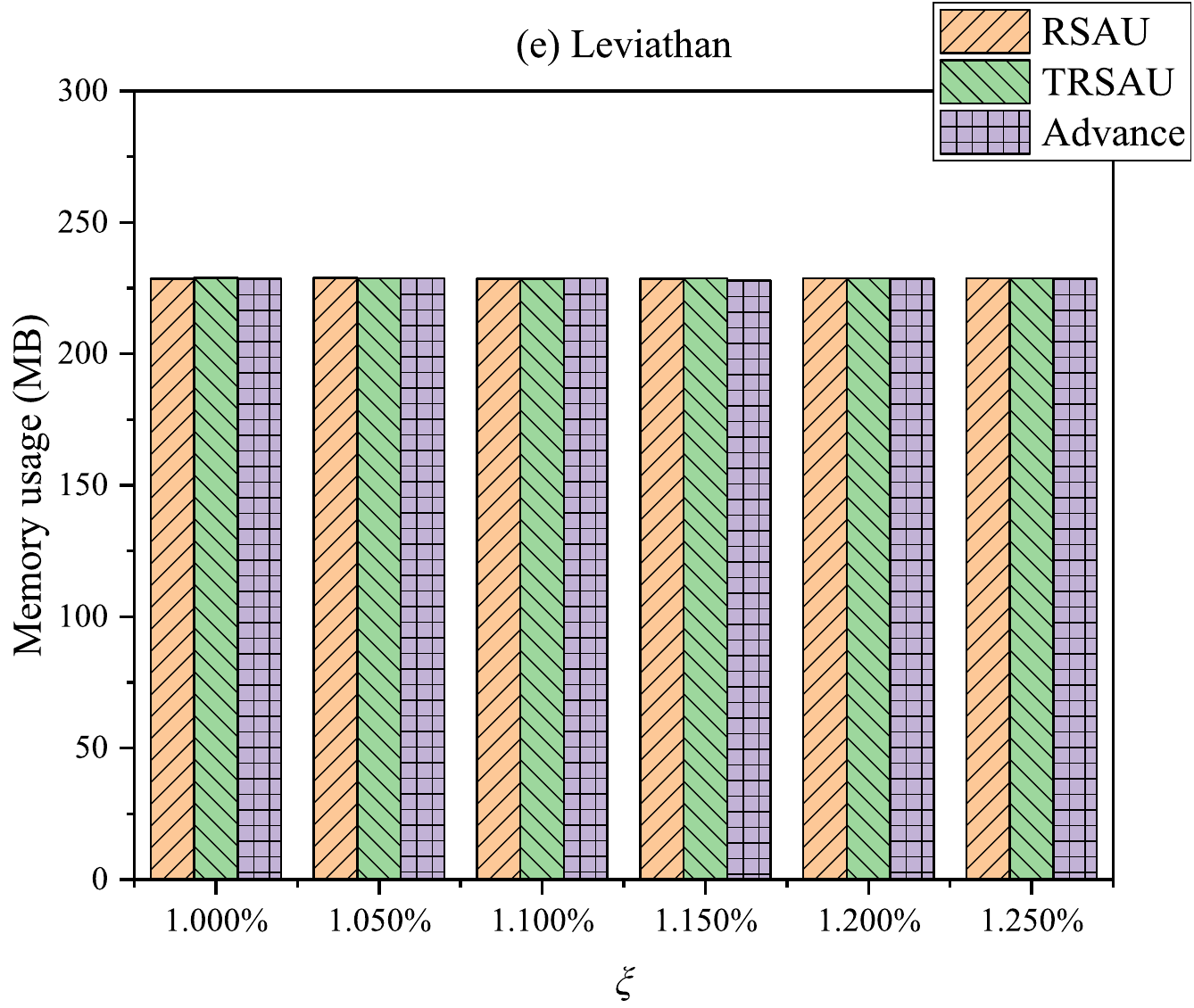}
			\label{fig: 11e}
			\includegraphics[width=0.31\linewidth]{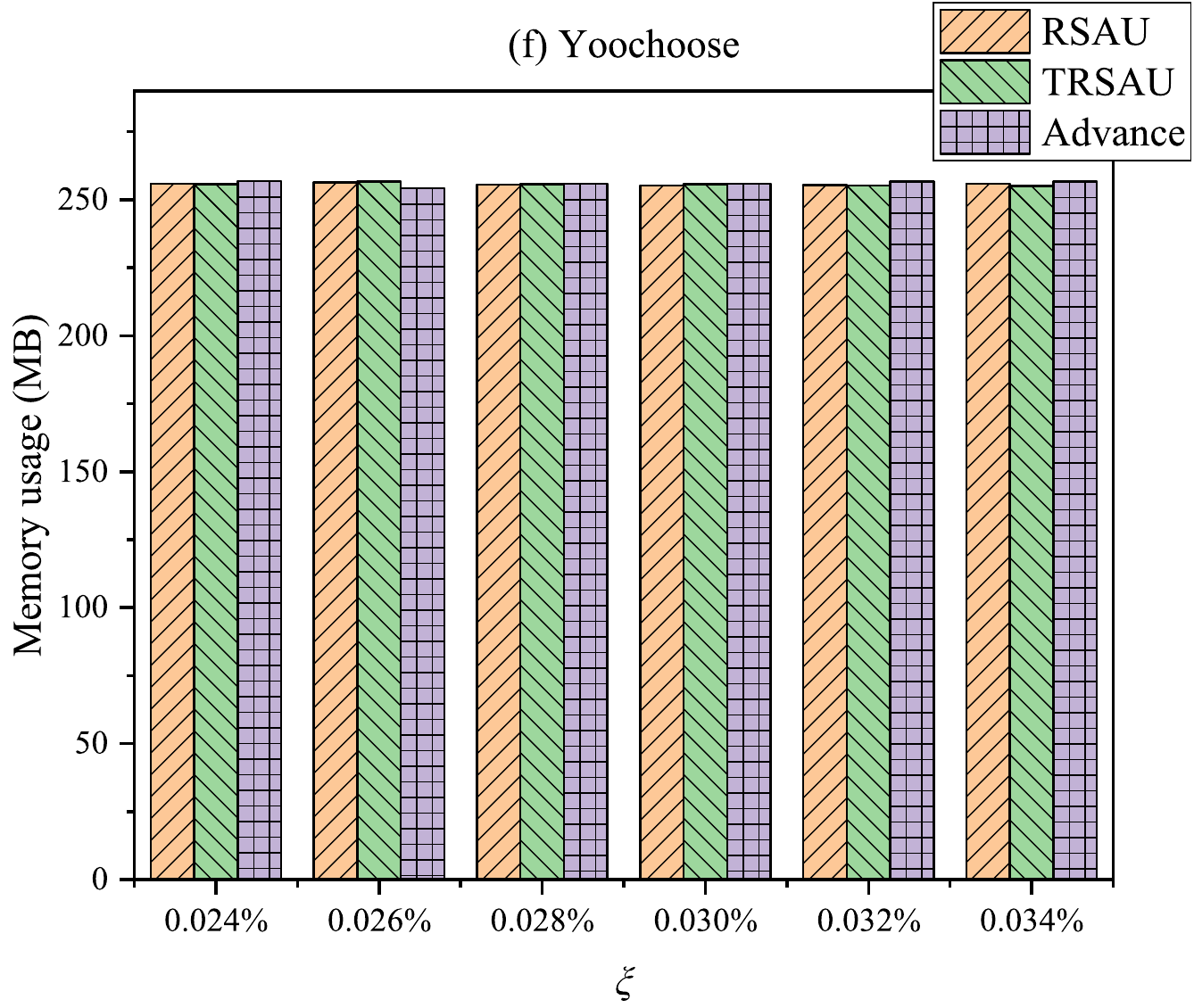}
			\label{fig: 11f}
	\end{minipage}
	\caption{Memory usage for various versions of upper bounds.}
	\label{fig: 11}
\end{figure*}

As shown in Fig. \ref{fig: 11}, in terms of memory usage, Advance does not absolutely outperform TRSAU and RSAU. Compact data structures store essential information, and all methods adopt the same projection mechanism and data structure. Notably, the improved strategies introduced in this study require updates to the existing data structures to support filtering operations. Memory consumption is typically correlated with the number of generated sequences and usually shows similar variation trends. While more complex strategies enable earlier pruning, they also involve additional necessary filtering operations and computational overhead. Hence, the differences in memory consumption among different methods are not significant in Fig. \ref{fig: 11}(b), Fig. \ref{fig: 11}(c), and Fig. \ref{fig: 11}(d). Even the obvious advantage in the number of generated sequences, as shown in Fig. \ref{fig: 10}(a), is weakened by the underlying computational overhead. Meanwhile, this also reflects the need to balance efficiency and strategy complexity according to specific tasks. In Fig. \ref{fig: 11}(f), the minimal difference in strategy efficiency is primarily due to the low \textit{AvgLen}, a characteristic of the dataset. The shorter sequence length restricts the strategy from achieving better performance.

\subsection{Scalability Test}
\label{sec: scalability}

\begin{figure}
	\centering
	\includegraphics[width=0.47\linewidth]{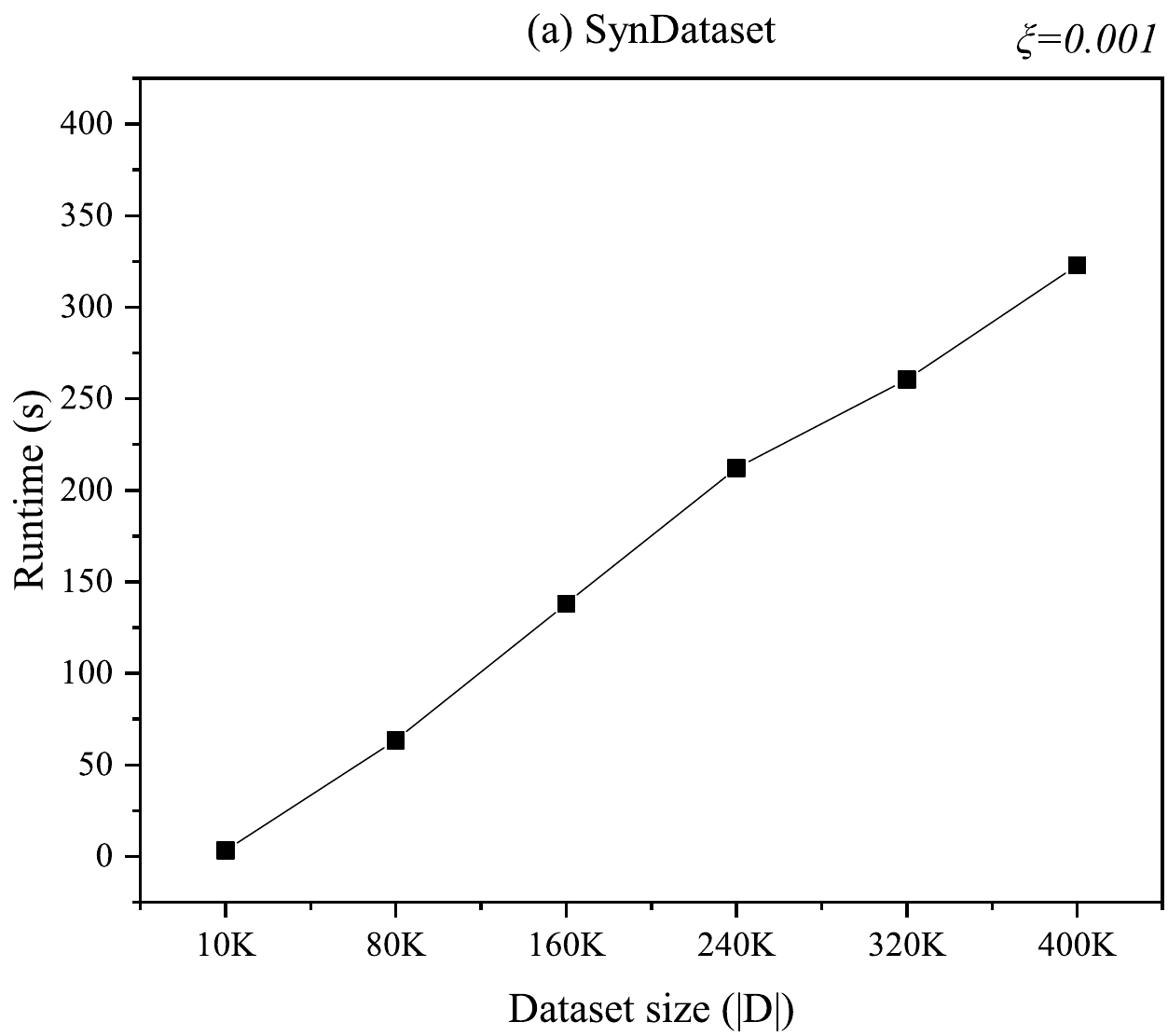}
	\label{fig: 12a}
	\includegraphics[width=0.47\linewidth]{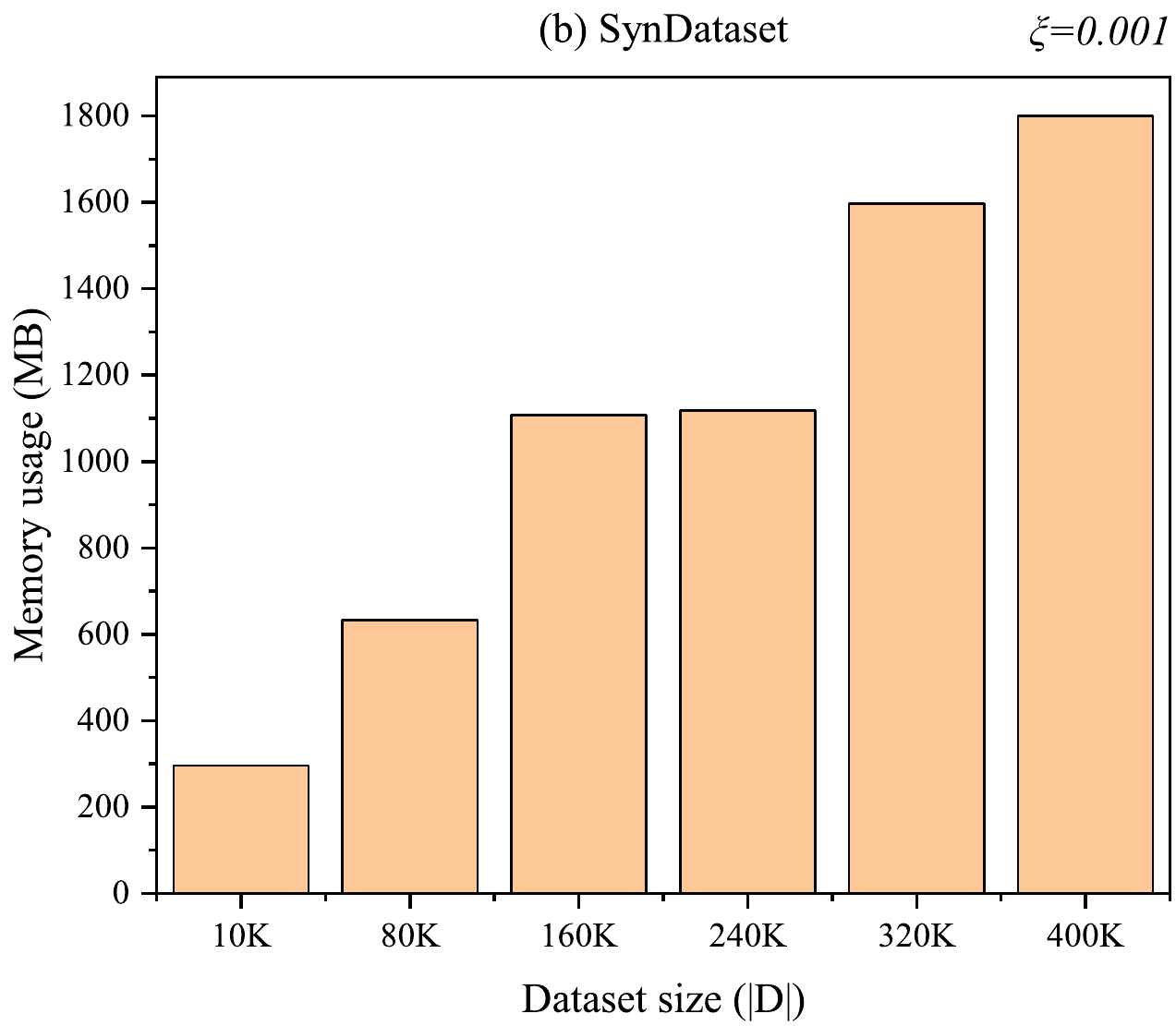}
	\label{fig: 12b}
	\caption{Scalability test.}
	\label{fig: 12}
\end{figure}

\begin{figure*}
	\centering
	\includegraphics[width=0.31\linewidth]{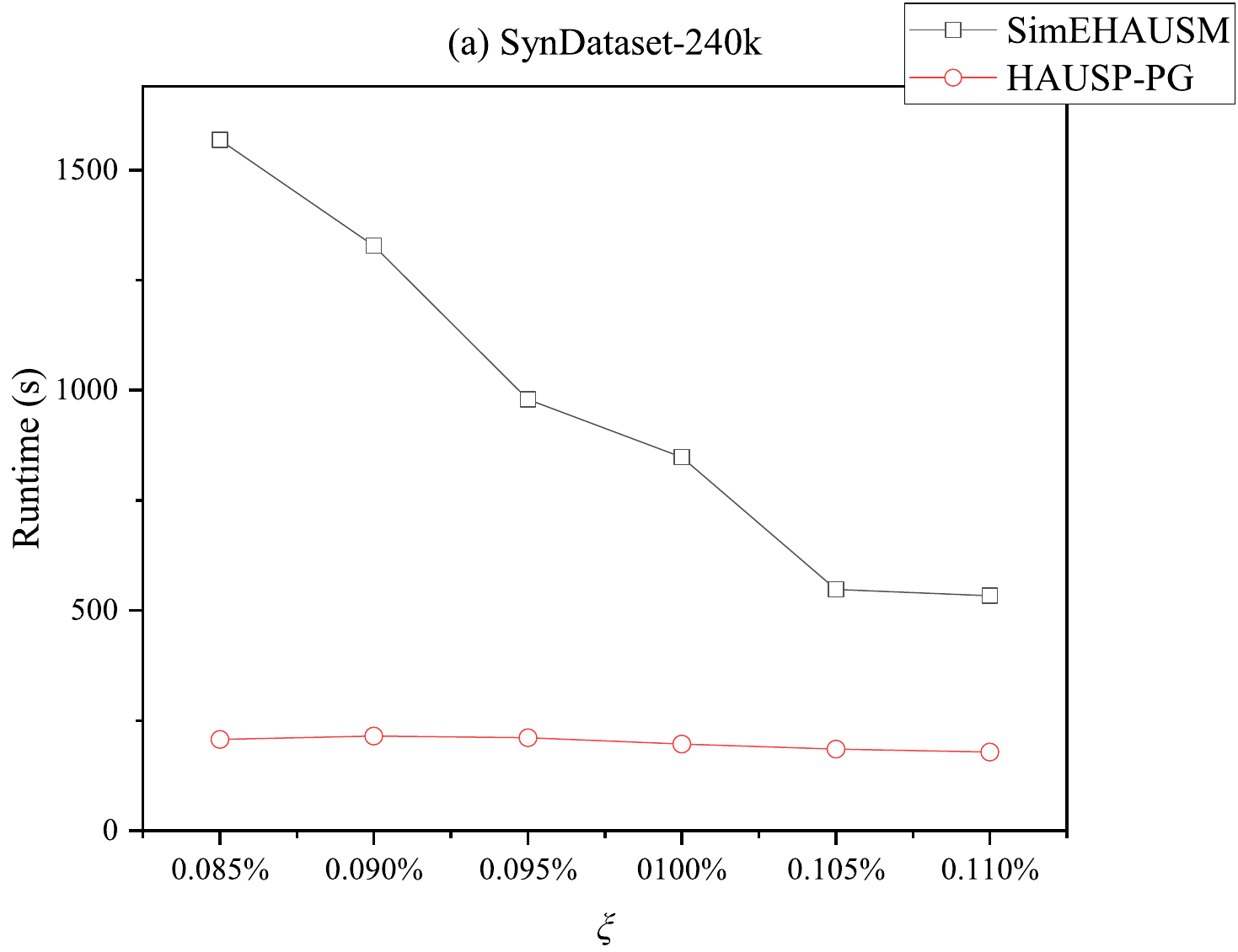}
	\label{fig: 13a}
	\includegraphics[width=0.31\linewidth]{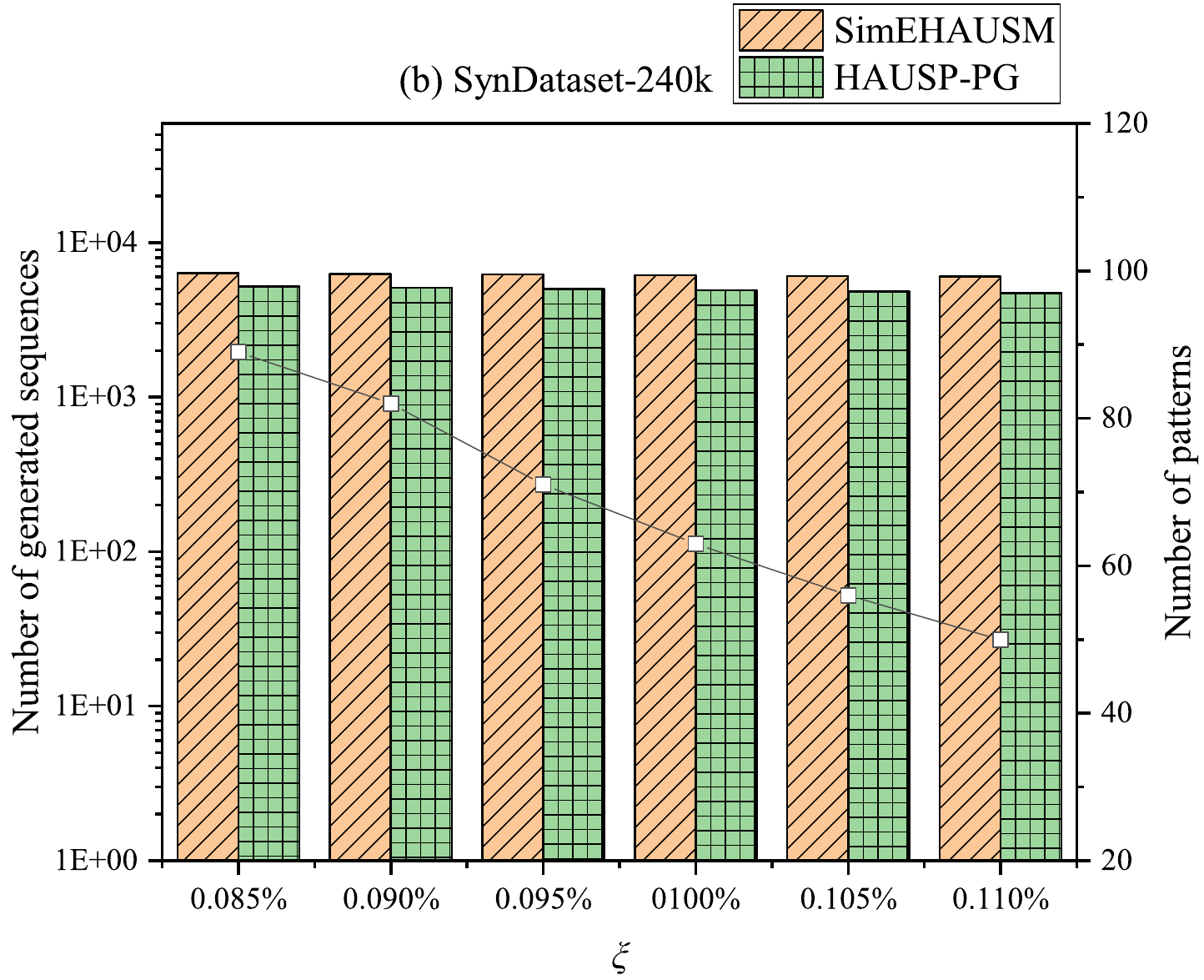}
	\label{fig: 13b}
	\includegraphics[width=0.31\linewidth]{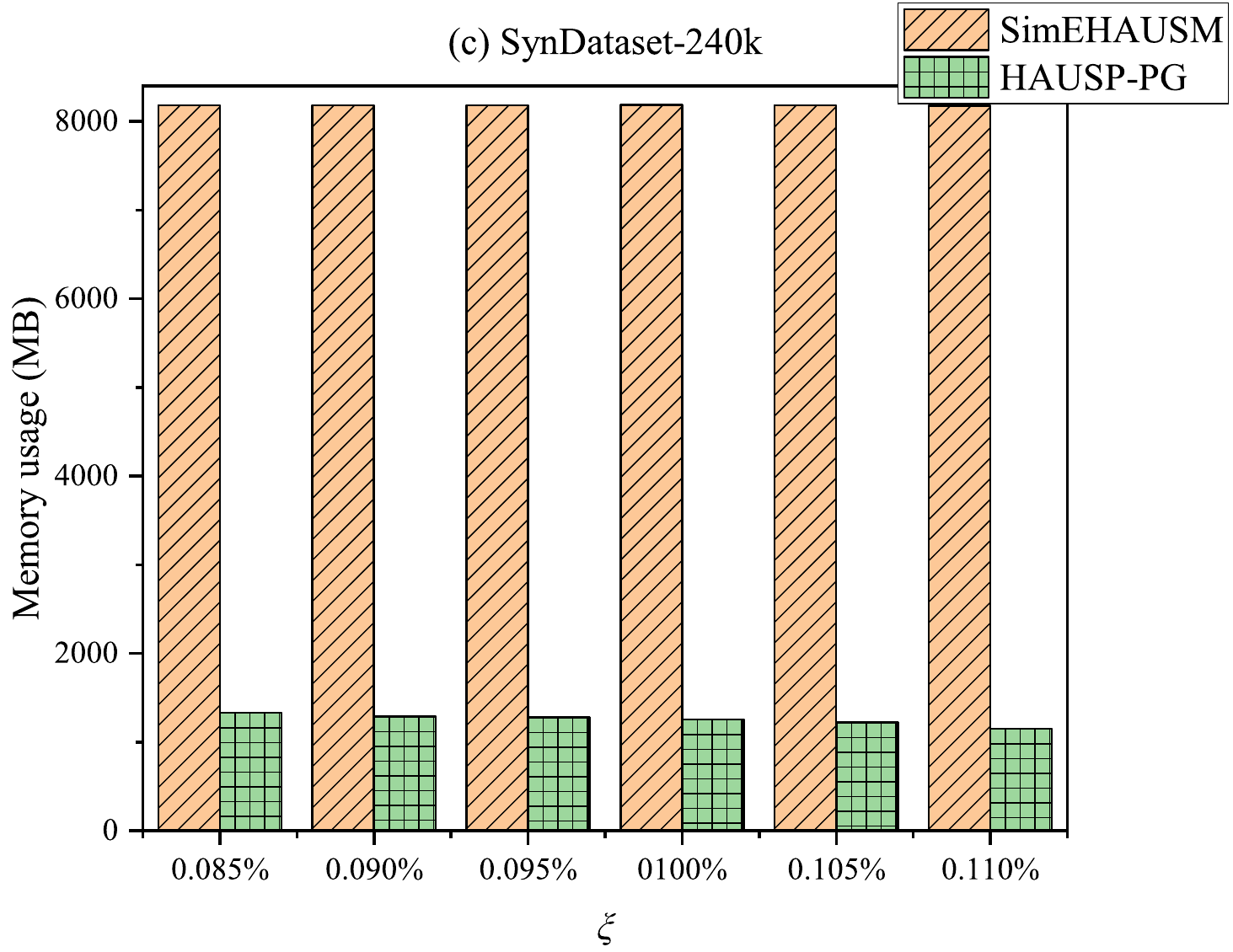}
	\label{fig: 13c}
	\caption{Performance comparison on 240K-scale dataset.}
	\label{fig: 13}
\end{figure*}

This subsection analyzes the scalability of the algorithm by experiment. Dataset sizes range from 10K to 400K by repeatedly combining duplicates of the dataset {SynDataset}\_{80K}. Note that the threshold $\xi$ is set to 0.001. Results in terms of runtime, generated sequences, and memory efficiency are shown in Fig. \ref{fig: 12}. As the number of dataset sequences increases, the corresponding memory usage and runtime display nearly linear growth. The number of generated sequences remains constant, except for {SynDataset}\_{10K}. The experiment results demonstrate that the proposed HAUSP-PG algorithm has excellent scalability. To further verify the performance in larger-scale data scenarios, we conducted additional comparative experiments on the 240K-scale dataset, comparing HAUSP-PG with the baseline algorithm SimEHAUSM. As shown in Fig. \ref{fig: 13}, the results indicate that even on this larger dataset, HAUSP-PG maintains advantages in runtime efficiency and memory usage compared to baseline methods, with the number of generated sequences being more effectively controlled. This further confirms the strong scalability of HAUSP-PG, validating its applicability in handling larger-scale data scenarios. As $\xi$ increases from 0.085\% to 0.110\%, SimEHAUSM's runtime surges exponentially, while HAUSP-PG maintains stably low latency. This widening gap confirms HAUSP-PG's optimized computational overhead for large datasets. The proposed HAUSP-PG generates far fewer sequences, and its sequence count remains nearly flat across $\xi$ variations. Its memory footprint is less than 1/5 of SimEHAUSM's peak usage, reflecting space-efficient design. These results collectively confirm HAUSP-PG has strong scalability, validating its applicability to large real-world scenarios.

\section{Conclusion} \label{sec: conclusion}

While HUSPM has taken into consideration frequent measures and utility measures, it remains limited in terms of sequence length, especially when sequences contain many low-utility items. HUSPM does not treat long sequences and short sequences fairly. The goal of the proposed algorithm is to address this issue. In contrast to the utility of each item in a sequence, HUSPM merely places more emphasis on the overall utility of a sequence. The HAUSPM method designed in this study takes a different approach from previous work. It introduces an efficient HUSPM algorithm and integrates the characteristics of average utility mining to formulate and optimize. The proposed algorithm adopts the pattern-growth method, utilizes novel upper bounds, and designs variants of the upper bounds to optimize algorithm performance. Based on the characteristic of average utility, a new measurement for the remaining sequence is proposed herein. The comparative experiment demonstrates certain advantages of the algorithm HAUSP-PG in terms of runtime, number of generated sequences, memory usage, and scalability. Currently, this proposed algorithm cannot guarantee uniform superiority across all datasets, and further refinement and optimization of the algorithm constitute promising directions for future work, aiming to enhance its performance and robustness across an even wider range of real-world applications.

\bibliographystyle{cas-model2-names}
\bibliography{eaai}

\end{document}